\documentclass[aps,prd,11pt, preprint,nofootinbib,showkeys]{revtex4-2}
\usepackage{amsmath,amssymb,amsthm,mathrsfs}
\usepackage{latexsym,amscd,amsbsy,amsfonts}
\usepackage{graphicx}
\usepackage{ragged2e}
\usepackage{caption}
\usepackage[utf8]{inputenc}
\usepackage{marvosym}
\usepackage{float}
\usepackage{slashed}
\usepackage{cancel}
\usepackage{multirow}
\usepackage{soul}
\usepackage{physics}
\usepackage{comment}
\usepackage[usenames,dvipsnames]{xcolor}
\usepackage[
      colorlinks=true,
      linktocpage=true,
      breaklinks=true,
      linkcolor=Aquamarine,
      urlcolor=Purple,
      filecolor=black,
      citecolor=Purple,
      menucolor=black,
      pdfstartview=FitV,
      bookmarksopen=true
      ]{hyperref}
\usepackage{tikz}

\makeindex
\begin{document}
%--------------------------------------------------------------------

\title{Robustness of analogue Hawking radiation in cavities with moving boundaries}

\author{Alberto Garc\'{i}a Mart\'{i}n-Caro}
\email{alberto.garcia.martin-caro@uvigo.es}
\affiliation{EHU Quantum Center and Department of Physics, University of the Basque Country UPV/EHU, 48940, Bilbao, Spain}
\affiliation{Instituto de Física e Ciencias Aeroespaciais (IFCAE), University of Vigo, Ourense E-32004, Spain}

\author{Javier Olmedo}
\email{javolmedo@ugr.es}
\affiliation{Departamento de F\'isica Te\'orica y del Cosmos, Universidad de Granada, Granada-18071, Spain}

\author{Jose M. S\'anchez Vel\'azquez}
\email{jmsvelazquez@ceiec.es}
\affiliation{CEIEC, Universidad Francisco de Vitoria, Pozuelo, 28223, Madrid, Spain}

%------------------------------------------------
\begin{abstract}
In this work we explore the limitations and robustness of thermal radiation in dynamical Casimir systems serving as analogs for Hawking radiation. Through detailed numerical analysis, we characterize particle production spectra in cavities with moving boundaries under various configurations, including expanding, collapsing, and rigidly accelerating scenarios. We find that thermal signatures emerge in specific expanding cavity configurations but are highly dependent on frequency bands and acceleration parameters. In those configurations of the cavity where there is thermal production, we derive fitting expressions that quantify deviations from idealized thermal spectra through gray-body factors, revealing oscillatory behaviors tied to acceleration duration. Our results identify which experimental setups can reliably simulate gravitationally-induced phenomena and quantify how finite-size effects and transient dynamics modify the expected thermal distributions, providing a comprehensive framework for distinguishing genuine Hawking-like radiation from experimental artifacts.
\end{abstract}
%------------------------------------------------

%------------------------------------------------
%------------------------------------------------

\maketitle

{

\hypersetup{citecolor=black, linkcolor=black,urlcolor=black}

\tableofcontents

}

%------------------------------------------------
\section{Introduction}
\label{sec:introduction}
%------------------------------------------------

%A thermal state in quantum field theories can arise from different scenarios when any interaction with a classical environment or additional quantum fields is considered. This phenomenon manifests itself as either due to phase transitions in the context of condensed matter or due to the presence of horizons in the subjacent spacetime when curved geometries are considered. In this latter scenario, the distinctive thermal structure of Hawking \cite{Hawking1975} and Unruh \cite{Unruh1976} radiation is well-established in the literature, providing fundamental connections between gravity, thermodynamics, and quantum theory.

Hawking radiation stands as one of the most studied theoretical predictions of quantum field theory in curved spacetimes. It arises from quantum fluctuations in the vacuum. Pairs of particle-antiparticle are generated in the vicinity of the horizon, one falling into the black hole and the other escaping towards infinity. The latter is found to be in a thermal (mixed) state described entirely by the mass, charge, and angular momentum of the black hole, and nothing else. This phenomenon (production of particles following a thermal distribution) was shown for the first time in the presence of black hole horizons \cite{Hawking1975}, and for observers following accelerated trajectories measuring quantum fields with Unruh detectors \cite{Unruh1976}. The distinctive thermal structure of the state of a quantum field on a classical curved background provides fundamental connections between gravity, thermodynamics, and quantum theory.

These studies kick-started research on potential experimental proposals, in analogue systems, to study gravitational-like particle production. There is a well-known analogue setting, proposed back in the days, where a quantum field in a vacuum state can be driven into a thermal state at late times: an accelerating mirror (boundary condition) \cite{Davies1977, Birrell1982}. However, its experimental realization has remained elusive so far. More recent examples of analogue systems dedicated to the study of Hawking-like radiation are Bose-Einstein condensates~\cite{Garay1999,Garay2000,Barcelo2001,Steinhauer2015,Nova:2019,Kolobov2019,Leonhardt2016,Steinhauer2018}; surface waves in water flows~\cite{Schutzhold2002,Rousseaux2007,Weinfurtner2010,Weinfurtner2013,Michel2014,Coutant2016,Euve2014,Euve2015,Euve2018}; and also nonlinear optics systems~\cite{Schutzhold:2004tv,Philbin:2007ji,Belgiorno2010,Rubino2011,Drori2018}; and quantum fluids of light~\cite{Carusotto2004,Nguyen2015,Falque:2023ctx}. 

Despite the success in both the theoretical and experimental aspects of all previous settings, we are interested in analogue systems based on flat spacetimes with moving boundaries, in the lines of Davies and Fulling \cite{Davies1977}. These systems also serve to study the dynamical Casimir effect \cite{Moore1970,Dodonov:2025rxz}. Besides, they can mimic some of the properties of black hole spacetimes \cite{Carlitz1987}, thereby establishing a fundamental correspondence between these (in principle very different) physical systems and their exploration in controlled laboratory settings. We also want to note that they have been considered for the study of relativistic quantum information ~\cite{Friis:2012tb,Alsing_2012,Bruschi:2012pd,Lindkvist:2013dha,Sabin:2015pwa,Romualdo:2019eur,DelGrosso:2020rbi}, entanglement harvesting \cite{Cong:2018vqx,Perche:2023nde} and potentially as Unruh detectors \cite{Ragula:2025mte}. Significant progress has been made in recent years in the implementation of the dynamical Casimir effect using optical cavities \cite{Chen2017} and superconducting circuits \cite{Wilson2011}, offering promising platforms for testing quantum field theory predictions in curved spacetimes. Indeed, it has been recently demonstrated that superconducting circuits also provide a feasible experimental setup for studying both classical and quantum black holes \cite{Terrones:2021bjp,Maceda:2025jti}. Besides, in the last few years, we have revisited this setting in Ref. \cite{GarciaMartin-Caro:2024qpk}, exploring the parameter space compatible with these experimental proposals. 

What is more interesting, as shown in Ref. \cite{GarciaMartin-Caro:2023jjq}, is that this setting of moving boundaries can actually produce a thermal spectrum of particles with a temperature determined by the acceleration of the boundary. However, due to finite-size effects and transient regimes inherent to these experimental systems, the obtained production spectrum for the field under study is not exactly thermal. This is in fact not surprising when working in realistic scenarios. The discrepancies between the idealized theoretical models and the realistic experimental configurations introduce significant challenges in identifying the genuine signatures of thermal radiation. As has already been studied \cite{Good:2013lca}, the radiation spectra of moving mirrors can exhibit deviations from perfect thermality depending on the specific trajectories and boundary conditions. Unfortunately, the number of trajectories in which one obtains closed-form expressions, even under certain approximations, is very limited. So, a more detailed analysis beyond the standard approximations and the unavoidable use of numerical tools seems to be needed.

In this work, we conduct a systematic numerical investigation of particle production in cavities with moving boundaries for various configurations, focusing specifically on the robustness of thermal radiation signatures. By developing improved fitting expressions that capture both the thermal and non-thermal components of the particle spectra, we provide a framework for distinguishing intrinsic features of Hawking-like radiation from spurious creation effects attributable to the actual dynamics of the mirrors. Our approach combines high-precision numerical simulations with analytical approximations to characterize the spectrum in expanding and collapsing cavity scenarios, with particular attention to experimentally relevant parameter regimes \cite{Lahteenmaki2013, Nation2012}.

The significance of this distinction extends beyond theoretical interest, as it bears implications for experimental verification of fundamental predictions in quantum field theory in curved spacetimes. Furthermore, precise characterization of deviations from perfectly thermal spectra may provide valuable insights into the robustness and limitations of analogies between condensed matter systems and gravitational phenomena, thereby contributing to the development of more accurate quantum simulators for the study of quantum gravity effects.

This manuscript is organized as follows. Sec.~\ref{sec:qft} is devoted to the description of the formalism to quantize a field in flat spacetime with moving boundaries and the equivalence between this setup and the theory in an effective acoustic metric with fixed boundaries. In Sec.~\ref{Sec:expanding-cav} we show our numerical results for boundaries that expand, for several configurations and following different trajectories. We also propose a set of fitting functions that approximate the numerical results and that are useful for discerning the spurious effects on the thermal spectra. In Sec.~\ref{Sec:collap-cav} we present a detailed numerical study of different scenarios for several collapsing configurations and trajectories of the boundaries, and in Sec. \ref{Sec:expand-collap-cav} we discuss the effects of concatenating these expanding and collapsing trajectories in the resulting spectrum. Finally, in Sec.~\ref{Sec:conclusions} we present the main conclusions of this work.

%------------------------------------------------
\section{Quantum fields with nonstationary boundaries}
\label{sec:qft}
%------------------------------------------------
We will start this section motivating our study of Hawking radiation with the analogue model provided originally by Davies and Fulling \cite{Davies1977} for a massless scalar field in a 1+1 spacetime with one Dirichlet boundary condition $\phi(t,x=f(t))=0$. Similar models, with more realistic trajectories, where the mirror becomes inertial at late times, yield similar results regarding the production of a thermal spectrum. Here, one typically assumes that $f(t)=0$ for $t<0$, $0\neq|\dot f(t)|<1$ for $t>0$ and $t<t_i$, and finally the mirror remains inertial with speed ${\rm const}=|\dot f(t)|\simeq 1$ for $t>t_i$. For concreteness, the trajectories typically take the form
\begin{equation}
    f(t) = -t -\frac{1}{2\kappa} e^{-2\kappa t}+\frac{1}{2\kappa},
\end{equation}
for $0<t<t_i$, with $\kappa$ some positive constant, and 
\begin{equation}
    f(t) = -(1-e^{-2\kappa t_i})(t-t_i)-t_i -\frac{1}{2\kappa} e^{-2\kappa t_i}+\frac{1}{2\kappa}
\end{equation} for $t>t_i$. These scenarios have been studied in detail, for example in \cite{Haro_2005,Haro:2007jr}. After some approximations, in the limit $\omega'\gg\omega$, with $\omega'$ and $\omega$ the $in$ and $out$ frequencies, and for $t_i\gg \kappa^{-1}$, the (moduli of the) Bogoliubov coefficients typically satisfy the celebrated \emph{detailed balance condition}:
\begin{equation}\label{eq:BD-thermal}
    |\alpha_{\omega',\omega}|^2=e^{2\pi \omega/\kappa}|\beta_{\omega',\omega}|^2\quad {\rm with}\quad|\beta_{\omega',\omega}|^2\simeq \frac{1}{2\pi \kappa \omega'}\frac{1}{e^{2\pi \omega/\kappa}-1}.
\end{equation}
This condition is usually considered a hallmark of thermal equilibrium  and guarantees that the rate of particle emission and absorption are related as in a thermal ensemble, ensuring that the created particle distribution is consistent with a thermal state in equilibrium at a temperature $T = \kappa/2\pi$. \footnote{More technically, the detailed balance condition \eqref{eq:BD-thermal} is usually presented as an equivalent relation for the (Fourier transformed) Wightman function, which implies the Kubo-Martin-Schwinger \cite{Kubo:1957mj,PhysRev.115.1342} condition of periodicity in imaginary time of the two point function for a thermal state \cite{Takagi:1986kn}, although the equivalence is lost in the case of non-thermal states, see \cite{RLEP}.}

In our study, we will focus our attention on a Klein-Gordon (massless) scalar field $\phi$ in a one-dimensional cavity represented by two Dirichlet boundary conditions $\phi(t,x=f(t))=\phi(t,x=g(t))=0$. A comprehensive and detailed analysis can be found in Refs. \cite{GarciaMartin-Caro:2023jjq,GarciaMartin-Caro:2024qpk}. The functions $f(t)$ and $g(t)$ represent the prescribed trajectories of the left and right boundaries, respectively. We will assume that the field propagates at the speed of light, which, as customary, we will set in what follows to $c=1$. Then, the field $\phi(t)$ admits a natural decomposition in modes given by
\begin{equation}\label{eq:fourier}
 \phi(t,x)= \sum_{n=1}^{\infty} \phi_n(t)\sin \left[\frac{n \pi}{g(t)-f(t)} \Big(x-f(t)\Big)\right].
\end{equation}
The dynamics of the Fourier modes of this Klein-Gordon field is determined by the Hamiltonian 
\begin{align}\nonumber
 H_T=&\sum_{n=1}^{\infty}\frac{1}{4 L}\left[\pi_{n}^{2}+(n \pi)^{2} \phi_{n}^{2}\right]-\frac{\dot L}{4 L} \pi_{n} \phi_{n}\\&+\sum_{m}\left(1-\delta_{n m}\right)\frac{n m}{n^{2}-m^{2}}\left[\frac{\dot f}{L}\left((-1)^{n+m}-1\right)+\frac{\dot L}{L}(-1)^{n+m}\right]  \pi_{m} \phi_{n},
\end{align}
where $L(t)=g(t)-f(t)$ is the length of the cavity, and 
\begin{align}
& \pi_{n}={L}\dot \phi_{n}+\frac{\dot L}{2} \phi_{n} -2 \sum_{m}\left(1-\delta_{m n}\right)\frac{m n}{m^{2}-n^{2}}\left[{\dot f}\left((-1)^{m+n}-1\right)+{\dot L}(-1)^{m+n}\right]  \phi_{m},
\end{align}
are the conjugate momenta of the Fourier modes $\phi_n$. They satisfy the Poison algebra
%------------------------------------------------
\begin{equation}\label{eq:fourier-poisson}
\left\{\phi_{n}, \pi_{m} \right\}=2 \delta_{n m}.
\end{equation}
%------------------------------------------------
The equations of motion can be computed via $\dot \phi_{n}=\left\{\phi_{n}, H_T \right\}$ and $\dot{\pi}_{n}=\left\{\pi_{n}, H_T \right\}$. We easily obtain
\begin{align}
& \dot \phi_{n}=\frac{1}{L} \pi_{n}-\frac{\dot L}{2 L} \phi_{n} +2 \sum_{m}\left(1-\delta_{m n}\right)\frac{m n}{m^{2}-n^{2}}\left[\frac{\dot f}{L}\left((-1)^{m+n}-1\right)+\frac{\dot L}{L}(-1)^{m+n}\right]  \phi_{m},\\
&\dot{\pi}_{n}=-\frac{1}{L}(n \pi)^{2} \phi_{n}+\frac{\dot L}{2 L} \pi_{n} +2 \sum_{m}\left(1-\delta_{n m}\right)\frac{n m}{m^{2}-n^{2}}\label{eq:eom}
 \left[\frac{\dot f}{L}\left((-1)^{n+m}-1\right)+\frac{\dot L}{L}(-1)^{n+m}\right]  \pi_{m}.
\end{align}
In the cases where the cavity is neither moving nor expanding, namely $\dot f(t)=0$ and $\dot L = 0$, the equations of motion reduce to a collection of decoupled harmonic oscillators. However, in general, whenever $\dot f(t)\neq 0$ and/or $\dot L \neq 0$, there will be couplings between different modes. 

In the complexified space of solutions $\cal{S}^{\mathbb{C}}$ there is a natural Klein-Gordon product that is preserved by evolution. Concretely, let us define a solution as 
\begin{eqnarray}\nonumber
    &&{\bf U}(t) = \Big(U_{1,\epsilon_1}(t),U_{2,\epsilon_2}(t), \ldots\Big),
\end{eqnarray}
such that $U_{n,\epsilon_n}(t)=\Big(\phi_n(t),\pi_n(t)\Big)$ and with $U_{n,\epsilon_n=0}(t)=\phi_n(t)$ and $U_{n,\epsilon_n=1}(t)=\pi_n(t)$. Hence, the Klein-Gordon product $\cal{S}^{\mathbb{C}} \times \cal{S}^{\mathbb{C}} \rightarrow \mathbb{C}$ is defined as the map 
%------------------------------------------------
\begin{align}\nonumber
&\langle {\bf U}^{(1)}(t), {\bf U}^{(2)}(t)\rangle=\frac{i}{2} \sum_{n=1}^{\infty} \bar{\phi}_{n}^{(1)}(t) \pi_{n}^{(2)}(t)-\bar{\pi}_{n}^{(1)}(t) \phi_{n}^{(2)}(t),
\end{align}
%------------------------------------------------
with ${\bf U}^{(1)}(t)$ and ${\bf U}^{(2)}(t)$ two solutions in $\cal{S}^{\mathbb{C}}$. It is important to remark that it is independent of $t$ (on solutions).

This inner product allows us to construct a basis of orthonormal solutions $\left({\bf u}^{(I)}, \bar{{\bf u}}^{(I)}(t)\right)$, with $I=1,2,\ldots$. Concretely, we can always find solutions such that 

%------------------------------------------------
\begin{equation}\label{eq:basis}
    \langle {\bf u}^{(I)}(t), {\bf u}^{\left(J\right)}(t)\rangle=\delta^{I J}, \;\langle {\bf u}^{(I)}(t), \bar{{\bf u}}^{\left(J\right)}(t)\rangle=0, \;\langle\bar{{\bf u}}^{(I)}(t), \bar{{\bf u}}^{\left(J\right)}(t)\rangle=-\delta^{I J},
\end{equation}
for all $t$. Here, ${\bf u}^{(I)}(t)$ is said to represent a positive frequency solution, while $\bar {\bf u}^{(I)}(t)$ is a negative frequency solution. 

In addition, given another basis of solutions $\left({\bf w}^{(I)}, \bar{{\bf w}}^{(I)}(t)\right)$, they are related by the well-known Bogoliubov transformation
%------------------------------------------------
\begin{equation}\label{eq:bogou}
    {\bf u}^{(I)}(t) = \sum_{J=1}^{\infty} \alpha_{IJ}{\bf w}^{(J)}(t)+\beta_{IJ}\bar {\bf w}^{(J)}(t).
\end{equation}
%------------------------------------------------
such that 
%------------------------------------------------
\begin{align}\nonumber
&\alpha_{IJ} = \langle {\bf w}^{(J)}(t), {\bf u}^{(I)}(t)\rangle=
%\\\nonumber
%&
\frac{i}{2} \sum_{n=1}^{\infty} {}^{w}\bar{\phi}_{n}^{(J)}(t) {}^{u}\pi_{n}^{(I)}(t)-{}^{w}\bar{\pi}_{n}^{(J)}(t) {}^{u}\phi_{n}^{(I)}(t),\\
&\beta_{IJ} = -\langle \bar {\bf w}^{(J)}(t), {\bf u}^{(I)}(t)\rangle=
%\\ \nonumber
%&
-\frac{i}{2} \sum_{n=1}^{\infty} {}^{w}{\phi}_{n}^{(J)}(t) {}^{u}\pi_{n}^{(I)}(t)-{}^{w}{\pi}_{n}^{(J)}(t) {}^{u}\phi_{n}^{(I)}(t), 
\end{align}
are the Bogoliubov coefficients.\footnote{These Bogoliubov coefficients satisfy the conditions
%------------------------------------------------
\begin{align}\label{eq:bogo1}
&\sum_{K=1}^{\infty} \alpha_{I K} \bar{\alpha}_{J K}-\beta_{I K} \bar{\beta}_{J K}=\delta_{I J}, 
%\\\label{eq:bogo2}
%&
\quad \sum_{K=1}^{\infty} \alpha_{I K} \beta_{J K}-\beta_{I K} \alpha_{J K}=0.\\%\label{eq:bogo3}
&\sum_{K=1}^{\infty} \bar\alpha_{K I} {\alpha}_{KJ}-\beta_{KI} \bar{\beta}_{ KJ}=\delta_{I J}, %\\\label{eq:bogo4}
%&
\quad \sum_{K=1}^{\infty} \bar\alpha_{KI} \beta_{ K J}-\beta_{KI} \bar\alpha_{K J}=0.
\end{align}
}
Conversely, the equation relating the new basis of solutions with the first one reads
\begin{equation}
    {\bf w}^{(I)}(t) = \sum_{J=1}^{\infty} \bar{\alpha}_{JI}{\bf u}^{(J)}(t) - \beta_{JI}\bar {\bf u}^{(J)}(t).
\end{equation}

For a correct interpretation of these coefficients, it is convenient to jump into the quantum theory. The Fourier modes collected in ${\bf U}(t)$ must satisfy the reality condition $\bar{\bf U}(t)={\bf U}(t)$. Hence, if we express it in terms of the complex basis $\left({\bf u}^{(I)}(t), \bar{{\bf u}}^{(I)}(t)\right)$, they must be written as 
\begin{equation}\label{eq:sol}
{\bf U}(t)=\sum_{I=1}^\infty a_I {\bf u}^{(I)}(t)+\bar a_I \bar{{\bf u}}^{(I)}(t)=\sum_{I=1}^\infty b_I {\bf w}^{(I)}(t)+\bar b_I \bar{{\bf w}}^{(I)}(t), 
\end{equation}
or, as in the last equality, in terms of the solutions in the basis $\left({\bf w}^{(I)}(t), \bar{{\bf w}}^{(I)}(t)\right)$. The complex amplitudes $\left(a_I, \bar{a}_I\right)$ and $\left(b_I, \bar{b}_I\right)$ represent the annihilation and creation variables of each basis. In the classical theory, they satisfy the Poisson algebra $\left\{a_{I}, \bar{a}_{J}\right\}=-i\delta^{IJ}$, $\left\{a_{I}, {a}_{J}\right\}=0=\left\{\bar a_{I}, \bar{a}_{J}\right\}$, and similarly for $\left(b_I, \bar{b}_I\right)$. One can easily see that the annihilation variables $\left(b_I, \bar{b}_I\right)$ are related to $\left(a_I, \bar{a}_I\right)$ by
%------------------------------------------------
\begin{equation}\label{eq:bogob2a}
    b_{I} = \sum_{J=1}^{\infty} \alpha_{JI}a_{J}+\bar\beta_{JI}\bar a_{J}.
\end{equation}
One can easily invert this relation to express $\left(a_I, \bar{a}_I\right)$ in the basis
$\left(b_I, \bar{b}_I\right)$. The result is 
\begin{equation}\label{eq:inv-bogo}
    a_{I} = \sum_{J=1}^{\infty} \bar\alpha_{IJ}b_{J}-\bar\beta_{IJ}\bar b_{J}.
\end{equation}
In the quantum theory, the classical algebra is promoted to
\begin{equation}\label{eq:amu-commut}
\left[\hat a_{I}, \hat a^{\dagger}_{J}\right]=\delta_{IJ}\hat{\bf I},\quad \left[\hat a_{I}, \hat {a}_{J}\right]=0=\left[a^{\dagger}_{I}, a^{\dagger}_{J}\right],
\end{equation}
and again, we have similar expressions for $(\hat b_{I}, \hat b^{\dagger}_{I})$. Now, we must note that each collection of annihilation variables defines a different vacuum state if $\beta_{IJ}\neq 0$. Concretely, the vacuum state is determined by 
%------------------------------------------------
\begin{equation}
\hat a_{I}|0\rangle = 0,\quad I=1,2,\ldots.
\end{equation}
%------------------------------------------------
Hence, we immediately note that 
\begin{equation}
\hat b_{I}|0\rangle = \sum_{J=1}^{\infty} \bar\beta_{JI}\hat a^{\dagger}_{J}|0\rangle\neq 0,\quad I=1,2,\ldots.
\end{equation}
Therefore, the coefficients $\beta_{IJ}$ codify the information about particle production in the quantum theory. They are nonvanishing when there is mixing of positive and negative frequency solutions. In quantum optics, this is equivalent to a squeezing operator. Now, if $\beta_{IJ}=0$ for all $I,J$, and $\alpha_{IJ}$ is nonvanishing for $I\neq J$, it implies a symplectic transformation without mixing positive and negative solutions. The two bases are different but, define the same vacuum state. This is typically identified with beam splitters in quantum optics.

In Ref. \cite{GarciaMartin-Caro:2023jjq} we noted that there are configurations of the boundaries for which it is possible to obtain a nearly thermal spectrum, with superposed oscillations attributed to a graybody factor consequence of the finiteness of the motion of the trajectories of the mirrors. We will extend the analysis carried out there by studying several configurations in order to understand the robustness of the thermal production of particles due to accelerating boundaries and also if it is possible to obtain a thermal spectrum when the cavity collapses, as it was claimed in Ref. \cite{CASTAGNINO19841}. We will see that, for the trajectories we consider here, the answer to this last question is in the negative. In addition to these trajectories, we will also discuss particle production when the cavity either collapses or expands, reaching its original configuration at the end of the process. These configurations are time symmetric, and can serve, for instance, after repeating the process several times, to reproduce the dynamical Casimir effect of vibrating cavities. 

In order to carry out our computations, we will solve Eqs. \eqref{eq:eom} numerically, using an explicit embedded Prince-Dormand $(8,9)$ method of the GNU scientific library, which belongs to the family of Runge-Kutta methods. It carries two types of errors (the absolute and the relative errors) whose value can be specified. In our simulations, we set the absolute error to be between $[10^{-10}, 10^{-12}]$, and the relative error to zero. We also introduce a cut of $N$ in the total number of modes we evolve, but we allow it to vary just as $N=256$, $N=512$, and $N=1024$, so that we can take a Richardson extrapolation in order to probe the limit $N\to\infty$. Since we always assume that the cavity is static in the past, we always have a preferred $in$ basis that we denote by $({}^{(in)}{\bf u}^{(I)}(t),{}^{(in)}\bar{\bf u}^{(I)}(t))$. Likewise,
we will always assume that the cavity becomes static in the future. Hence, we will have a natural $out$ basis $({}^{(out)}{\bf u}^{(I)}(t),{}^{(out)}\bar{\bf u}^{(I)}(t))$ there.

%------------------------------------------------
\section{Numerical analysis for expanding cavities}
\label{Sec:expanding-cav}
%------------------------------------------------

In this section, we will analyze several configurations of accelerating mirrors. We will consider relatively large changes in the size of the cavity with small accelerations and relatively small changes in the size of the cavity with large accelerations. These are two interesting asymptotic regimes corresponding to adiabatic and sharp dynamical configurations of the mirrors, respectively. Each regime shows advantages from a theoretical and experimental perspective. We will discuss them in the following. 

We will compare our numerical results with some fitting expressions for $|\alpha_{IJ}|^2$ and $|\beta_{IJ}|^2$, which contain the most physically relevant information that we want to discuss in this manuscript. Concretely,
we consider a (modified) Fulling-Davies spectrum of particle production of the form
%------------------------------------------------
\begin{equation}\label{eq:betas-fit}
|\beta^{\text{(f)}}_{IJ}|^2 = \frac{N_\beta \Delta \omega_I\Delta \omega_J}{\pi\kappa\omega_I}\frac{\Gamma_\beta (\epsilon,\omega_J)}{(e^{2\pi\omega_J/\kappa}-1)},
\end{equation}
%------------------------------------------------
where $\omega_I= \pi I/ L_0$ are the {\it in} frequencies and $\Delta\omega_I=\pi/ L_0$ the corresponding gap (which in this model equals the fundamental {\it in} frequency); $\omega_J$ and $\Delta\omega_J$ are the out frequencies that depend on the trajectory of the plates and hence will be specified below for each case. Note that in the continuum limit, $\Delta\omega_I\Delta\omega_J\to d\omega d\omega'$. Moreover, $N_\beta$ is a normalization factor that will also depend on the concrete dynamical configuration of the mirrors. Its typical value is 2, in agreement with the usual Fulling-Davies spectrum, but it can also be 4 in some cases, as we will justify below. Finally, 
%------------------------------------------------
\begin{equation}\label{eq:Gamma}
\Gamma_\beta(\epsilon,\omega_J)=\left[A_\beta+B_\beta\sin^2\big({\cal T}_\beta \, \omega_J\big)\right],
\end{equation}
%------------------------------------------------
is a graybody factor that shows a strong dependence on $\epsilon$ (the change in the size of the cavity) and the $out$ frequency. It accounts for the finiteness of the duration of the acceleration, producing the oscillations described in very good approximation by a sinusoidal function with a frequency characterized by ${\cal T}_\beta=(1+C_\beta)\epsilon$, with $C_\beta$ a very small dimensionless parameter with a weak dependence on the trajectory parameters and the $in$ and $out$ frequencies. $A_\beta$ and $B_\beta$ are also dimensionless parameters, with $A_\beta$ small and $B_\beta$ order unit, both of which are weakly dependent on the trajectory parameters and the $in$ and $out$ frequencies. This parametrization is valid only for the sets of modes that reach a nearly thermal final state. These sets of thermal modes will depend on the trajectories of the mirrors. An exact thermal state will correspond to $\Gamma_\beta(\epsilon,\omega_J)=1$. 

We have also found a fitting expression for the $\alpha$ coefficients, although it is not as simple as the one for the $\beta$ coefficients. Concretely,
\begin{widetext}
\begin{align}
    |\alpha^{\text{(f)}}_{IJ}|^2 = \begin{cases}
      \displaystyle{\frac{N_\alpha \Delta \omega_I\Delta \omega_J}{\pi\kappa\omega_I}\frac{1}{1+D_1}\left(1+D_1\frac{(F\omega_I)^2}{|F\omega_I-\omega_J|^2}\right)
\frac{\Gamma_\alpha (\epsilon,\omega_J)e^{2\pi\omega_J/\kappa}}{(e^{2\pi\omega_J/\kappa}-1)}},& \omega_J<F\omega_I, \\
      &\\
      \displaystyle{\frac{N_\alpha \Delta \omega_I\Delta \omega_J}{\pi\kappa\omega_I}\left(1+D_2\frac{(F\omega_I)^2}{|F\omega_I-\omega_J|^2}\right)
\frac{\Gamma_\alpha(\epsilon,\omega_J)}{(e^{2\pi(\omega_J-F\omega_I)/\tilde\kappa}-1)}},& \omega_J>F\omega_I.
    \end{cases} 
\label{eq:alphas-fit}
\end{align}
\end{widetext}
Here, the dimensionless parameters $D_1$, $D_2$ and $F$ take values that are ${\cal O}(1)$ or smaller, which means that $|\alpha_{IJ}|^2$ reaches a maximum value when $\omega_J\simeq F\omega_I$. In addition, $\Gamma_\alpha(\epsilon,\omega_J)$ has the same functional form as $\Gamma_\beta(\epsilon,\omega_J)$ in Eq. \eqref{eq:Gamma} but with dimensionless constants $A_\alpha$, $B_\alpha$ and $C_\alpha$ (the latter defining ${\cal T}_\alpha$). Besides, $N_\alpha$ is a normalization factor that depends on the trajectories of the mirrors.

Let us now compare these fitting expressions in more detail with concrete numerical examples. In all cases, we choose several parameters of the trajectories of the mirrors within the range of several experimental settings Ref. \cite{Lahteenmaki2013, Nation2012}.

%------------------------------------------------
\subsection{One expanding mirror}
\label{Sec:1plt}
%------------------------------------------------

The trajectories of the boundaries of the cavity are given by:
%------------------------------------------------
\begin{align}\nonumber
   f(t)= & 0, \\ 
   g(t)= & 1 +\frac{s}{2\kappa}+\frac{1}{2\kappa}\bigg[\log\Big(\cosh\big(\kappa(t-t_0)\big)\Big)  -\log\Big(\cosh\big(s-\kappa(t-t_0)\big)\Big)\bigg],
   \label{eq:Traj-1plt}
\end{align}
%------------------------------------------------
In this configuration, at $t \ll t_0$, the left boundary remains at $x^f=0$ at all times, while the right boundary is nearly static at the initial position $x^g_{in}=1$, while at $t\gg T\simeq t_0+\epsilon$ (with $\epsilon=s/\kappa$) its final position will be $x^g_{out} =(1+\epsilon)$. In the interval $[t_0,t_0+\epsilon]$ the right boundary follows an acceleration close to the speed of light. Besides, regarding Eqs. \eqref{eq:betas-fit} and \eqref{eq:alphas-fit}, the normalization in this case is just $N_\beta=2=N_\alpha$ and the out frequencies are given by $\omega_J= \pi J/ (L_0+\epsilon)$ and their gap by $\Delta\omega_J=\pi/ (L_0+\epsilon)$.

\begin{figure}[ht]
{\centering     
  \includegraphics[width = 0.48\textwidth]{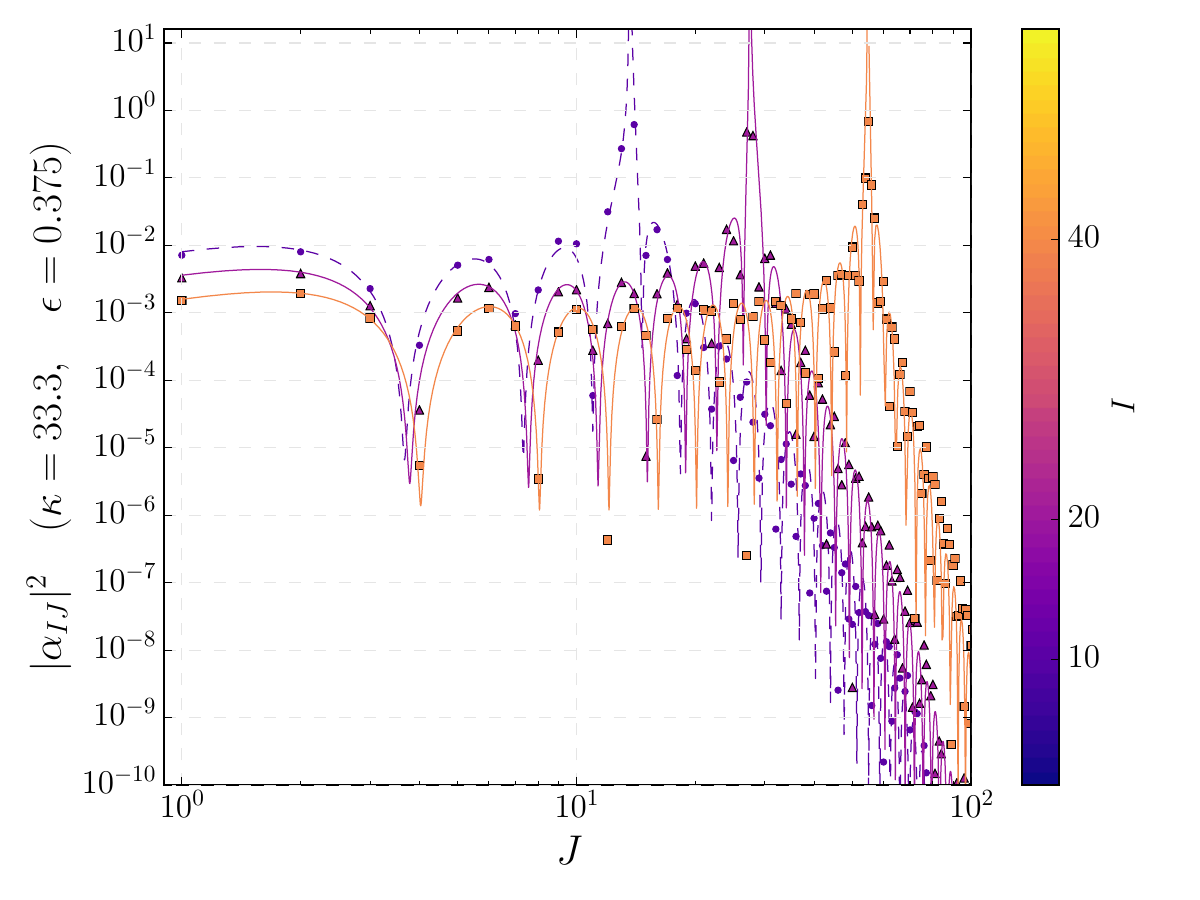}\includegraphics[width = 0.48\textwidth]{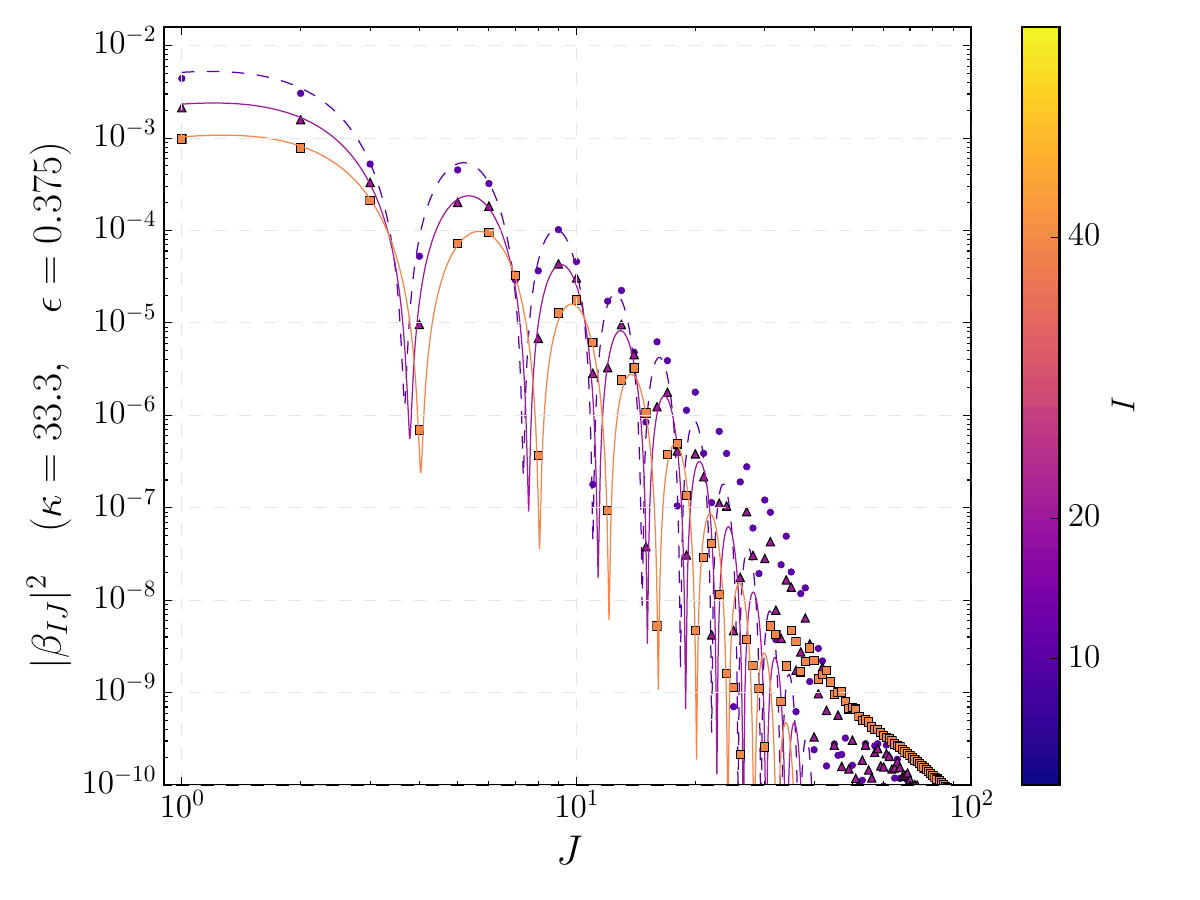}\\\includegraphics[width = 0.48\textwidth]{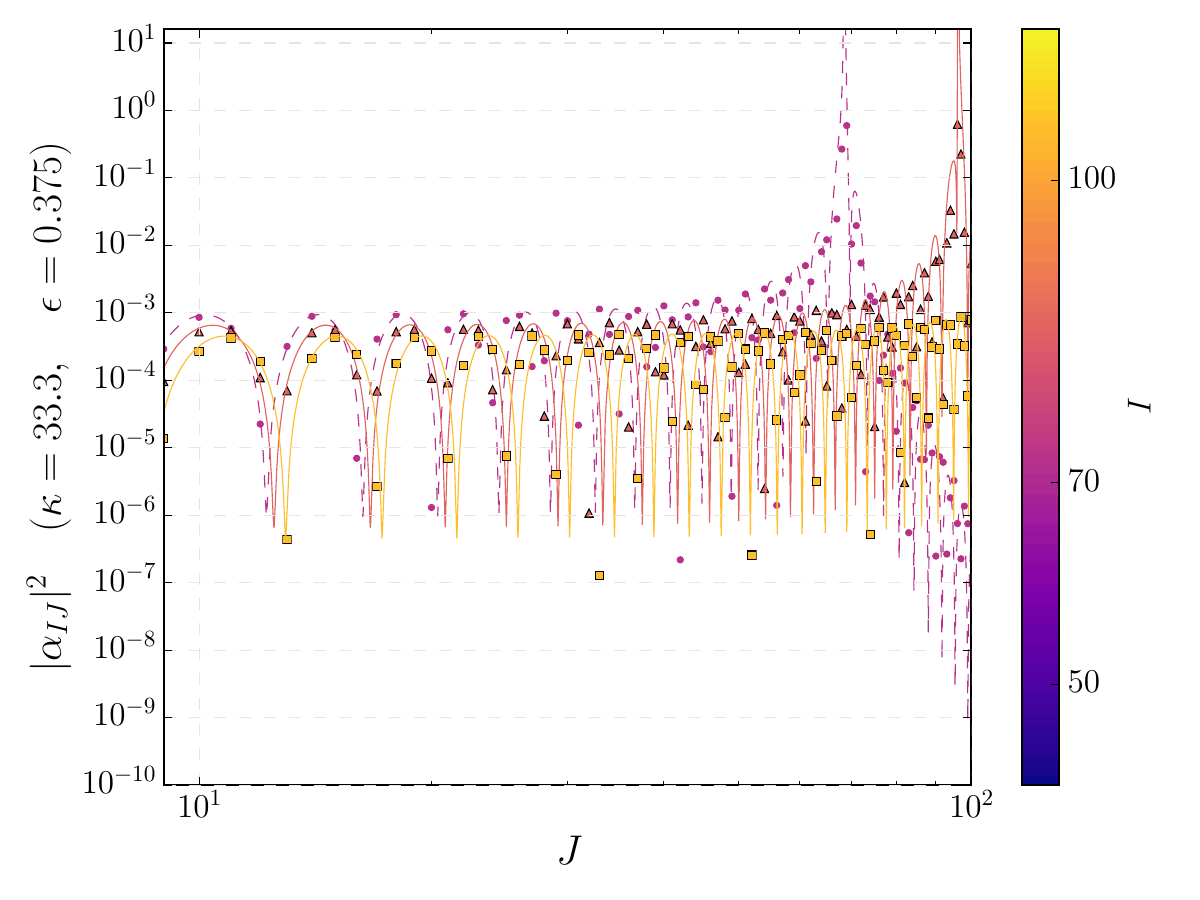}
  \includegraphics[width = 0.48\textwidth]{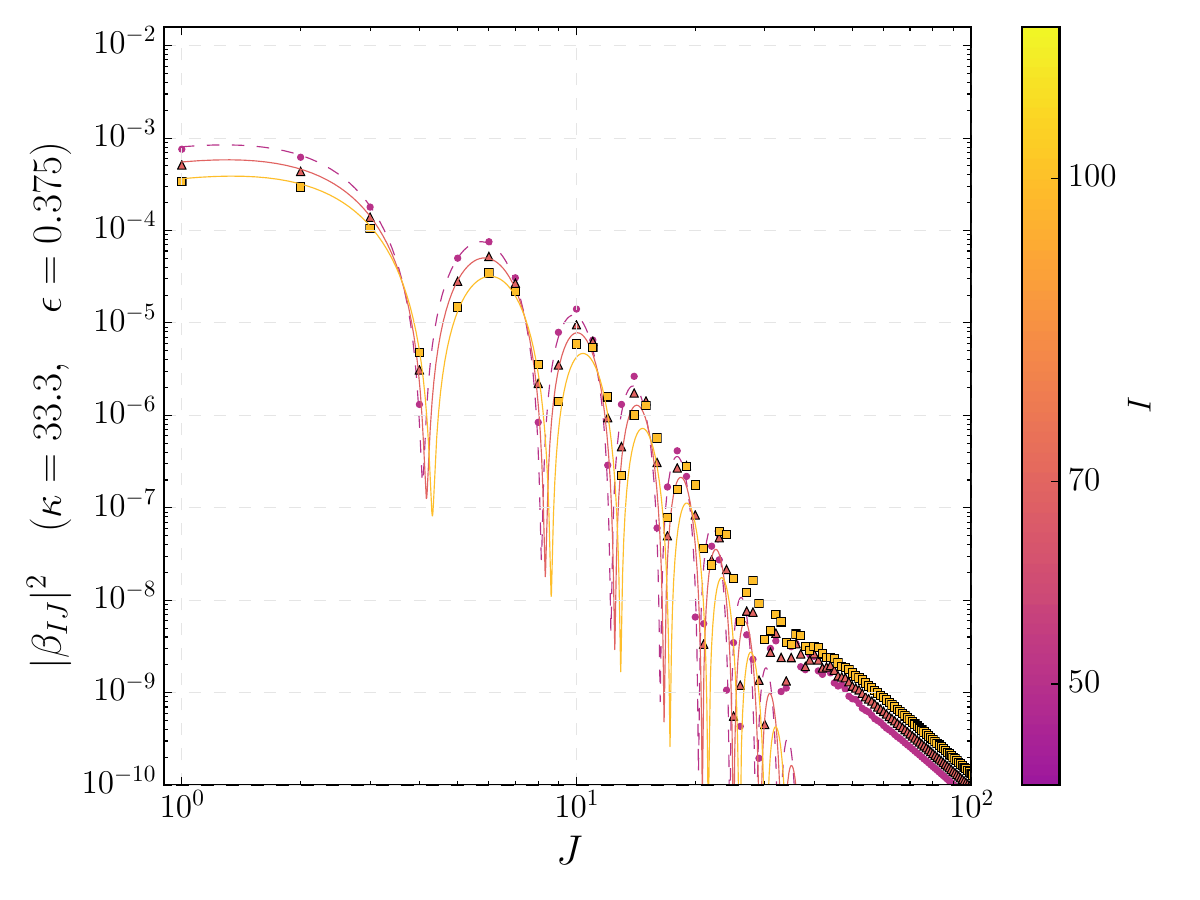}
}
\caption{\justifying Bogoliubov coefficients: These plots correspond to the trajectories given by Eq. \eqref{eq:Traj-1plt}, with $\epsilon=0.375$ and $\kappa=33.3$. The upper panel for the infrared {\it in} modes. The lower panel for the ultraviolet {\it in} modes. The lines interpolating points correspond to the fitting expressions in Eqs. \eqref{eq:betas-fit} and \eqref{eq:alphas-fit} for the right and left panels, respectively.}
\label{fig:1plt}
 \end{figure}

{\bf Small accelerations:} We have studied several configurations for small accelerations. In all cases, we have found qualitatively similar results. Let us consider a concrete realization given by $\epsilon=0.375$ and $\kappa=33.3$. In Fig. \ref{fig:1plt} we show the behavior of Bogoliubov coefficients for some frequency bands and compare them with the fitting expressions of Eqs. \eqref{eq:betas-fit} and \eqref{eq:alphas-fit}. Here, we fix $A_\alpha=1=A_\beta$, and $B_\alpha=10^{-3}=B_\beta$, $D_1=0.1=D_2$ and $\tilde\kappa = 1.6 \kappa$. The upper panel shows several (relatively) infrared {\it in} modes $\omega_I$ as functions of the {\it out} frequencies $\omega_J$. We see that for low $\omega_J$, the computed Bogoliubov coefficients agree very well with Eqs. \eqref{eq:betas-fit} and \eqref{eq:alphas-fit}, with $C_\alpha=C_\beta$ and $F$ suitable functions of $\omega_I$. For infrared $in$ modes, we see that $C_\alpha\to 0$ and $C_\beta\to 0$. We show the behavior of $C_\alpha$, $C_\beta$ and $F$ in Fig. \ref{fig:1plt-CDF-plot} below. At very high $\omega_J$ we have checked numerically that the Bogoliubov coefficients decay as $\omega_J^{-n}$ with $n>3$. On the other hand, the lower panel shows several (relatively) ultraviolet {\it in} modes $\omega_I$ as functions of the {\it out} frequencies $\omega_J$. We see that for low $\omega_J$ the computed beta coefficients still agree well with Eq. \eqref{eq:betas-fit}. Nevertheless, the beta coefficients reach quicker the behavior $\omega_J^{-n}$, departing from Eq. \eqref{eq:betas-fit} at lower {\it out} frequencies. This indicates that these {\it in} modes did not have enough time to thermalize. 

\begin{figure}[ht]
{\centering     
  \includegraphics[width = 0.48\textwidth]{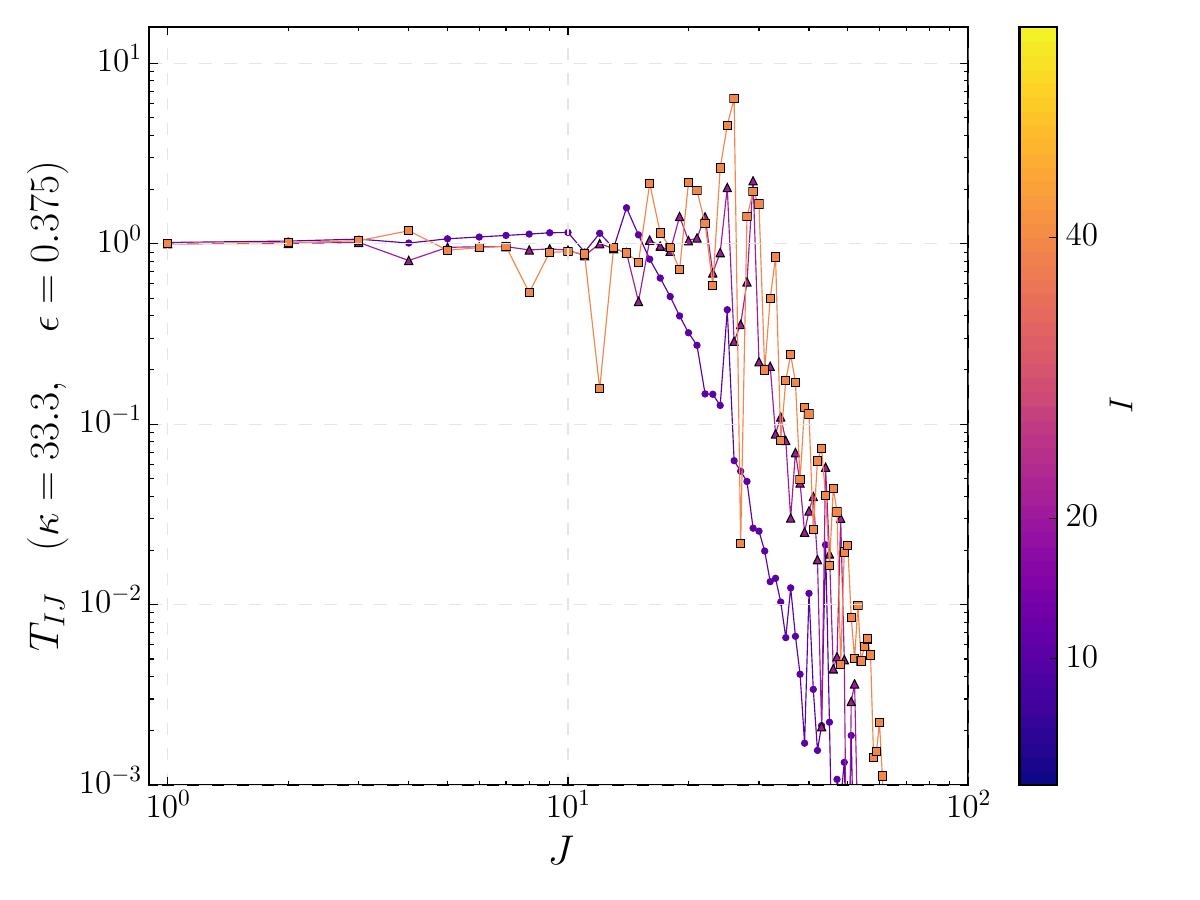}\includegraphics[width = 0.48\textwidth]{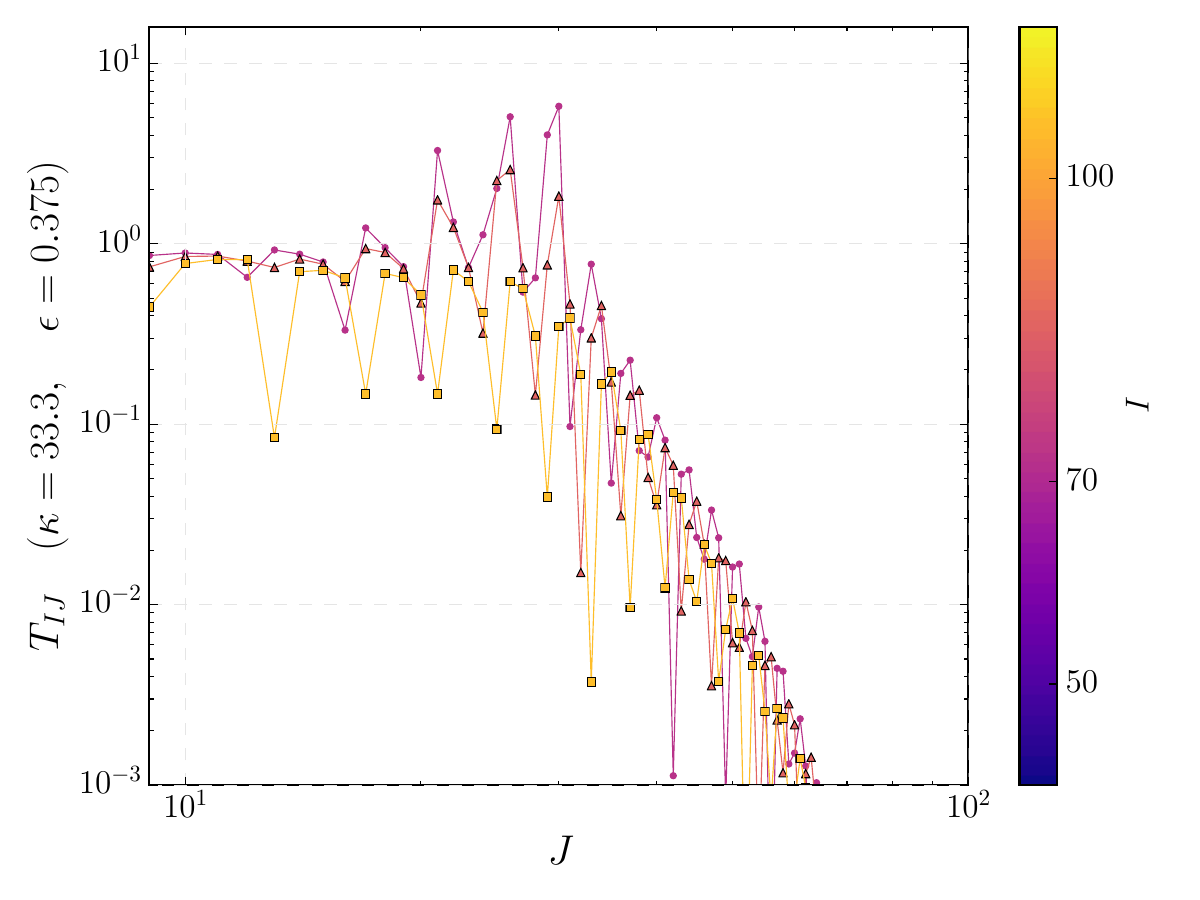}
}
\caption{\justifying Thermal relation: These plots correspond to the trajectories given by Eq. \eqref{eq:Traj-1plt}, with $\epsilon=0.375$ and $\kappa=33.3$. The upper panel for the infrared modes. The lower panel for the ultraviolet modes. The lines joining points do not represent any interpolation and have been included only for visualization purposes. }
\label{fig:thermality-1plt}
\end{figure}

To gain insight into the thermal properties of these modes, one could plot the ratio $|\alpha_{IJ}|/|\beta_{IJ}|$ and check if it is equal to $e^{\pi\omega_J/\kappa}$, in other words, whether the detailed balance condition \eqref{eq:BD-thermal} is satisfied. In our case, one concludes that this is the case for $\omega_J\ll F\omega_I$ to a good approximation, provided that the singular behavior of $|\alpha_{IJ}|$ is negligible, namely, in the regime in which 
\begin{equation}
    \frac{1}{1+D_1}\left(1+D_1\frac{(F\omega_I)^2}{|F\omega_I-\omega_J|^2}\right)\simeq 1.
\end{equation}
However, this limit might be extremely difficult to reach experimentally. Instead, our point of view is that one should check whether the following thermality function
\begin{equation}\label{eq:thermal-func}
T_{IJ}=\frac{|\alpha_{IJ}|}{|\beta_{IJ}|  e^{\pi\omega_J/\kappa}} \frac{(1+D_1)^{1/2}}{\left(1+D_1\frac{(F\omega_I)^2}{|F\omega_I-\omega_J|^2}\right)^{1/2}},
\end{equation}
equals unit. It will agree with the standard one in the limit $F\omega_I\gg \omega_J$ anyways. In Fig. \ref{fig:thermality-1plt} we show $T_{IJ}$ of several values of the in modes $I$. As we can see, for this trajectory of the mirrors, in the limit of infrared $in$ and $out$ modes, the thermality condition is near the unit and, hence, it is approximately fulfilled. However, for UV $in$ modes (large $I$), this condition is met only for very infrared $out$ modes (small $J$). 

\begin{figure}[ht]
{\centering     
  \includegraphics[width = 0.48\textwidth]{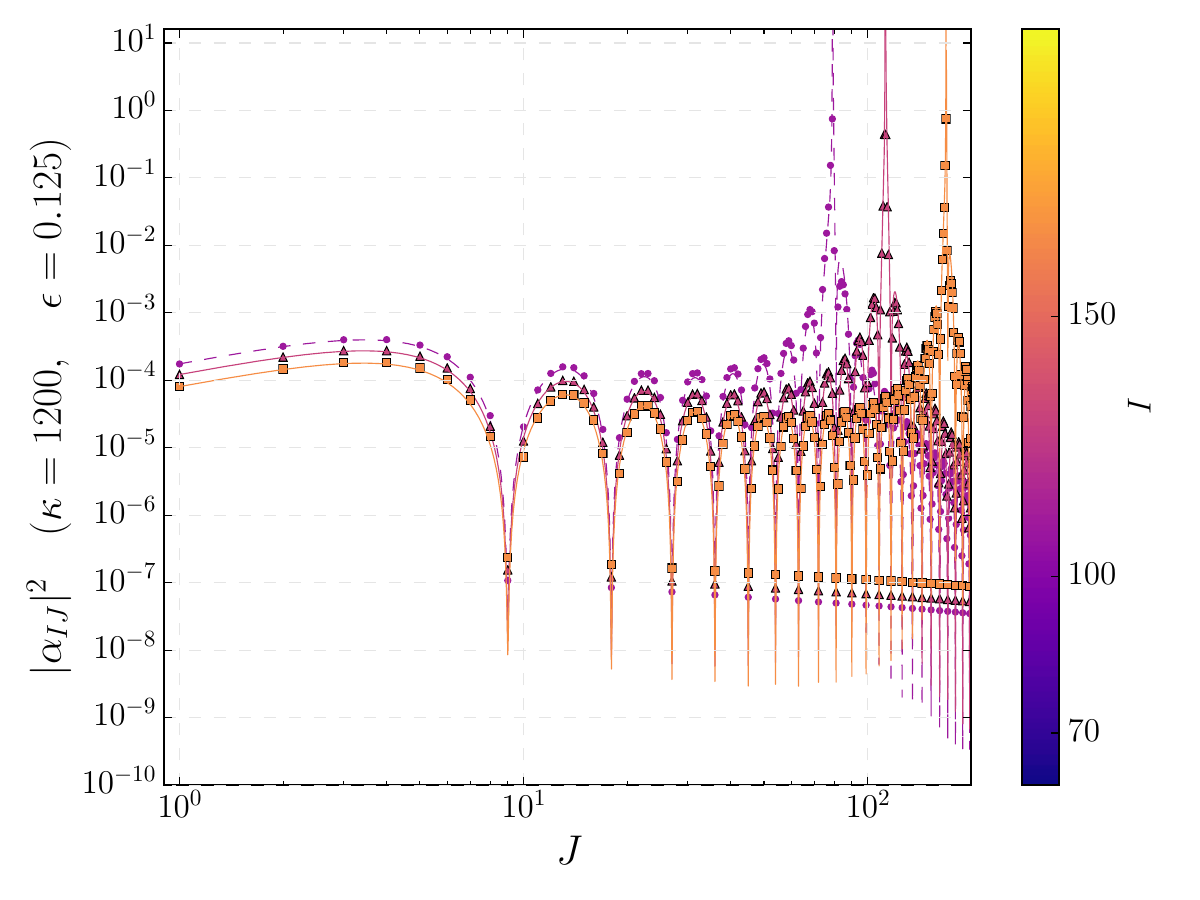}
  \includegraphics[width = 0.48\textwidth]{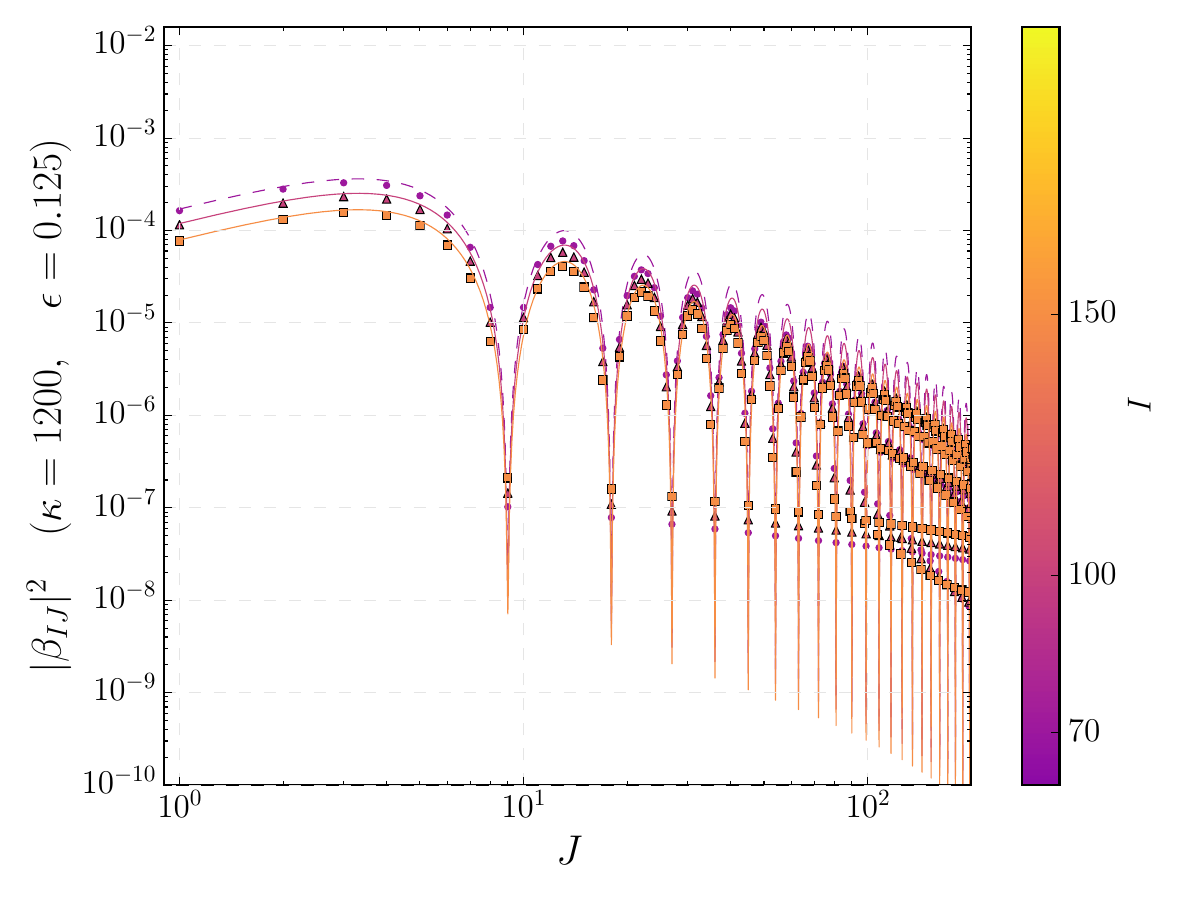}
}
\caption{\justifying Bogoliubov coefficients: These plots correspond to the trajectories given by Eq. \eqref{eq:Traj-1plt}, with $\epsilon=0.125$ and $\kappa=1200$. The lines interpolating points in right and left panels correspond to the fitting expressions in Eqs. \eqref{eq:betas-fit} and \eqref{eq:alphas-fit}, respectively.}
\label{fig:1plt-sharp}
\end{figure}

{\bf Large accelerations:} We have also studied several sharp configurations, namely, for large accelerations $\kappa$ and small changes $\epsilon$ in the size of the cavity. Here, the right mirror transitions very quickly from a steady configuration to a new one, sharply approaching the speed of light. The concrete configuration we will show here is given by $\epsilon=0.125$ and $\kappa=1200$. Other configurations with large accelerations show similar results. In Fig.~\ref{fig:1plt-sharp} we show the behavior of the Bogoliubov coefficients for modes in the frequency band $I\in(70,150)$ and $J\in(1,150)$, and compare them with the fitting expression of Eq.\eqref{eq:betas-fit}. In these simulations, we set $A=1.0$ and $B=10^{-4}$. We show the Bogoliubov coefficients for several {\it in} modes $\omega_I$ as functions of the {\it out} frequencies $\omega_J$. We see that for very low $\omega_I$, the computed beta coefficients agree very well with Eqs. \eqref{eq:alphas-fit} and \eqref{eq:betas-fit}. The parameters $D_1$, $D_2$, $F$ and $\tilde \kappa$ are plotted in Fig. \ref{fig:1plt-CDF-plot-b} as functions of $\omega_J$ for the two trajectories shown here.  

\begin{figure}[ht]
{\centering     
  \includegraphics[width = 0.48\textwidth]{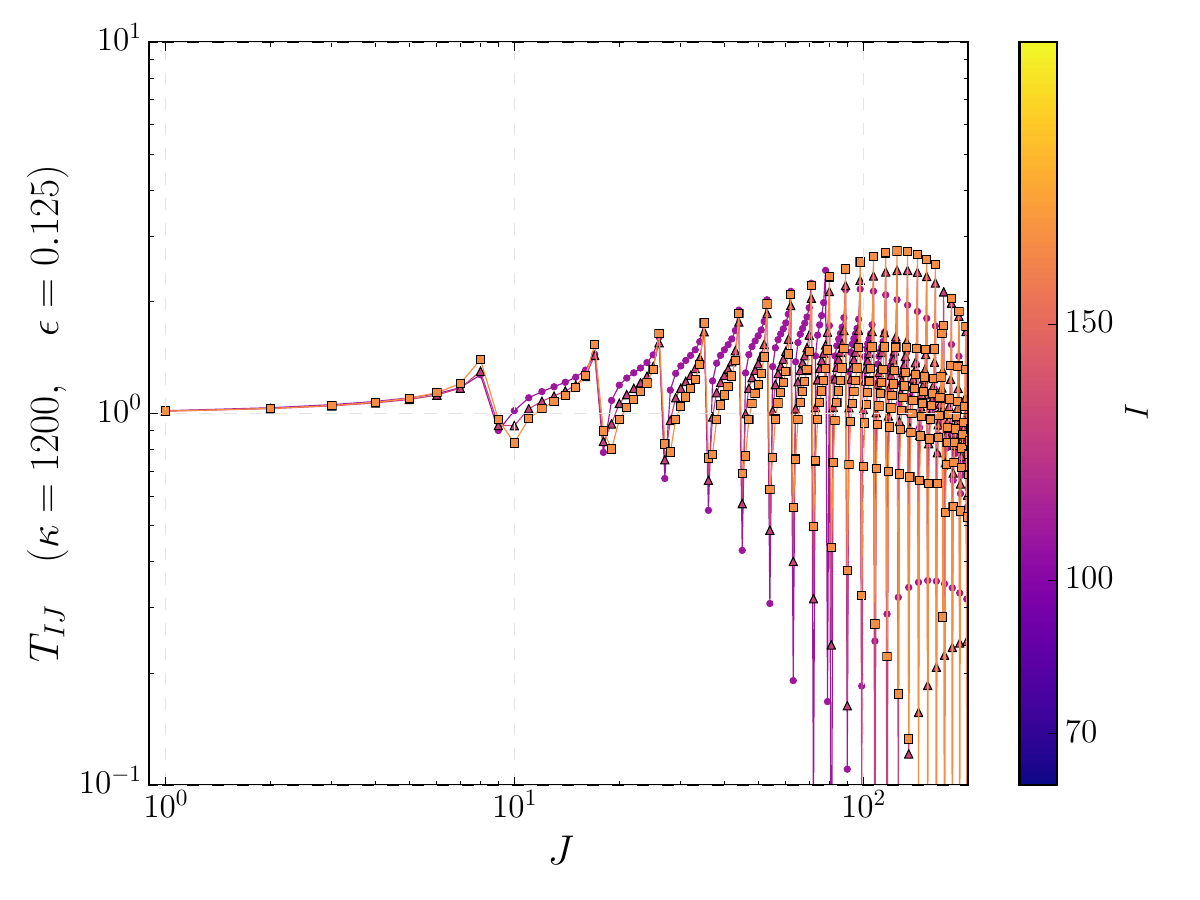}
}
\caption{\justifying Thermal relation: This plot corresponds to the trajectories given by Eq. \eqref{eq:Traj-1plt}, with $\epsilon=0.125$ and $\kappa=1200$.  The lines joining points do not represent any interpolation and have been included only to help visualizing $T_{IJ}$ for each fixed $I$.}
\label{fig:thermality-1plt-b}
\end{figure}

For lower or higher $in$ modes compared to those shown here, the fitting expression for $|\alpha_{IJ}|^2$ still agrees very well, but the one for $|\beta_{IJ}|^2$ shows an amplitude that disagrees with the fitting expression. Hence, those modes depart strongly from a thermal distribution and, therefore, we do not consider them here. On the other hand, in Fig. \ref{fig:thermality-1plt-b} we show the thermal relation $T_{IJ}$ in Eq. \eqref{fig:thermality-1plt} for several values of the $in$ modes $I$. As we can see, the most infrared $out$ modes yield $T_{IJ}\simeq 1$ and, hence, follow a thermal distribution. However, we also see departures from thermality in those modes where either $|\alpha_{IJ}|$ or $|\beta_{IJ}|$ reach their minima. In consequence, they do not follow a thermal distribution. The lower or higher $in$ modes show a stronger departure of $T_{IJ}$ from the unit.
%------------------------------------------------
\subsection{Two symmetrically expanding mirrors}
\label{Sec:expand}
%------------------------------------------------
The trajectories of the boundaries of the cavity are now given by:
%------------------------------------------------
\begin{align}\nonumber
f(t)= &-\frac{s}{2\kappa}-\frac{1}{2\kappa}\bigg[\log\Big(\cosh\big(\kappa(t-t_0)\big)\Big)
   -\log\Big(\cosh\big(s-\kappa(t-t_0)\big)\Big)\bigg],\\
   g(t)= & 1 +\frac{s}{2\kappa}+\frac{1}{2\kappa}\bigg[\log\Big(\cosh\big(\kappa(t-t_0)\big)\Big)-\log\Big(\cosh\big(s-\kappa(t-t_0)\big)\Big)\bigg].
   \label{eq:Traj-expand}
\end{align}
%------------------------------------------------
In this configuration, at $t \ll t_0$, both mirrors are nearly static at the initial positions $x^f_{in}=0$ and $x^g_{in}=1$, respectively, while at $t\gg T\simeq t_0+\epsilon$ (with $\epsilon=s/\kappa$) their final positions are $x^f_{out} =(-\epsilon)$ and $x^g_{out} =(1+\epsilon)$, respectively. In the interval $[t_0,t_0+\epsilon]$ both mirrors accelerate reaching a maximum speed (very close to the speed of light), and then they decelerate until they reach their respective final (stationary) positions. Moreover, regarding the spectrum in Eqs. \eqref{eq:betas-fit} and \eqref{eq:alphas-fit}, we will have {\it out} frequencies given by $\omega_J= \pi J/ (L_0+2\epsilon)$ and their gap by $\Delta\omega_J=\pi/ (L_0+2\epsilon)$. The normalization of the coefficients in the fitting expressions is given by $N_\alpha=4=N_\beta$. This value has a simple explanation. The trajectories are symmetric with respect to the center of the cavity. This results in a decoupling between the odd and even modes of the field. Therefore, particle production will double on each mode since the number of modes that can be occupied with particles now reduces to half the number one has when all modes are coupled, e.g., compared with the previous case of one moving mirror.  

\begin{figure}[ht]
{\centering     
  \includegraphics[width = 0.48\textwidth]{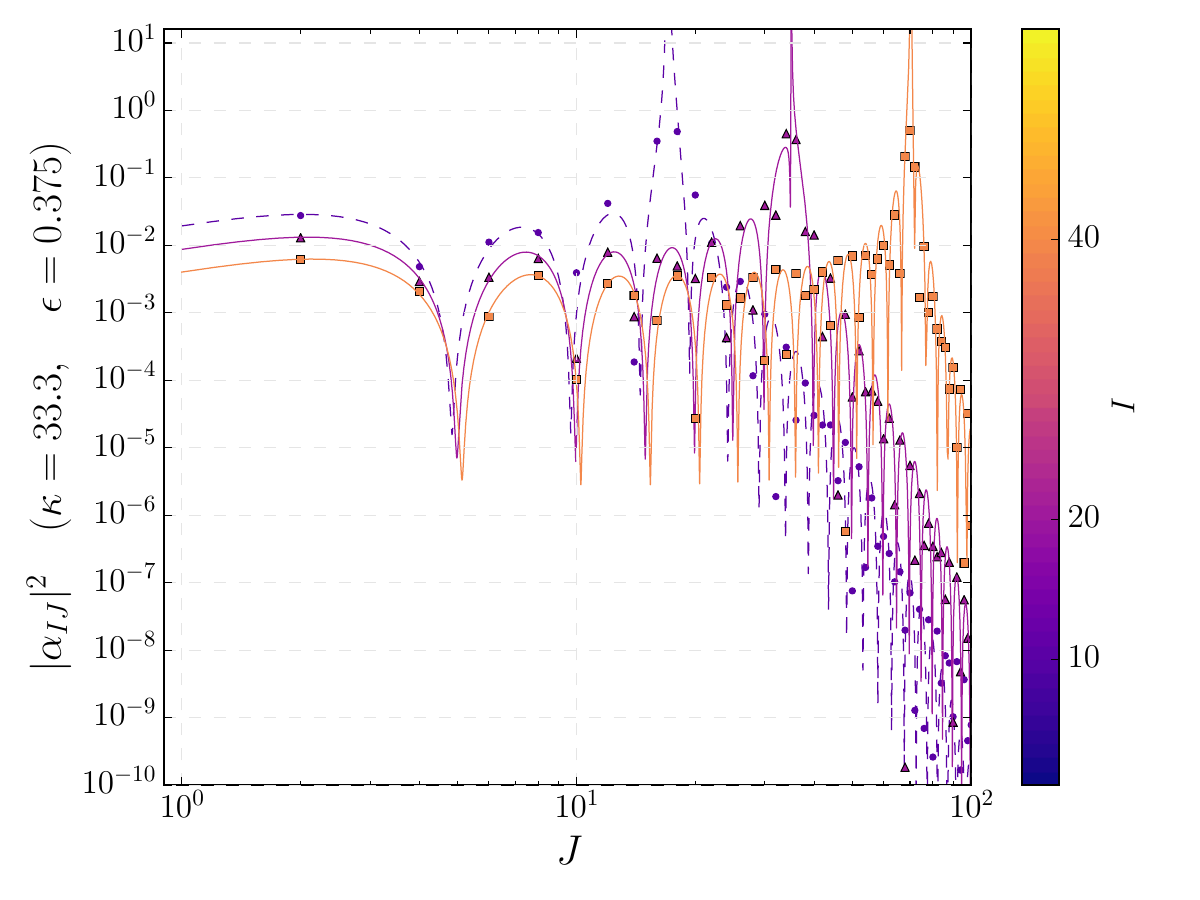}\includegraphics[width = 0.48\textwidth]{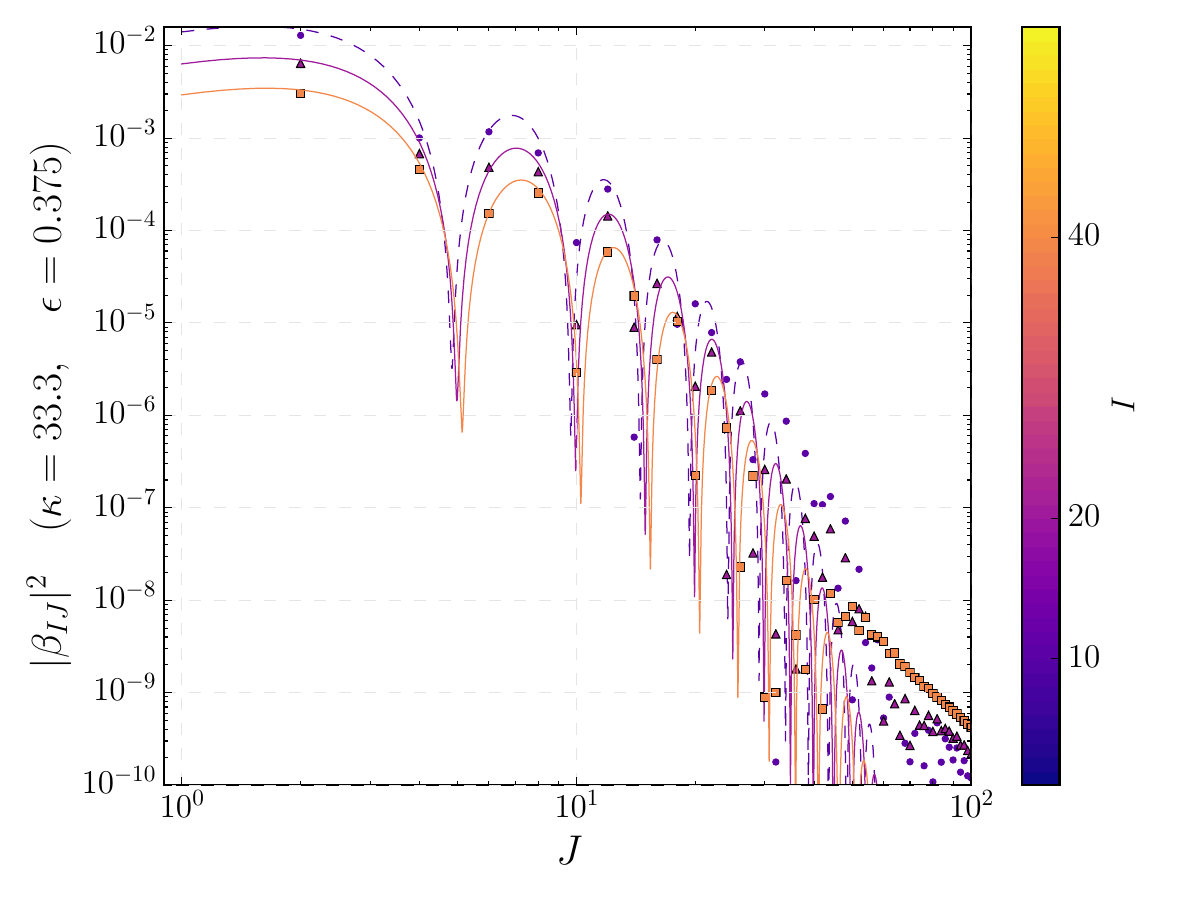}\\\includegraphics[width = 0.48\textwidth]{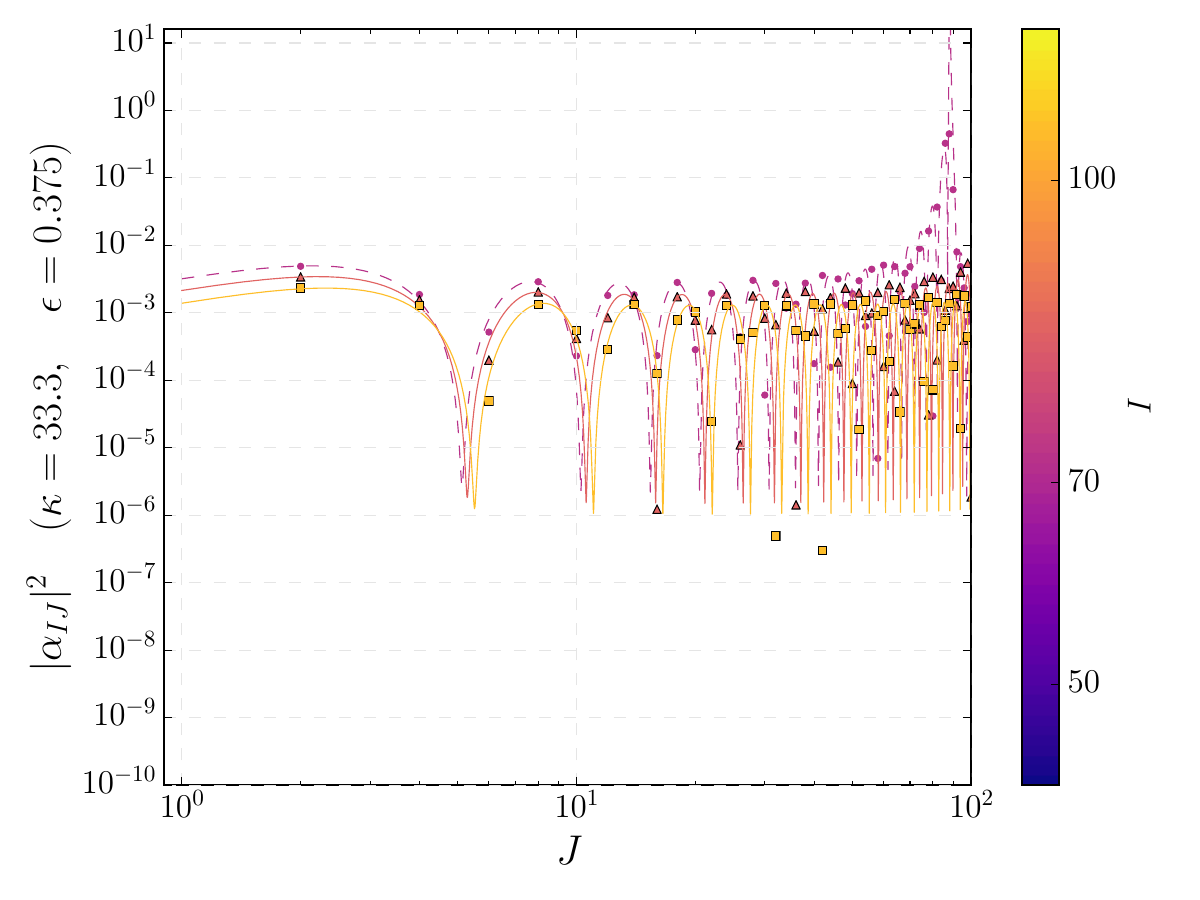}
  \includegraphics[width = 0.48\textwidth]{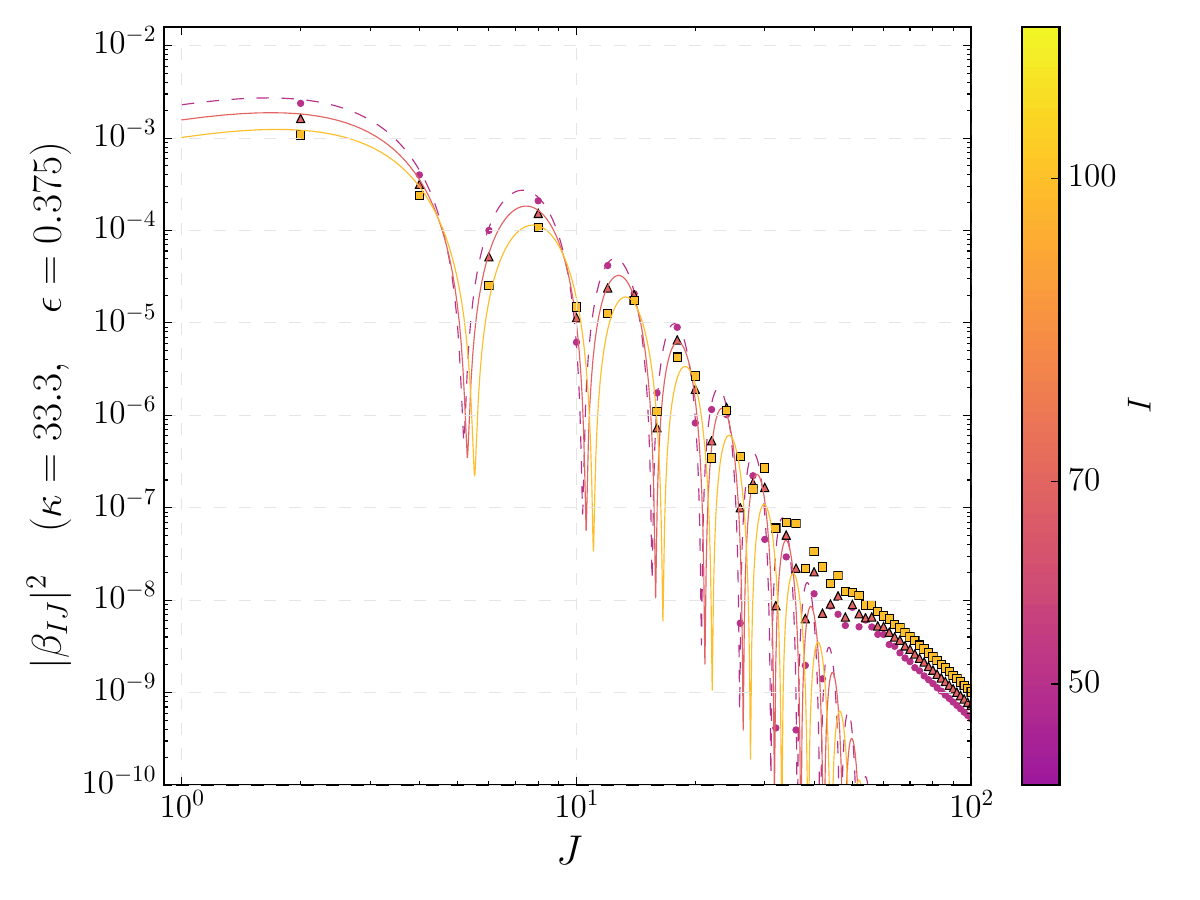}
}
\caption{\justifying Bogoliubov coefficients: These plots correspond to the trajectories given by Eq. \eqref{eq:Traj-expand}, with $\epsilon=0.375$ and $\kappa=33.3$. The upper panel for the infrared modes. The lower panel for the ultraviolet modes. The lines interpolating points correspond to the fitting expressions in Eqs. \eqref{eq:betas-fit} and \eqref{eq:alphas-fit}.}
\label{fig:small-expand}
\end{figure}

{\bf Small accelerations:} As in the previous configuration, we have small accelerations, and we found qualitatively similar results in all cases. Let us consider the same concrete realization given by $\epsilon=0.375$ and $\kappa=33.3$. In Fig. \ref{fig:small-expand} we show the numerically computed Bogoliubov coefficients in the band frequency $(I,J)\in(20,100)$ and the fitting expressions of Eqs. \eqref{eq:betas-fit} and \eqref{eq:alphas-fit}. In all cases, we fix $A_\alpha=1=A_\beta$, and $B_\alpha=10^{-3}=B_\beta$. The upper panel includes several (relatively) infrared {\it in} modes $\omega_I$ with respect to the {\it out} frequencies $\omega_J$. For low $\omega_J$ we see very good agreement between the computed Bogoliubov coefficients and the fitting expressions in Eqs. \eqref{eq:betas-fit} and \eqref{eq:alphas-fit}, with $C_\alpha$, $C_\beta$, $\tilde \kappa$, $D_1$, $D_2$ and $F$ suitable functions of $\omega_I$. For the $in$ modes considered here, we see that $|C_\alpha|\ll 1$ and $|C_\beta|\ll 1$. We show the behavior of $C_\alpha$, $C_\beta$ $\tilde \kappa$, $D_1$, $D_2$ and $F$ in Fig. \ref{fig:expand-CDF-plot}. At very high $\omega_J$ we have checked numerically that the Bogoliubov coefficients decay as $\omega_J^{-n}$ with $n>3$ (although we do not show here the results for the $\alpha$ coefficients). On the other hand, the lower panel shows several (relatively) ultraviolet {\it in} modes $\omega_I$ as functions of the {\it out} frequencies $\omega_J$. We see that for low $\omega_J$ the computed $\beta$ coefficients still agree well with Eq. \eqref{eq:betas-fit}. Nevertheless, the $\beta$ coefficients reach quicker the behavior $\omega_J^{-n}$, departing from Eq. \eqref{eq:betas-fit} at lower frequencies. This indicates that these {\it in} modes did not have enough time to thermalize. All of this qualitatively agrees with the results obtained for one moving mirror, namely, the trajectory \eqref{eq:Traj-1plt}.

\begin{figure}[ht]
{\centering     
  \includegraphics[width = 0.48\textwidth]{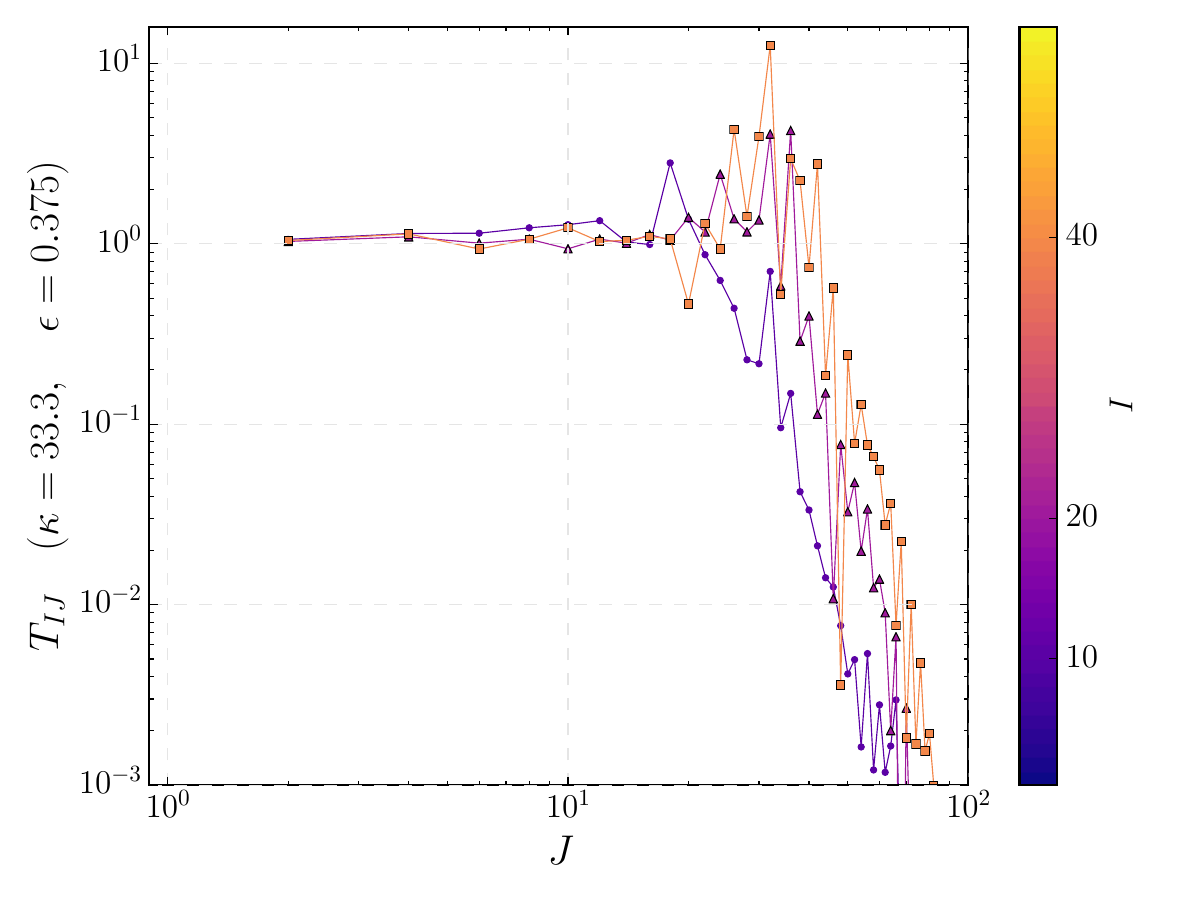}\includegraphics[width = 0.48\textwidth]{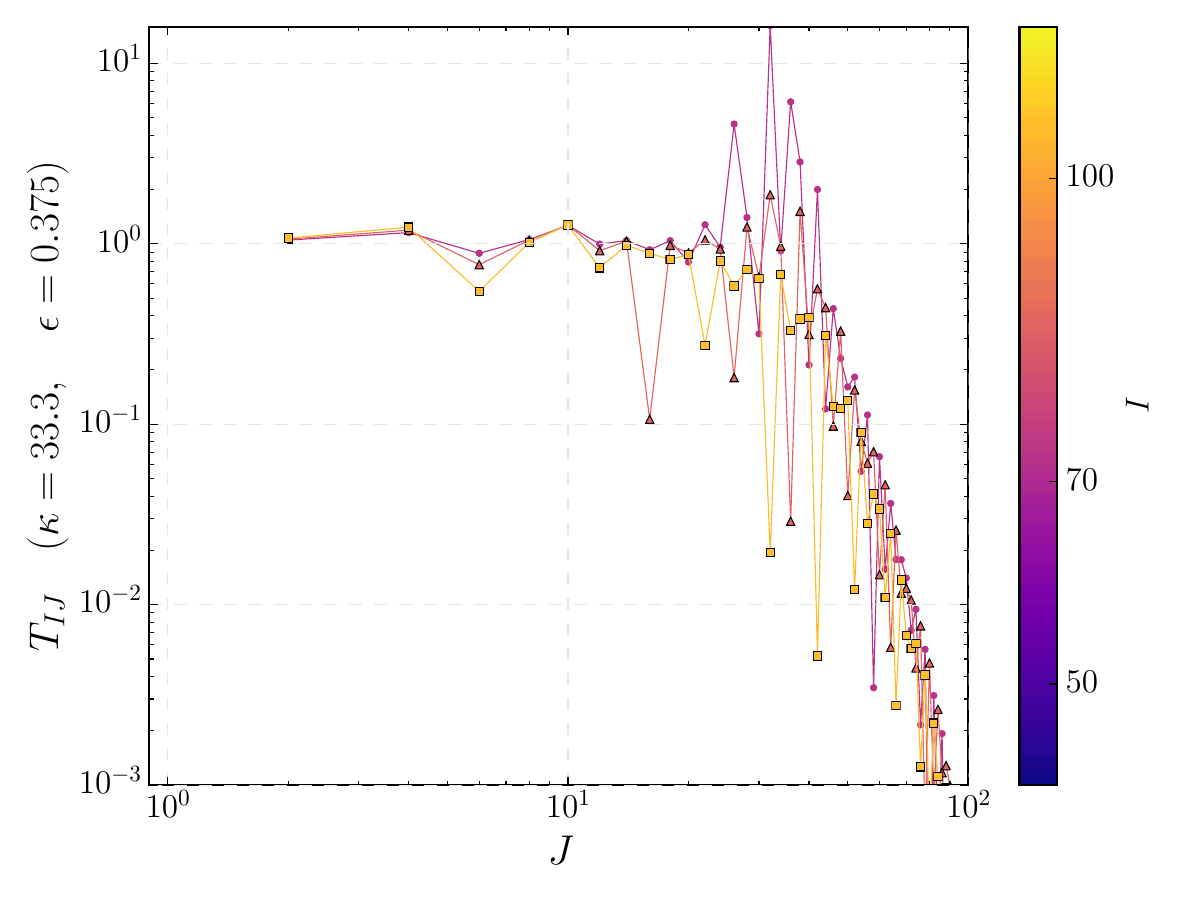}
}
\caption{\justifying Thermal relation: These plots correspond to the trajectories given by Eq. \eqref{eq:Traj-expand}, with $\epsilon=0.375$ and $\kappa=33.3$. The upper panel for the infrared modes. The lower panel for the ultraviolet modes. For low values of the $out$ modes, they reach values close to unit. The lines joining points do not represent any interpolation and have been included with the purpose of helping with the visualization of $T_{IJ}$ for fixed $I$.}
\label{fig:thermality-expand}
\end{figure}

As before, we can probe the thermal properties of these modes. We have again computed the thermality function given in Eq. \eqref{eq:thermal-func}. In Fig. \ref{fig:thermality-expand} we show $T_{IJ}$ for several values of the {\it in} modes $I$. As we can see, for this trajectory of the mirrors, in the limit of infrared $in$ and $out$ modes, the thermality condition is near unit and, hence, it is approximately fulfilled. However, for UV $in$ modes (large $I$), this condition is met only for very infrared $out$ modes. 

\begin{figure}[ht]
{\centering     
  \includegraphics[width = 0.48\textwidth]{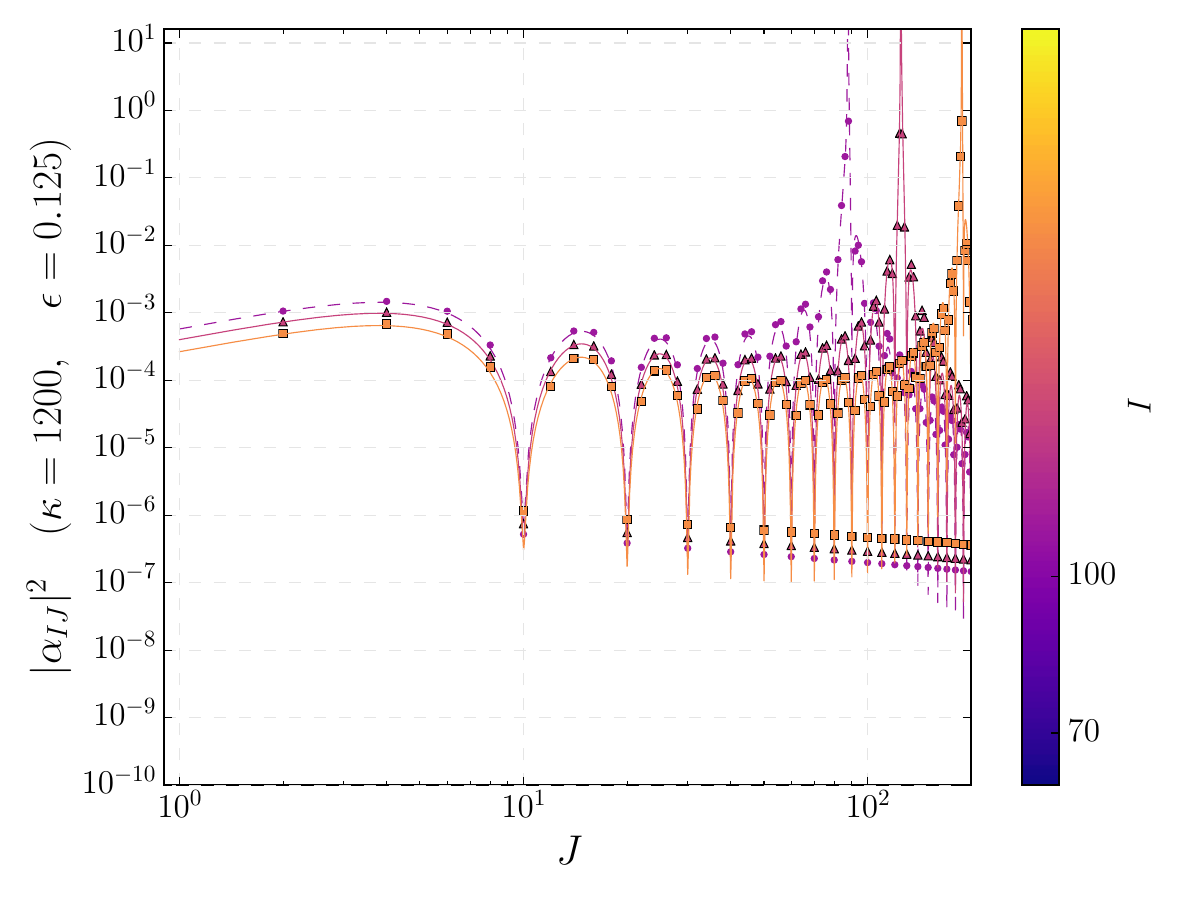}
  \includegraphics[width = 0.48\textwidth]{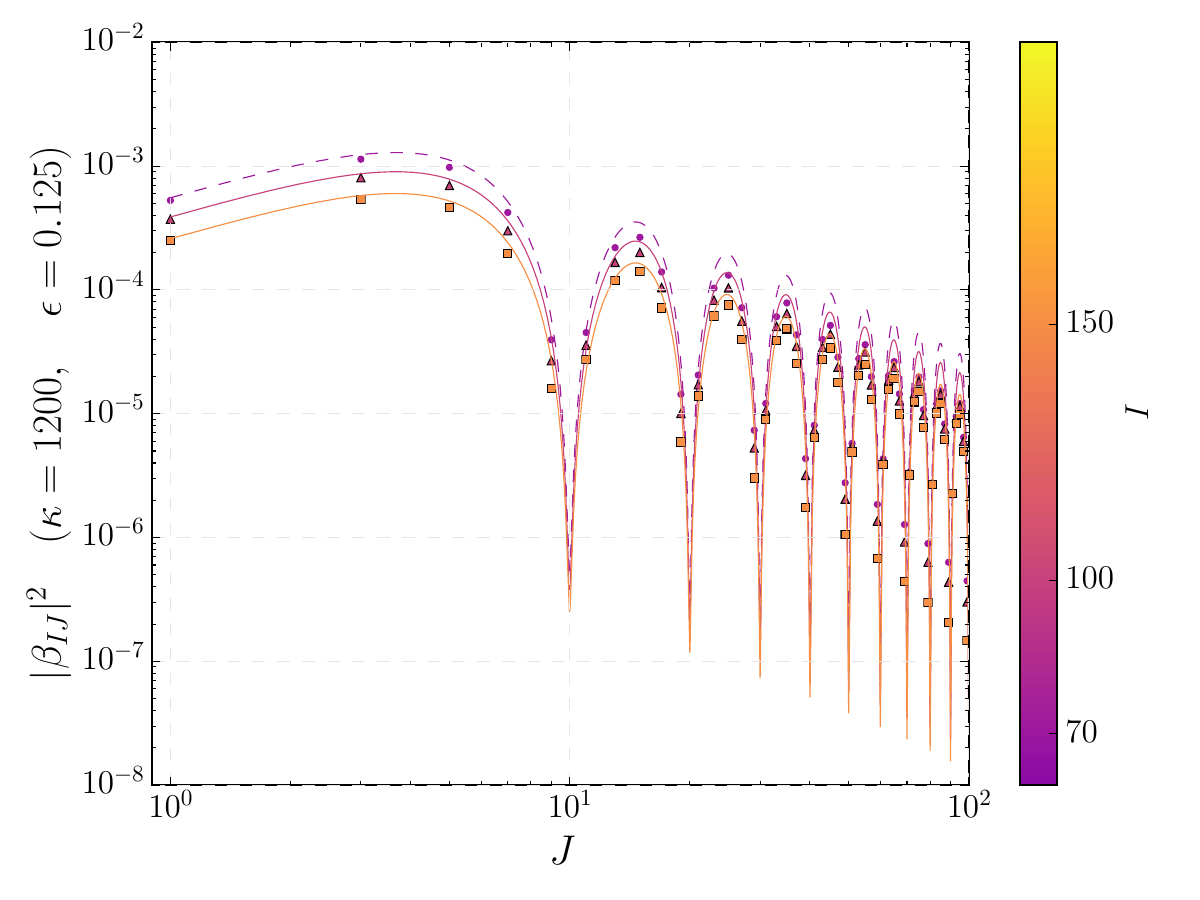}
}
\caption{\justifying Bogoliubov coefficients: These plots correspond to the trajectories given by Eq. \eqref{eq:Traj-expand}, with $\epsilon=0.125$ and $\kappa=1200$. The upper panel for the infrared modes. The lower panel for the ultraviolet modes. The lines interpolating points correspond to the fitting expressions in Eqs. \eqref{eq:betas-fit} and \eqref{eq:alphas-fit}.}
\label{fig:large-expand}
\end{figure}

{\bf Large accelerations:} For the case of large accelerations, we have analyzed several configurations. The mirrors transition very quickly from a steady configuration to a new one sharply approaching the speed of light for some period of time to suddenly break and stop in new stationary positions. The concrete configuration we will show here is given by $\epsilon=0.125$ and $\kappa=1200$. Other configurations with large accelerations show similar results. In Fig. \ref{fig:large-expand} we show the behavior of the modes in the frequency band ($(I,J)\in(70,150)$) and compare them with the fitting expression of Eqs. \eqref{eq:betas-fit} and \eqref{eq:alphas-fit}. In our simulations, we set $A_\alpha=1.0=A_\beta$ and $B_\alpha=10^{-3}=B_\beta$. Besides, $C_{\alpha}=1.0=C_{\beta}$. We show the Bogoliubov coefficients for several {\it in} modes $\omega_I$ as functions of the {\it out} frequencies $\omega_J$. We see that for relatively low $\omega_I$, the computed beta coefficients agree very well with Eqs. \eqref{eq:alphas-fit} and \eqref{eq:betas-fit}. The parameters $D_1$, $D_2$, $F$ and $\tilde \kappa$ are plotted in Fig. \ref{fig:expand-CDF-plot-b} as functions of $\omega_I$.  

\begin{figure}[ht]
{\centering     
  \includegraphics[width = 0.48\textwidth]{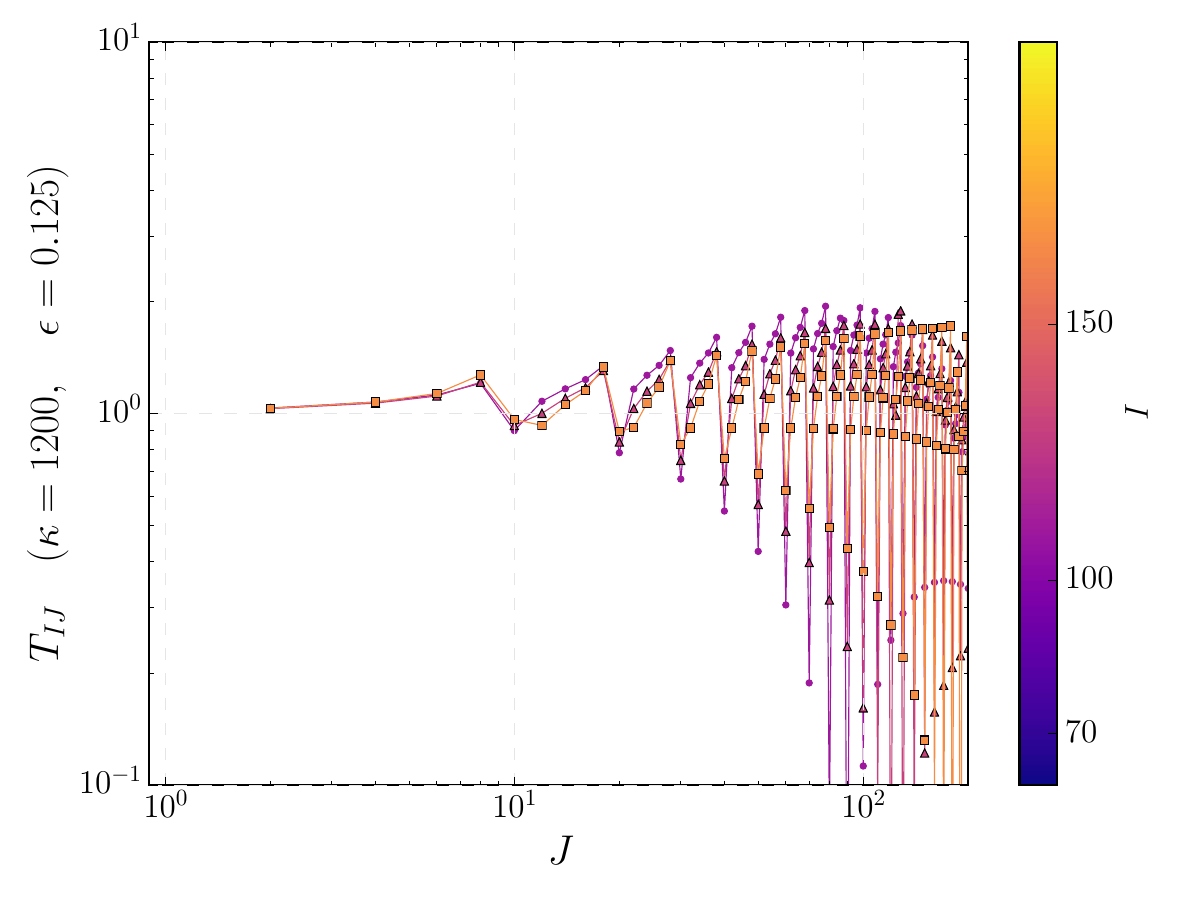}
}
\caption{\justifying Thermal relation: These plots correspond to the trajectories given by Eq. \eqref{eq:Traj-expand}, with $\epsilon=0.125$ and $\kappa=1200$.  For all values of the $in$ modes shown here, only for low values of the $out$ modes approach a thermal relation $T_{IJ}\simeq 1$. The lines joining points do not represent any interpolation and have been included only for visualization purposes. }
\label{fig:thermality-large-expand}
\end{figure}

For lower or higher $in$ modes compared to those shown here, the fitting expression for $|\alpha_{IJ}|^2$ still shows good agreement with the numerical results, but for $|\beta_{IJ}|^2$ we see important departures from the fitting expression. In conclusion, those modes depart strongly from a thermal distribution, and hence we do not consider them here. On the other hand, in Fig. \ref{fig:thermality-large-expand} we show the thermal relation $T_{IJ}$ in Eq. \eqref{eq:thermal-func} for several values of the $in$ modes $I$. As we can see, the most infrared $out$ modes yield $T_{IJ}\simeq 1$, following a thermal distribution. However, we also see departures from thermality in those modes where either $|\alpha_{IJ}|$ or $|\beta_{IJ}|$ reach their minima, departing from a thermal distribution. The lower or higher $in$ modes show a stronger departure of $T_{IJ}$ from the unit.

%------------------------------------------------
\subsection{Numerical analysis for an accelerating rigid cavity}
\label{Sec:2plt}
%------------------------------------------------
The trajectories of the boundaries of a rigid cavity (in the reference frame of the laboratory) are given by:
%------------------------------------------------
\begin{align}\nonumber
   f(t)= & \frac{s}{2\kappa}+\frac{1}{2\kappa}\bigg[\log\Big(\cosh\big(\kappa(t-t_0)\big)\Big)
   -\log\Big(\cosh\big(s-\kappa(t-t_0)\big)\Big)\bigg], \\ 
   g(t)= & 1 +\frac{s}{2\kappa}+\frac{1}{2\kappa}\bigg[\log\Big(\cosh\big(\kappa(t-t_0)\big)\Big)
   -\log\Big(\cosh\big(s-\kappa(t-t_0)\big)\Big)\bigg],
   \label{eq:Traj-2plt}
\end{align}
%------------------------------------------------
As in previous configurations, at $t \ll t_0$, both mirrors are nearly static at initial positions $x^f_{in}=0$ and $x^g_{in}=1$. Then, at $t\gg T\simeq t_0+\epsilon$ (with $\epsilon=s/\kappa$) their final position will be $x^f_{out} =(1+\epsilon)=x^g_{out}$. In the interval $[t_0,t_0+\epsilon]$ the mirrors accelerate, reaching a speed close to the speed of light. Besides, regarding Eqs. \eqref{eq:betas-fit} and \eqref{eq:alphas-fit}, and the \emph{out} frequencies are given by $\omega_J= \pi J/ L_0$ and their gap by $\Delta\omega_J=\pi/ L_0$. In this configuration, we were not able to find accurate fitting expressions for $|\alpha_{IJ}|$ or $|\beta_{IJ}|$.

\begin{figure}[ht]
{\centering     
  \includegraphics[width = 0.48\textwidth]{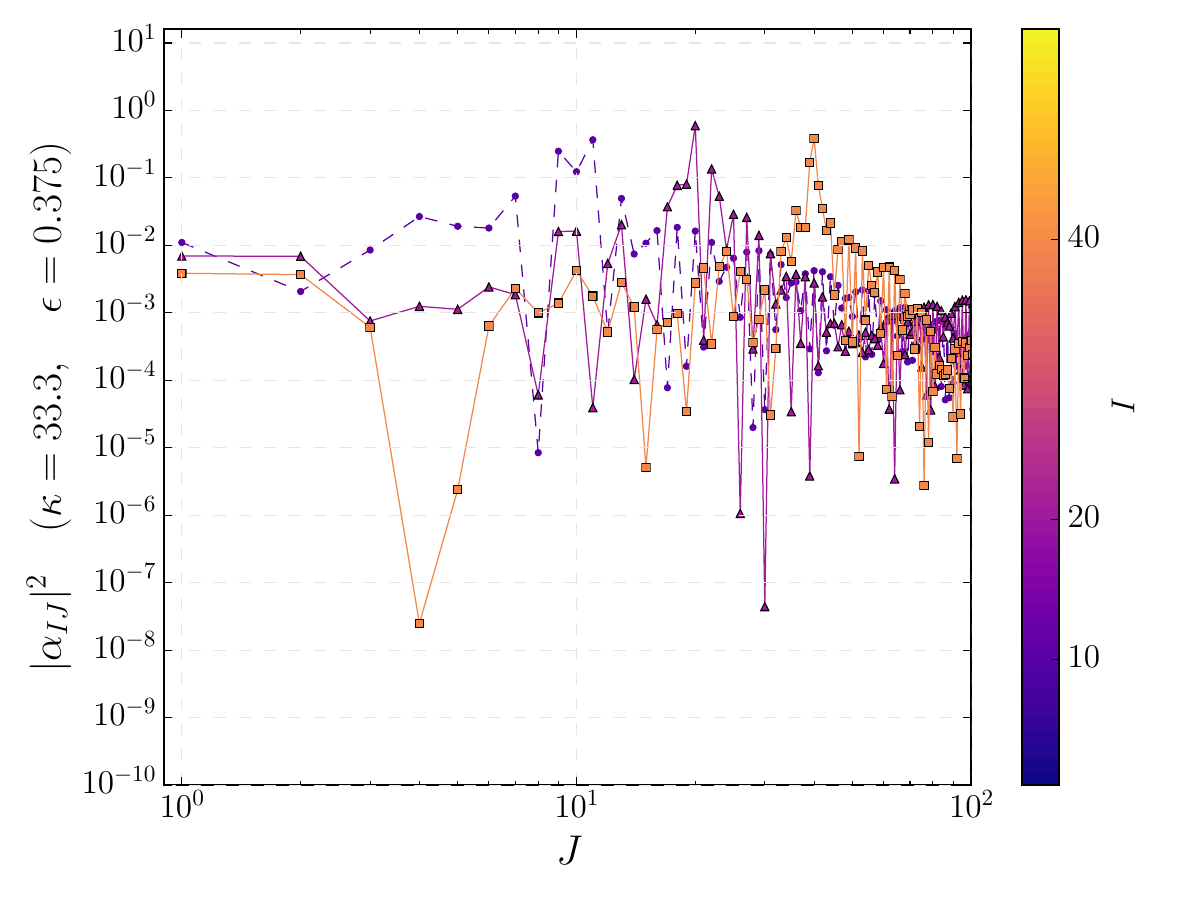}\includegraphics[width = 0.48\textwidth]{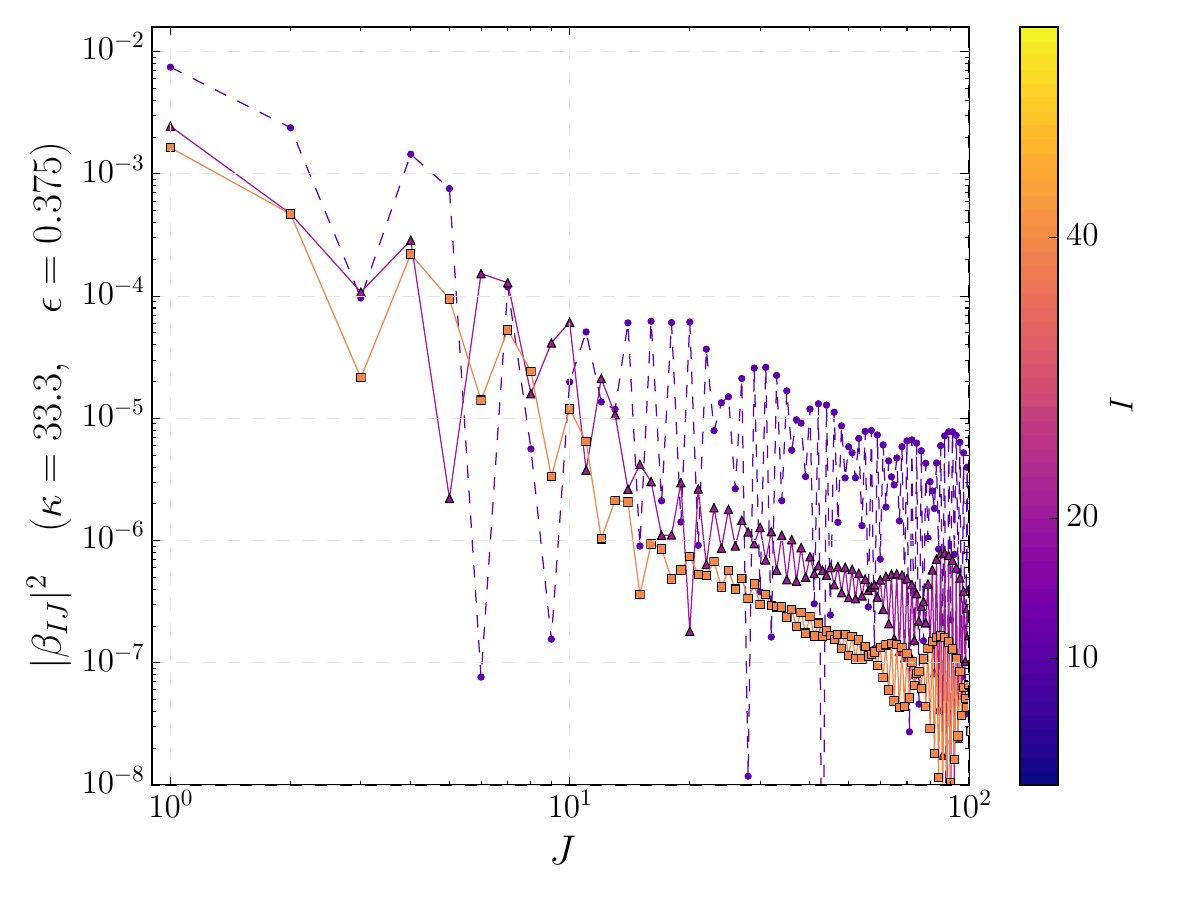}\\\includegraphics[width = 0.48\textwidth]{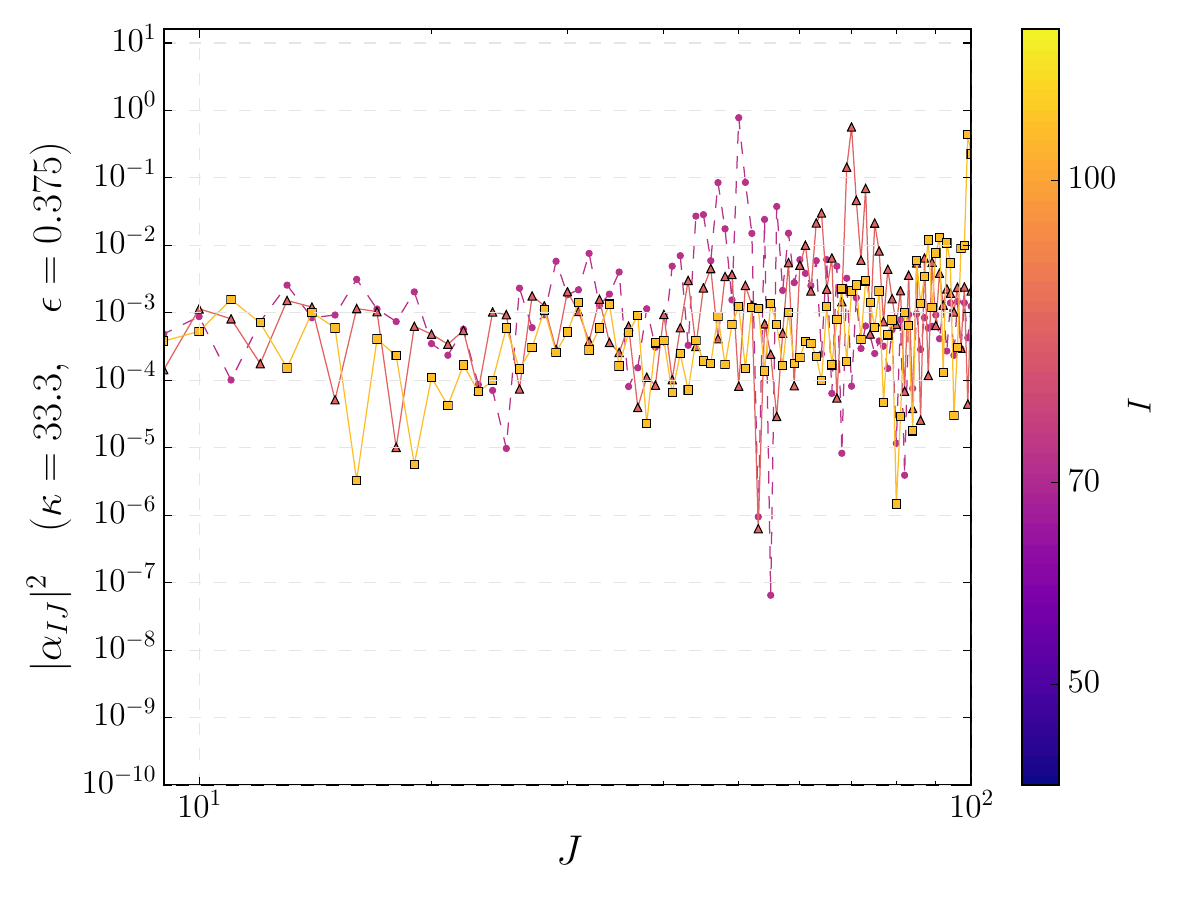}
  \includegraphics[width = 0.48\textwidth]{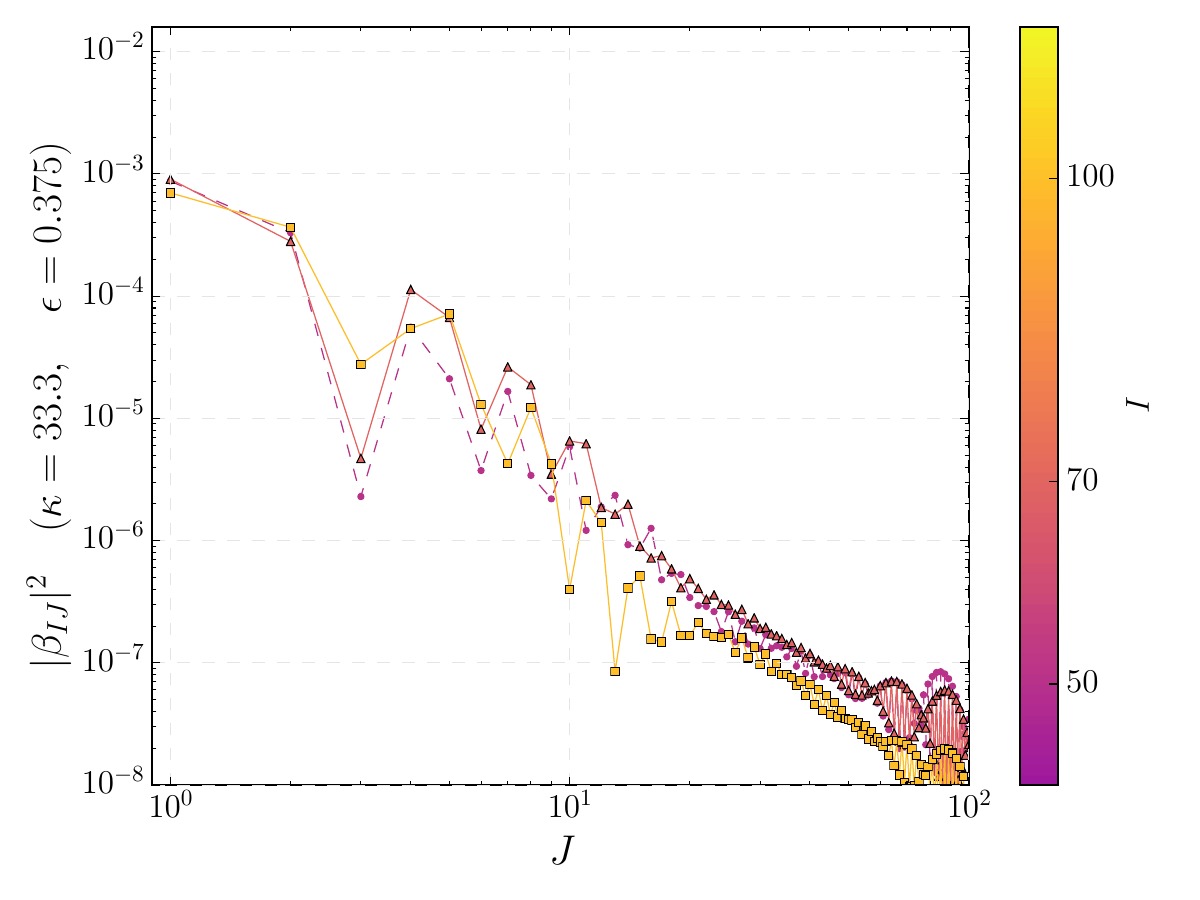}
}
\caption{\justifying Bogoliubov coefficients: These plots correspond to the trajectories given by Eq. \eqref{eq:Traj-2plt}, with $\epsilon=0.375$ and $\kappa=33.3$. The upper panel for the infrared modes. The lower panel for the ultraviolet modes. The lines joining points do not correspond to any fitting expression and are included only for visualization purposes.}
\label{fig:2plt-small}
\end{figure}

{\bf Small accelerations:} Following the same strategy as in previous sections, we have performed several simulations for small accelerations. Let us consider the concrete case of $\epsilon=0.375$ and $\kappa=33.3$. Fig. \ref{fig:2plt-small} includes the numerically computed Bogoliubov coefficients in the frequency band $(I,J)\in(20,100)$. The upper panel shows several (relatively) infrared {\it in} modes $\omega_I$ with respect to the {\it out} frequencies $\omega_J$. At very high $\omega_J$ we have checked numerically that the Bogoliubov coefficients do not decay as $\omega_J^{-n}$ with $n>0$. On the other hand, the lower panel shows several (relatively) ultraviolet {\it in} modes $\omega_I$ as functions of the {\it out} frequencies $\omega_J$. We have checked that the $\beta$ coefficients do not reach the behavior $\omega_J^{-n}$. Compared with previous configurations of the boundaries, we see a similar qualitative behavior regarding the overall amplitude and the oscillatory behavior of the coefficients. 

\begin{figure}[ht]
{\centering     
  \includegraphics[width = 0.48\textwidth]{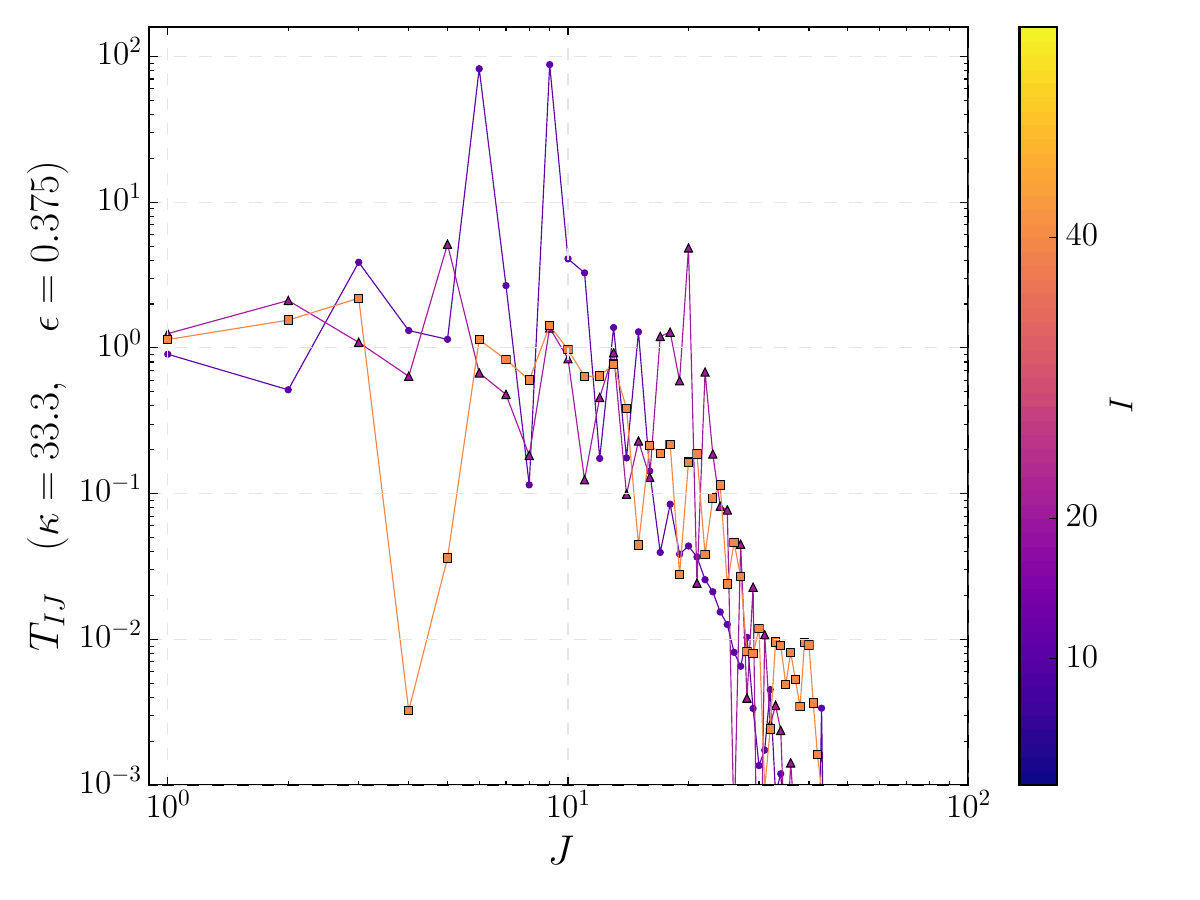}\includegraphics[width = 0.48\textwidth]{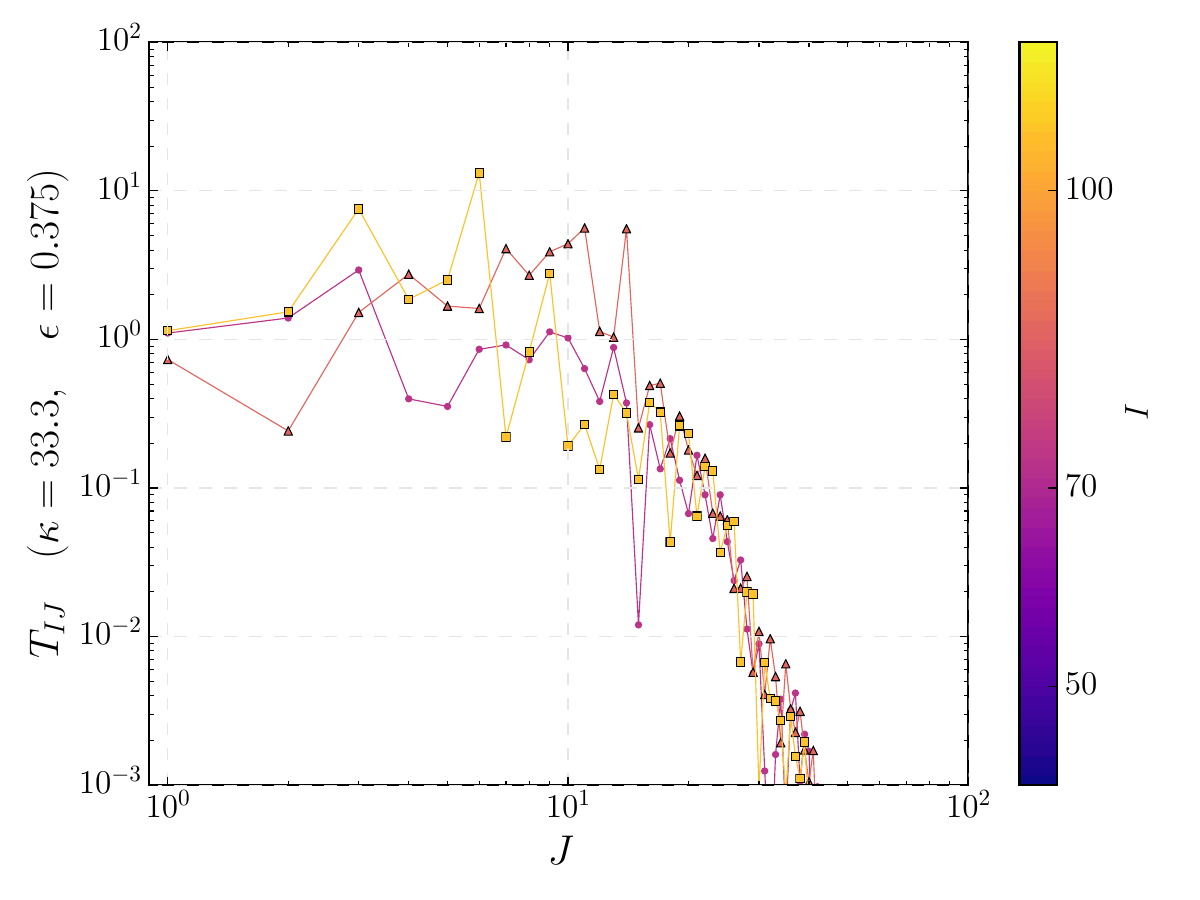}
}
\caption{\justifying Thermal relation: These plots correspond to the trajectories given by Eq. \eqref{eq:Traj-2plt}, with $\epsilon=0.375$ and $\kappa=33.3$. The left panel for the infrared \emph{in} modes. The right one for the ultraviolet \emph{in} modes. The lines joining points do not correspond to any fitting expression and are included for visualization purposes.}
\label{fig:thermality-2plt}
\end{figure}

Although the Bogoliubov coefficients do not follow a fitting expression that indicates some thermal aspects of the particle production, we will still analyze whether they agree with the thermality function given in Eq. \eqref{eq:thermal-func}. In Fig. \ref{fig:thermality-2plt} we show $T_{IJ}$ for several values of the {\it in} modes $I$. We set $D_1=0$ because we do not have a suitable fitting expression in this case. We see that although for some frequencies this function is of order 1 for low {\it out} modes, it considerably departs from the unit. Actually, for some {\it out} modes it can be of the order of $10^1$ to $10^2$. For higher modes, we see a similar behavior. 
Hence, we conclude that the particle production for these trajectories is not associated with a thermal state of the field. The reason behind this non-thermality must be the particle production on the left boundary, that contaminates the spectrum with ultraviolet modes, those that are reflected by a surface that is producing a blue-shift, rather than a red-shift, like the right boundary.

\begin{figure}[ht]
{\centering     
\includegraphics[width = 0.48\textwidth]{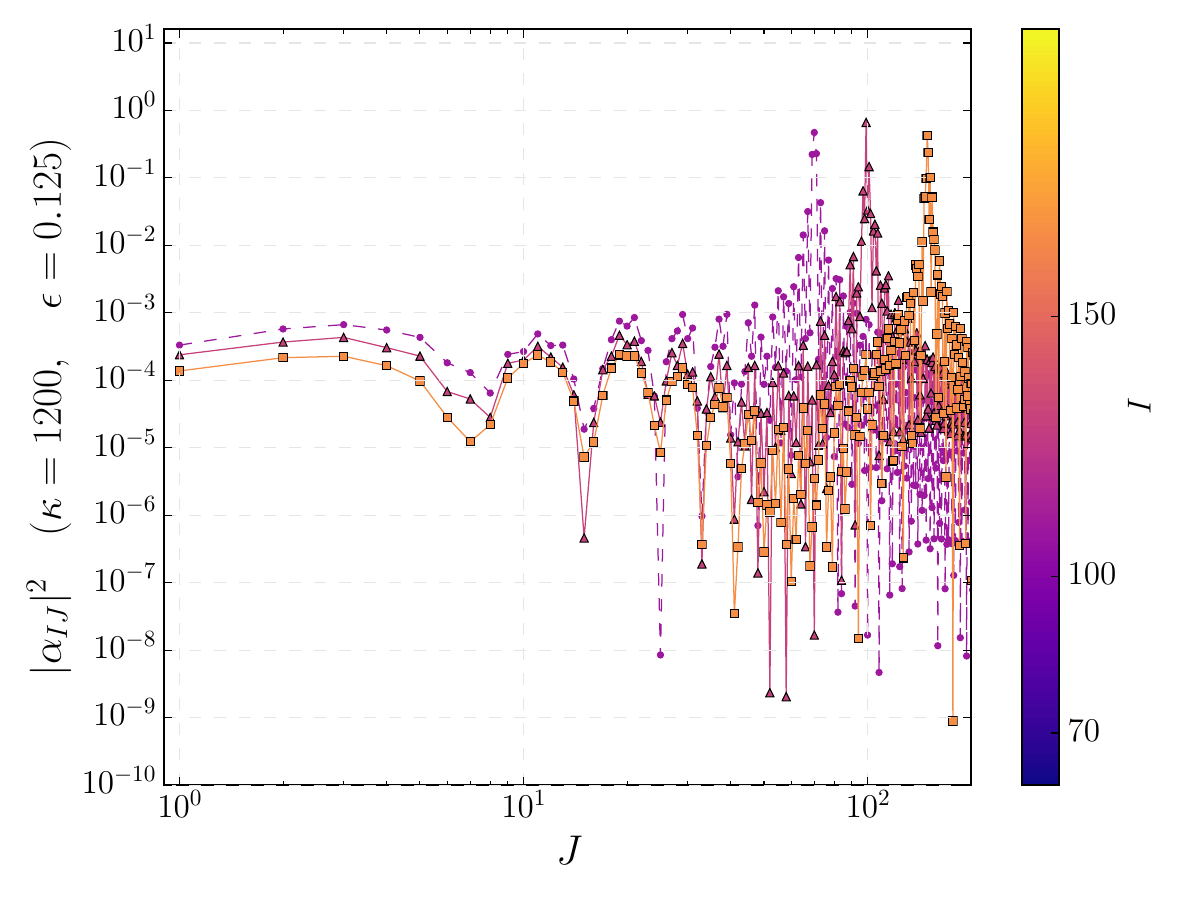}
  \includegraphics[width = 0.48\textwidth]{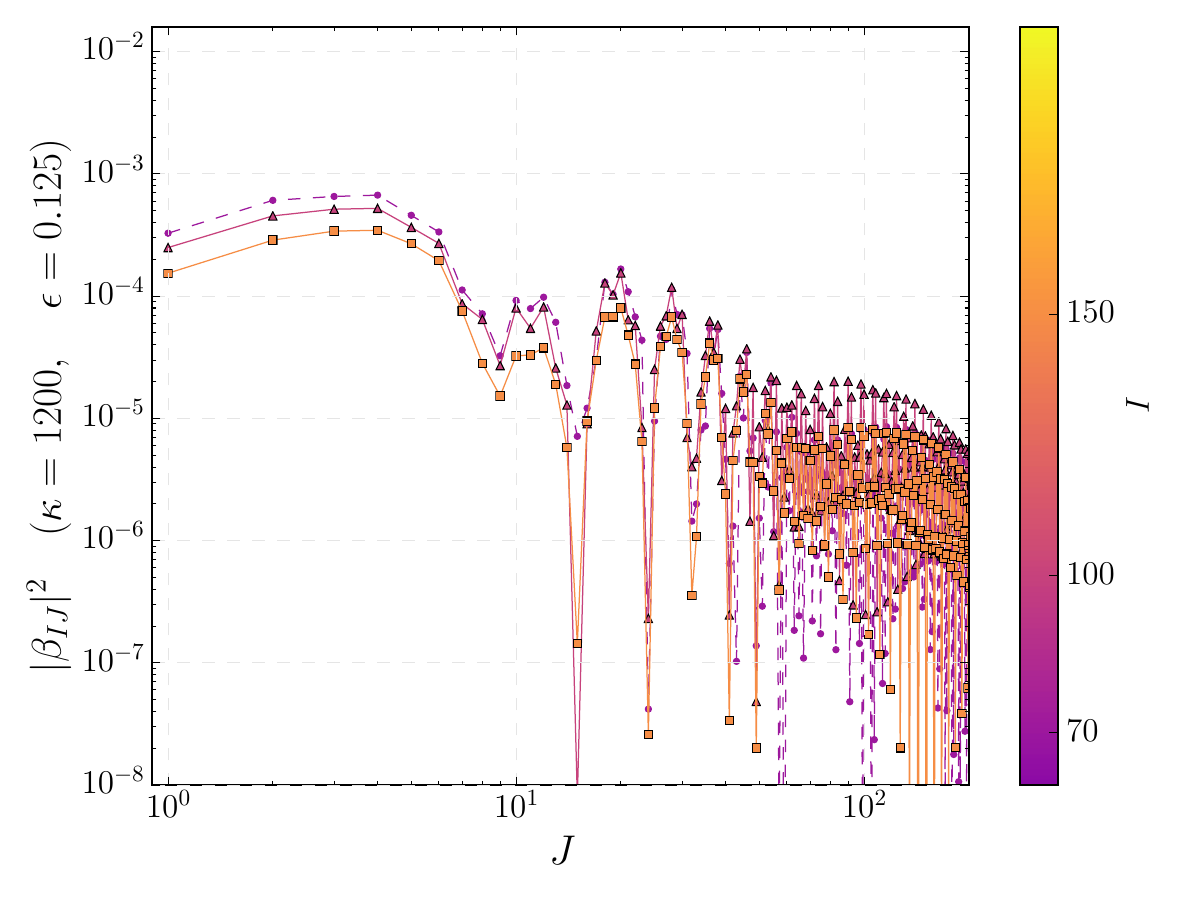}
}
\caption{\justifying Bogoliubov coefficients: These plots correspond to the trajectories given by Eq. \eqref{eq:Traj-2plt}, with $\epsilon=0.125$ and $\kappa=1200$. The lines joining points do not correspond to any fitting expression and are included for visualization purposes.}
\label{fig:2plt-large}
\end{figure}

{\bf Large accelerations:} We have also studied several simulations for large accelerations. The concrete case will show here is for $\epsilon=0.125$ and $\kappa=1200$. In Fig. \ref{fig:2plt-large} we show the resulting Bogoliubov coefficients in the band frequency $(I,J)\in(20,100)$. In the upper panel, we show several (relatively) infrared {\it in} modes $\omega_I$ versus the {\it out} frequencies $\omega_J$. At very high $\omega_J$ we have checked numerically that the Bogoliubov coefficients do not decay as $\omega_J^{-n}$ for any $n>0$. Actually, they do not show an amplitude that behaves monotonically. On the other hand, the lower panel includes several (relatively) ultraviolet {\it in} modes $\omega_I$ as functions of the {\it out} frequencies $\omega_J$. We see that the $\beta$-coefficients do not reach the behavior $\omega_J^{-n}$. Nevertheless, there is a qualitatively similar behavior in the overall amplitude and the oscillatory behavior of the coefficients (with no similar frequencies and amplitudes) compared to the ones obtained for the previous configurations of the mirrors.

\begin{figure}[ht]
{\centering     
  \includegraphics[width = 0.48\textwidth]{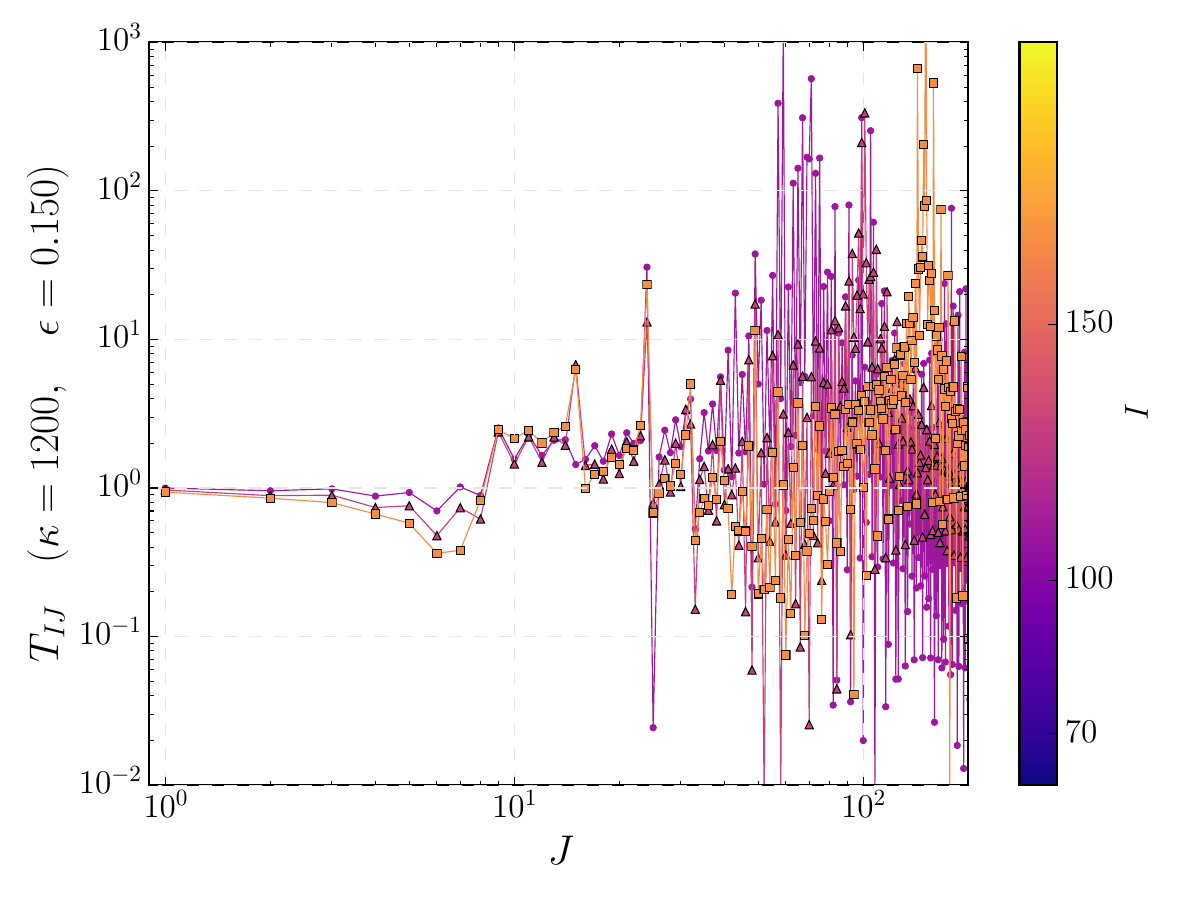}
}
\caption{\justifying Thermal relation: These plots correspond to the trajectories given by Eq. \eqref{eq:Traj-expand}, with $\epsilon=0.125$ and $\kappa=1200$. For low values of the $out$ modes, they reach values order one. However, as we approach $J\simeq 5$ and higher $out$ frequencies we see significant deviations from unit by several orders of magnitude. The lines joining points do not correspond to any fitting expression and are included for visualization purposes only.}
\label{fig:thermality-large-2plt}
\end{figure}

Although the Bogoliubov coefficients do not follow a fitting expression that indicates some thermal aspects of the particle production, we will still analyze if they agree with the thermality function given in Eq. \eqref{eq:thermal-func}. In Fig. \ref{fig:thermality-large-2plt} we show $T_{IJ}$ for several values of the {\it in} modes $I$ for $D_1=0$ since we do not have an appropriate fitting expression. We see that although for some frequencies this function is of order 1 for low {\it out} modes, it considerably departs from the unit for other modes. Actually, for some {\it out} modes it can be of order $10^1$ to $10^2$. For higher modes, see a similar behavior. 
Hence, we conclude that the particle production for these trajectories does not follow a thermal distribution. Again, the blue-shift in the left boundary seems to contaminate the spectrum with ultraviolet modes. 

%------------------------------------------------
\section{Numerical analysis for collapsing cavities}
\label{Sec:collap-cav}
%------------------------------------------------

In the previous section, we have discussed some properties related to particle production for several configurations that involve cavities that either expand or move rigidly, following trajectories that approach the Fulling-Davies accelerating mirror. In this section, we will discuss the effects of the corresponding time-reversed configurations. Namely, the case of collapsing cavities (the rigid configuration will not be discussed since it is already symmetric under $t\to (-t)$).

From a computational perspective, we do not need to carry out additional simulations. Concretely, if we have that the evolved {\it in} states are written on the {\it out} basis ($t\to+\infty$), as
\begin{equation}\label{eq:bogou}
   {}^{(in)}{\bf u}^{(I)}(t) = \sum_{J=1}^{\infty} \alpha_{IJ}\;{}^{(out)}{\bf u}^{(J)}(t)+\beta_{IJ}\;{}^{(out)}\bar {\bf u}^{(J)}(t),
\end{equation}
one can easily invert this relation and express the {\it out} basis evolved backwards in time ($t\to-\infty$) as
%------------------------------------------------
\begin{equation}\label{eq:bogow}
    {}^{(out)}{\bf u}^{(I)}(t) = \sum_{J=1}^{\infty} \sigma_{IJ}\;{}^{(in)}{\bf u}^{(J)}(t)+\lambda_{IJ}\;{}^{(in)}\bar {\bf u}^{(J)}(t),
\end{equation}
with $\sigma_{IJ} = \bar\alpha_{JI}$ and $\lambda_{IJ} = -\beta_{JI}$. Hence, particle production and mode-mixing can actually be studied through the very same numerical Bogoliubov coefficients of the previous section. In the following, we show some significant numerical results for these kind of trajectories.
%------------------------------------------------
\subsection{One collapsing mirror}
\label{Sec:1plt-collapse}
%------------------------------------------------
For the time-reversed trajectory considered in Eq. \eqref{eq:Traj-1plt}, the left cavity remains stationary at all times at $x^f_{in}=1$, while the right starts at $x^g_{in} =(1+\epsilon)$ and accelerates collapsing the cavity until it reaches $x^g_{out}=1$. \footnote{It is important to note the difference with previous trajectories, where the initial length of the cavity is $L=1$, while here is $L=1+\epsilon$.}

{\bf Small accelerations:} We have analyzed several simulations for relatively small accelerations. We summarize our findings with a concrete example given by $\epsilon=0.375$ and $\kappa=33.3$. We only show the numerical results, since we have not found analytical fitting expressions for them, since their behavior with respect to {\it in} modes for different given values of {\it out} modes is not trivial.

\begin{figure}[ht]
{\centering     
  \includegraphics[width = 0.48\textwidth]{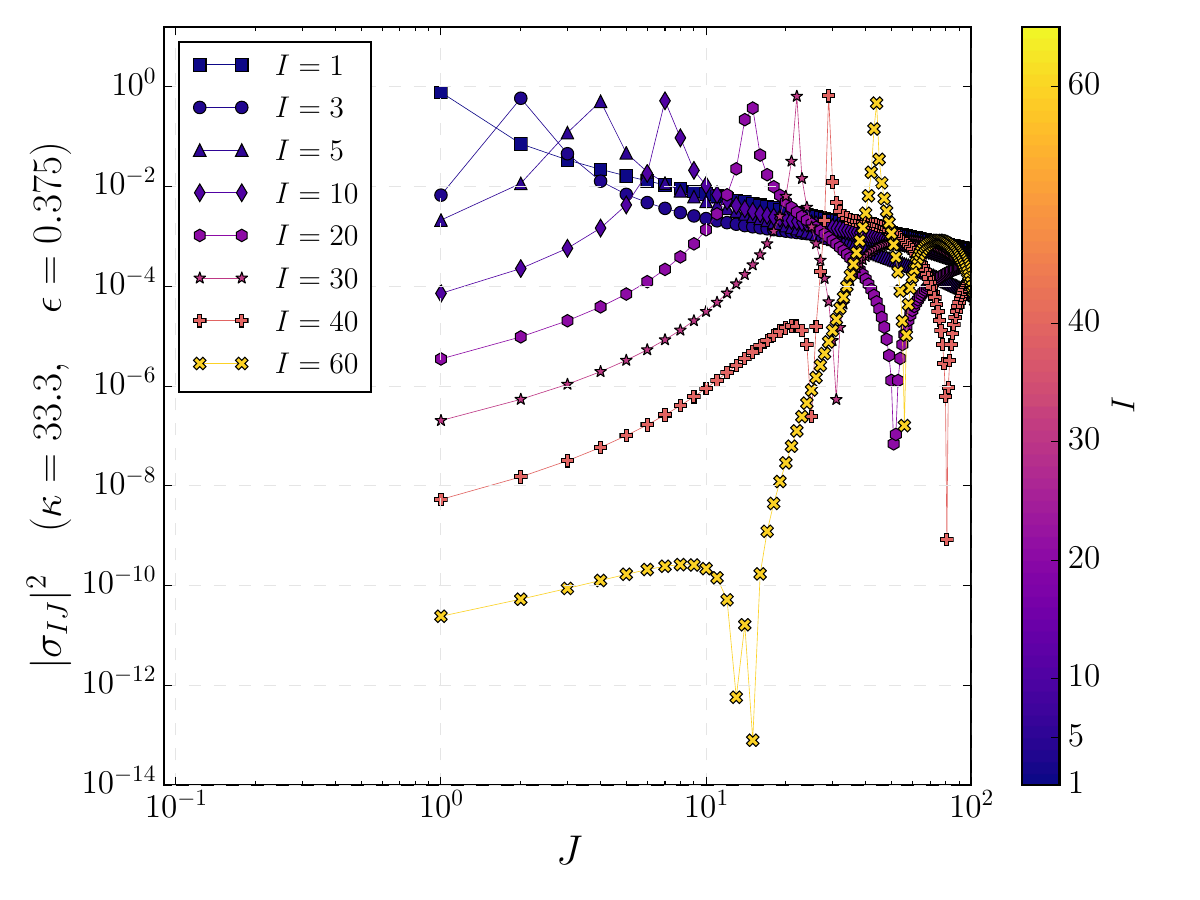}\includegraphics[width = 0.48\textwidth]{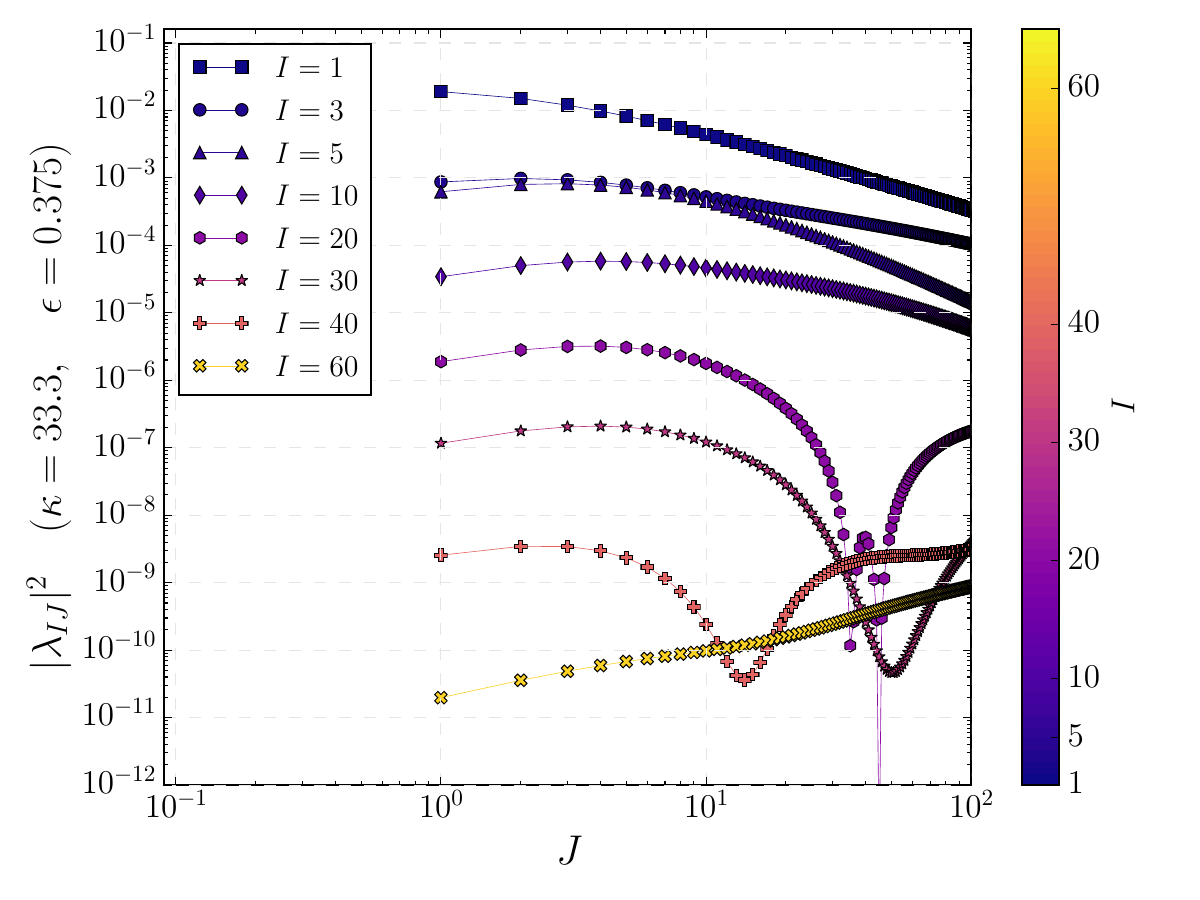}
}
\caption{\justifying Bogoliubov coefficients: These plots correspond to the trajectories given by a time reversed Eq. \eqref{eq:Traj-1plt}, with $\epsilon=0.375$ and $\kappa=33.3$. The lines interpolating points do not correspond to any fitting expression. The plots also show different markers for each value of $I$. }
\label{fig:1plt-collapse}
\end{figure}

For instance, let us consider the Bogoliubov coefficients for several values of the {\it in} modes $I$ for $|\sigma_{IJ}|^2$ and $|\lambda_{IJ}|^2$ (i.e. {\it in} modes $J$ for $|\alpha_{IJ}|^2$ and $|\beta_{IJ}|^2$). As we see in Fig. \ref{fig:1plt-collapse}, $|\sigma_{IJ}|^2$ grows as a function of $J$ until it reaches a peak at $J\simeq I$, and then it oscillates, resembling a Bessel function. This behavior of $|\sigma_{IJ}|^2$ involves a strong mode mixing. On the other hand, $|\lambda_{IJ}|^2$ shows a nontrivial behavior as a function of $J$ ({\it out} modes) for fixed values of $I$. We also see that the higher the $J$, the smaller is the amplitude of $|\lambda_{IJ}|^2$, indicating that there is less particle creation in the ultraviolet sector.    

We have also seen that the Richardson extrapolation for $|\sigma_{IJ}|^2$ shows relatively good results, in the sense that there is convergence at low values of $I$ and $J$, but the convergence is lost for high $I$ and very small and high values of $J$ (for intermediate values of $J$ the convergence is still preserved). On the other hand, the convergence of $|\lambda_{IJ}|^2$ is less robust, specially for high values of $J$. It does not seem to be a problem of our numerical methods, but a consequence of the cutoff we introduce on each simulation.

{\bf Large accelerations:} In the case of sharp accelerations of the right mirror, we have carried out several simulations. All our findings can be summarized in a concrete representative example: $\epsilon=0.125$ and $\kappa=1200$. Again, we only show the numerical results since we were unable to find simple analytic fitting expressions for this scenario.

\begin{figure}[ht]
{\centering     
  \includegraphics[width = 0.48\textwidth]{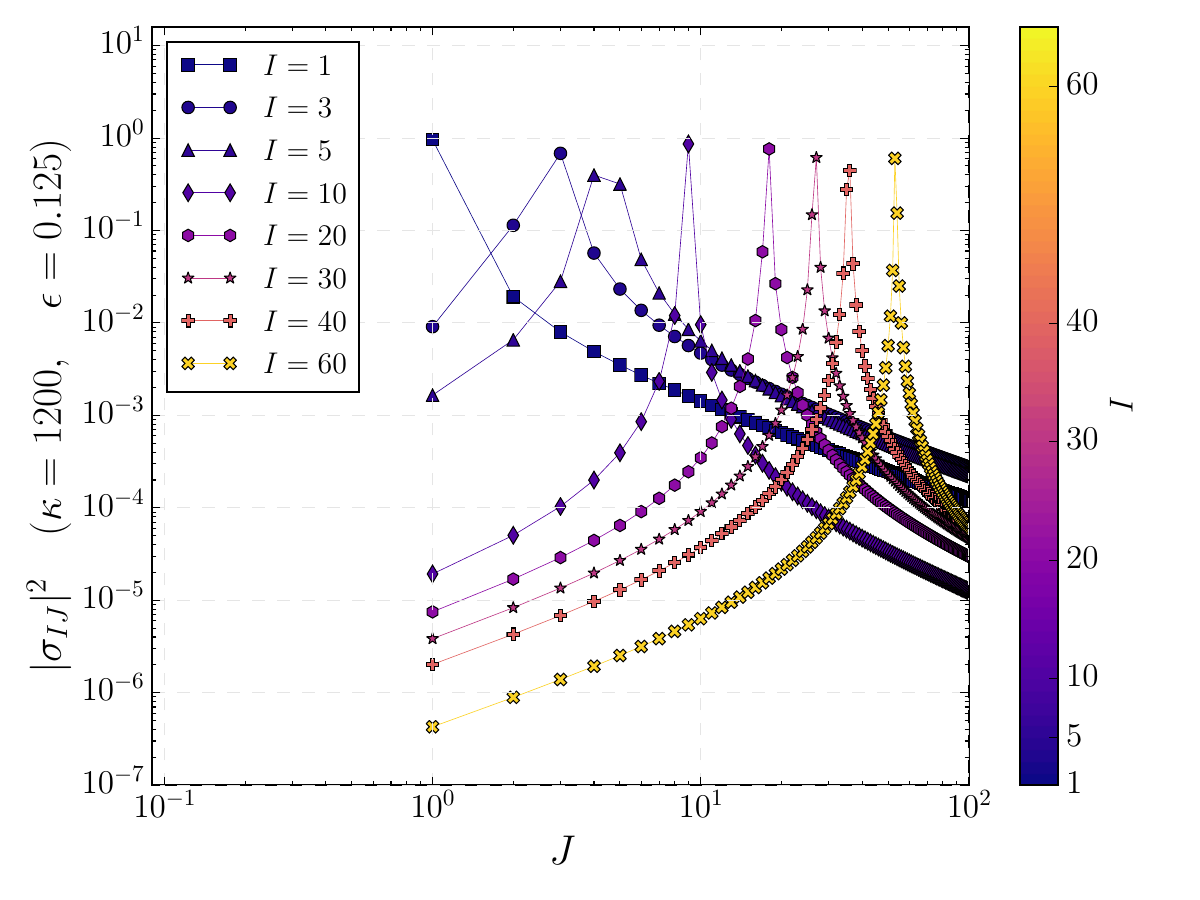}\includegraphics[width = 0.48\textwidth]{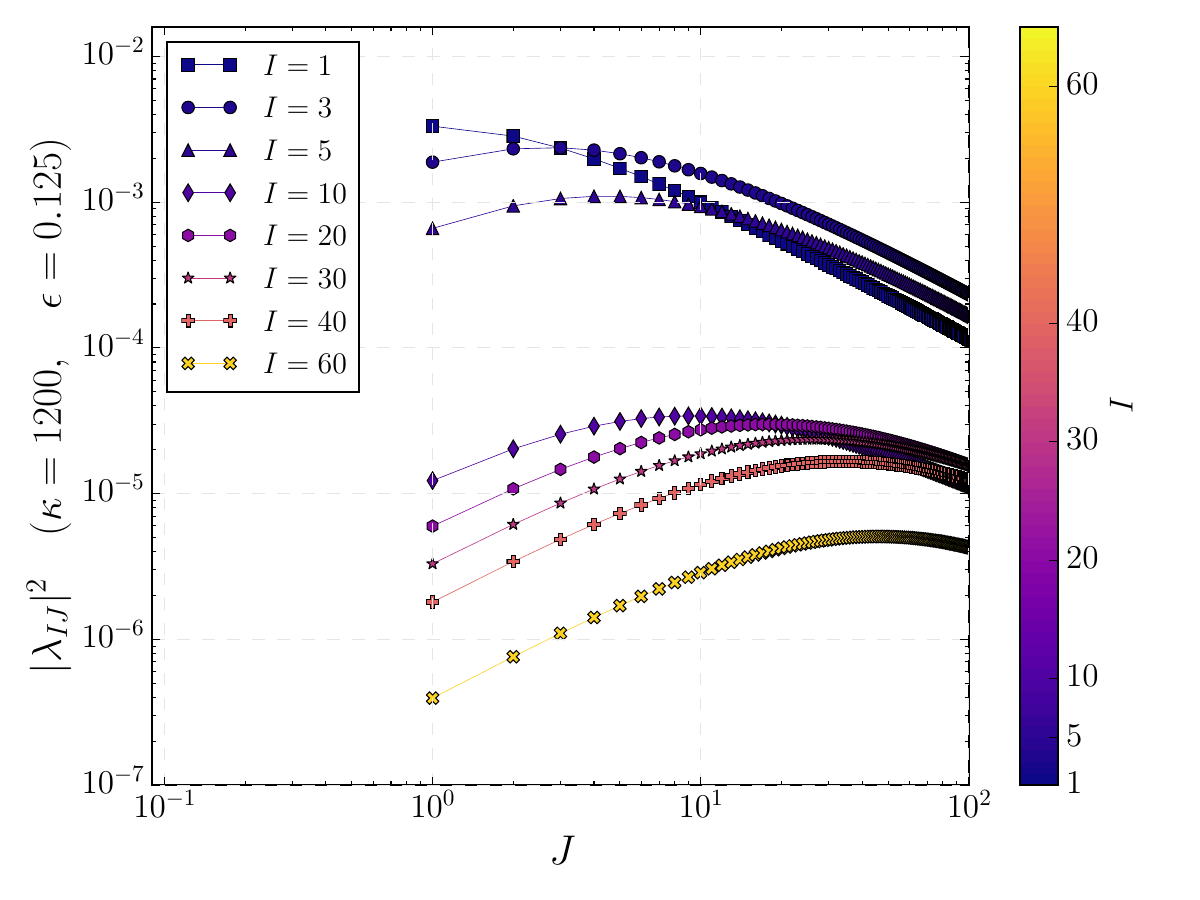}
}
\caption{\justifying Bogoliubov coefficients: These plots correspond to the trajectories given by a time reversed Eq. \eqref{eq:Traj-1plt}, but now with $\epsilon=0.125$ and $\kappa=1200$. The lines interpolating points do not correspond to any fitting expression. The plots also show different markers for each value of $I$. }
\label{fig:1plt-collapse-sharp}
\end{figure}

Let us again consider several values of the {\it in} modes $I$ for $|\sigma_{IJ}|^2$ and $|\lambda_{IJ}|^2$. As we see in Fig. \ref{fig:1plt-collapse-sharp}, $|\sigma_{IJ}|^2$ shows a similar behavior as for small accelerations: it grows as a function of $J$ until it reaches a peak at $J\simeq I$, and then it decreases. We do not see oscillations in the extrapolated data, but for each simulation (fixed number of modes $N$), we do see some oscillations in the most UV modes (largest values of $J$). This behavior of $|\sigma_{IJ}|^2$ involves a strong mode mixing. On the other hand, $|\lambda_{IJ}|^2$ shows a nontrivial behavior as a function of $J$ ({\it out} modes) for fixed values of $I$. We also see that the higher the $I$, the smaller is the amplitude of $|\lambda_{IJ}|^2$, indicating that there is less particle creation in the ultraviolet sector. We also see that they reach a maximum value around $J\simeq I$. 

Regarding the Richardson extrapolation for $|\sigma_{IJ}|^2$ and $|\lambda_{IJ}|^2$, we do see a very good convergence even for high values of $I$ and $J$.  

%------------------------------------------------
\subsection{Two symmetrically collapsing mirrors}
\label{Sec:expand-collapse}
%------------------------------------------------

Another (collapsing) trajectory we have analyzed corresponds to the time-reversed configuration of in Eq. \eqref{eq:Traj-expand}. Here, the left and right cavities start at $x^f_{in}=-(\epsilon)$ and $x^g_{in} =(1+\epsilon)$, respectively. Then, they accelerate towards each other, until they stop at positions $x^f_{out}=1=x^g_{out}$. 

{\bf Small accelerations:} In this case of smooth trajectories, all the simulations performed, for different values of the acceleration, show similar results. We summarize them with a concrete representative example given by $\epsilon=0.375$ and $\kappa=33.3$. As in the previous case, we report only the numerical results. No simple fitting expressions have been found.

\begin{figure}[ht]
{\centering     
  \includegraphics[width = 0.48\textwidth]{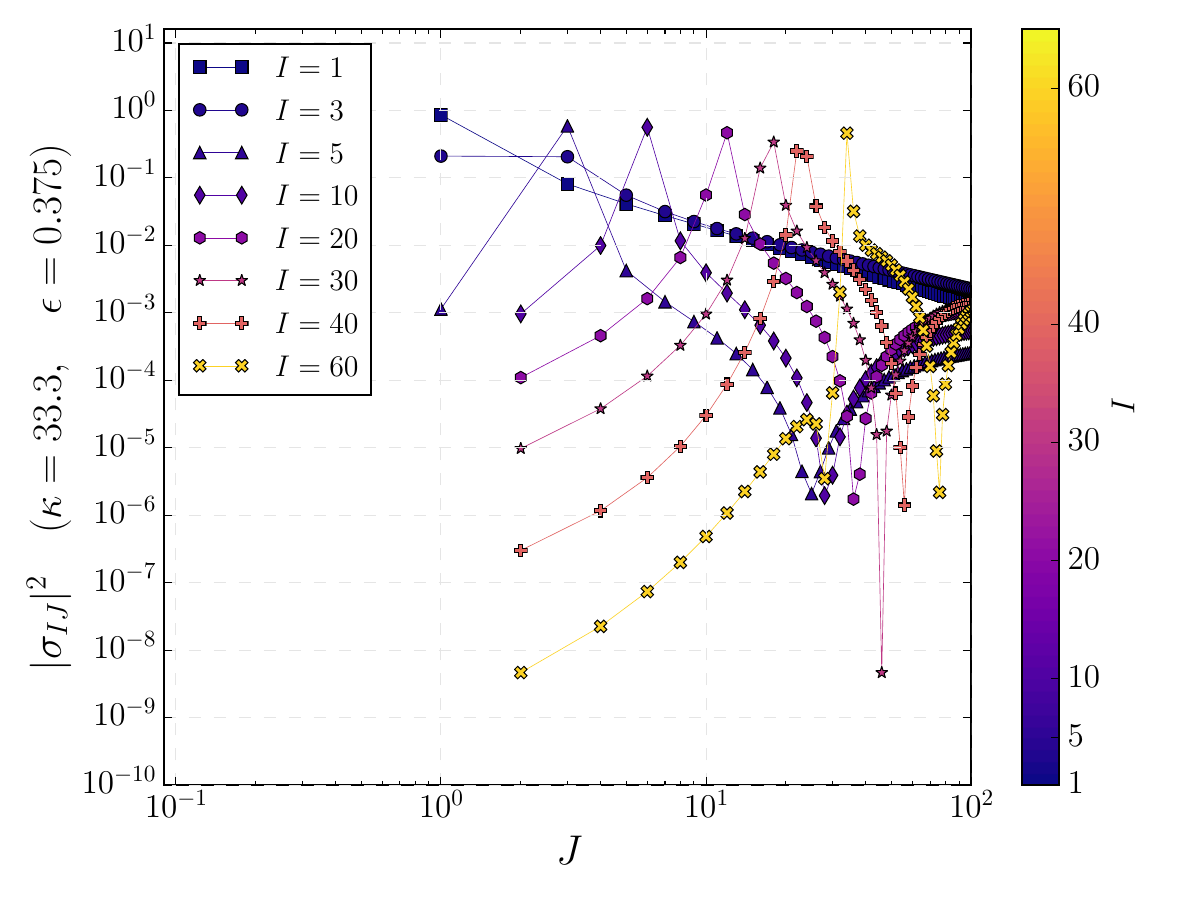}\includegraphics[width = 0.48\textwidth]{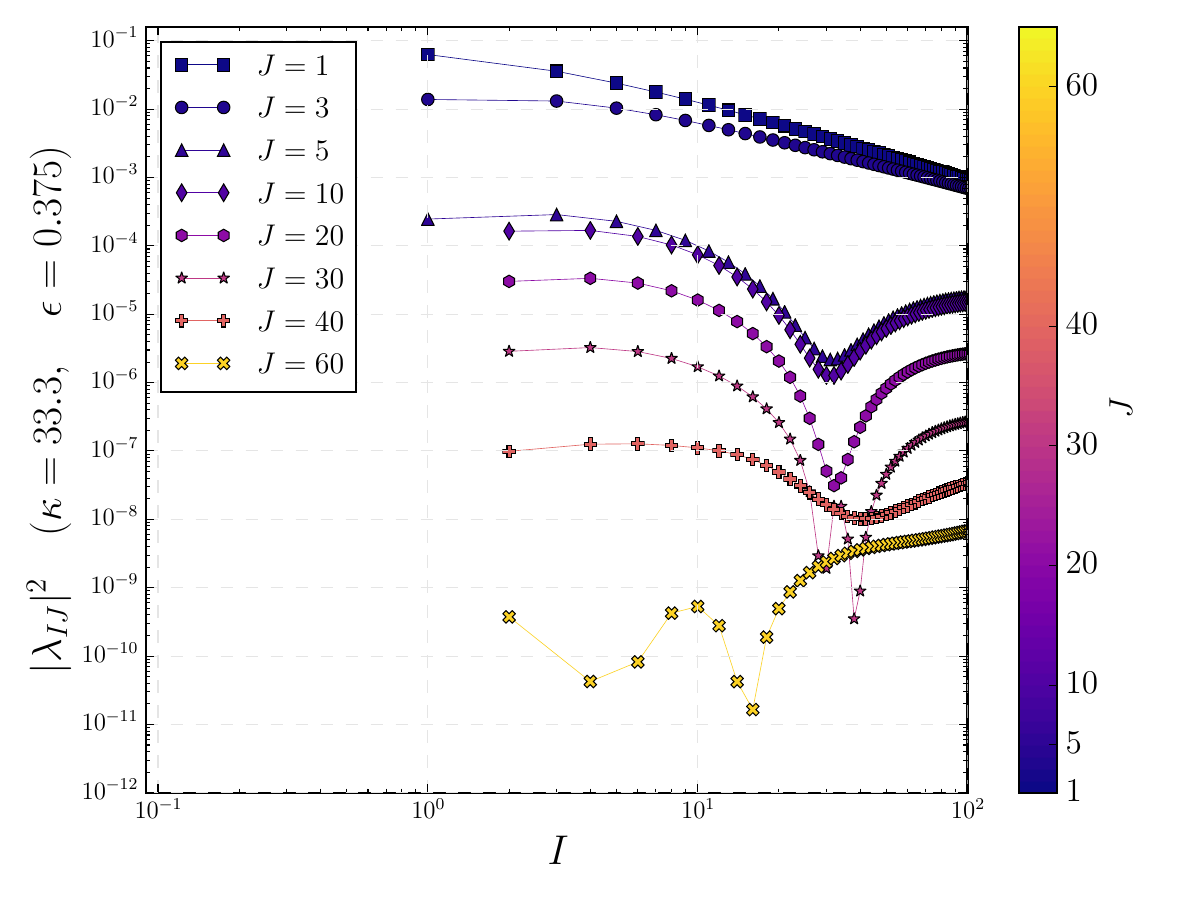}
}
\caption{\justifying Bogoliubov coefficients: These simulations correspond to the time-reversed trajectories given in Eq. \eqref{eq:Traj-expand}, with $\epsilon=0.375$ and $\kappa=33.3$. The lines interpolating points do not correspond to any fitting expression. The plots also show different markers for each value of $I$. }
\label{fig:expand-collapse}
\end{figure}

In Fig. \ref{fig:expand-collapse} we show $|\sigma_{IJ}|^2$ and $|\lambda_{IJ}|^2$ for several values of the {\it in} modes $I$. Concretely, $|\sigma_{IJ}|^2$ again starts growing as a function of $J$ until it reaches a peak at $J\simeq I$, and then it oscillates, with a behavior similar to a Bessel function. This behavior of $|\sigma_{IJ}|^2$ involves a strong mode mixing. On the other hand, $|\lambda_{IJ}|^2$ has a nontrivial behavior as a function of $J$ ({\it out} modes) for fixed values of $I$. In addition, its amplitude decreases with $J$, which indicates less particle creation in the UV sector. Moreover, for this trajectory, even and odd modes are decoupled. 

As in the previous configuration (one moving mirror and small accelerations), we have seen that the Richardson extrapolation for $|\sigma_{IJ}|^2$ is relatively good. We see convergence at low values of $I$ and $J$, but convergence worsens for high $I$, and very small and high values of $J$ (intermediate values of $J$ near the peak preserve convergence). Regarding the convergence of $|\lambda_{IJ}|^2$, for high values of $I$ and $J$, we see that it is lost (higher values in the cutoff in the number of modes yield smaller $|\lambda_{IJ}|^2$). Again, it does seem to be a consequence of the cutoff introduced on each simulation rather than a problem of our numerical methods.

{\bf Large accelerations:} We have also performed several simulations for sharp accelerations for the configuration of two moving (collapsing) mirrors. We have found similar results in all our simulations and our findings are summarized in the concrete case $\epsilon=0.125$ and $\kappa=1200$. We do not show any fitting expression together with the numerical results as there is no simple one that describes their behavior.

\begin{figure}[ht]
{\centering     
  \includegraphics[width = 0.48\textwidth]{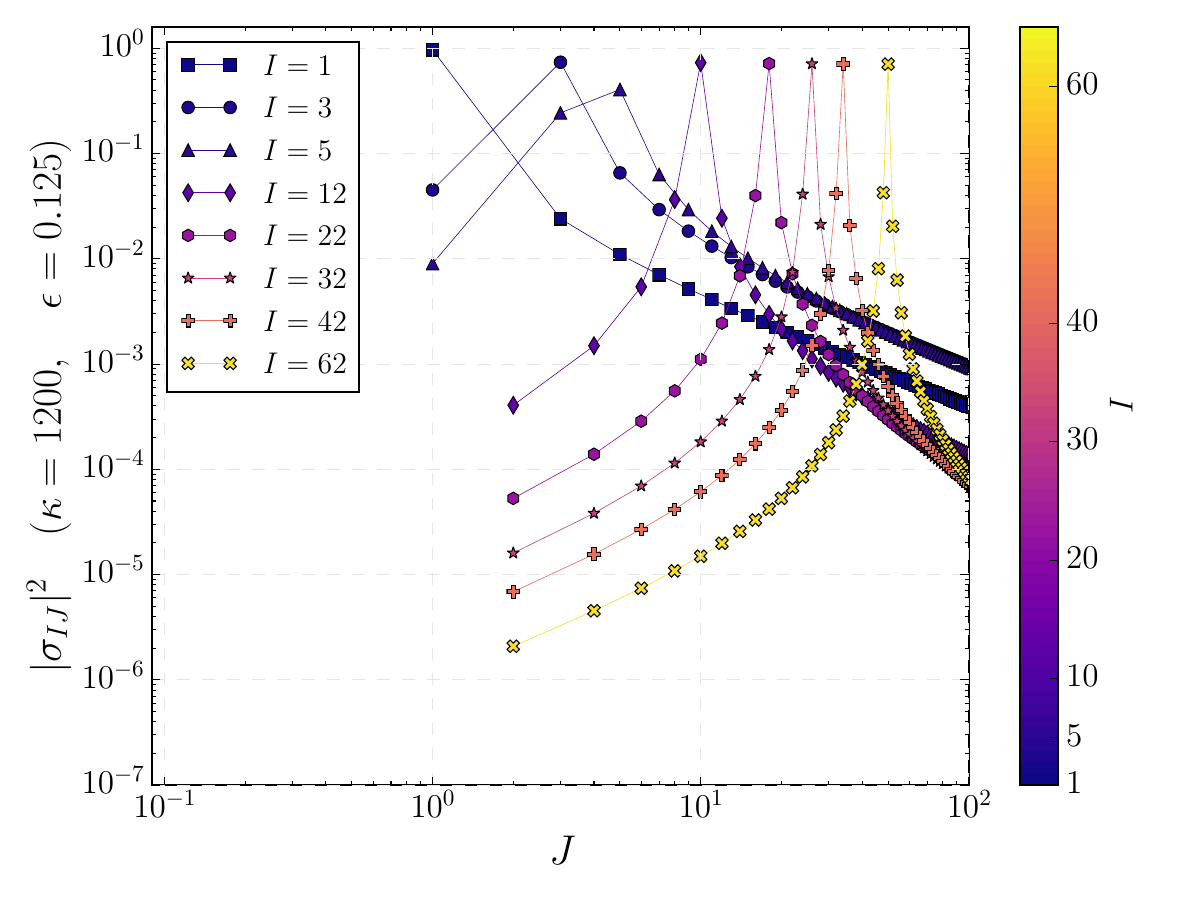}\includegraphics[width = 0.48\textwidth]{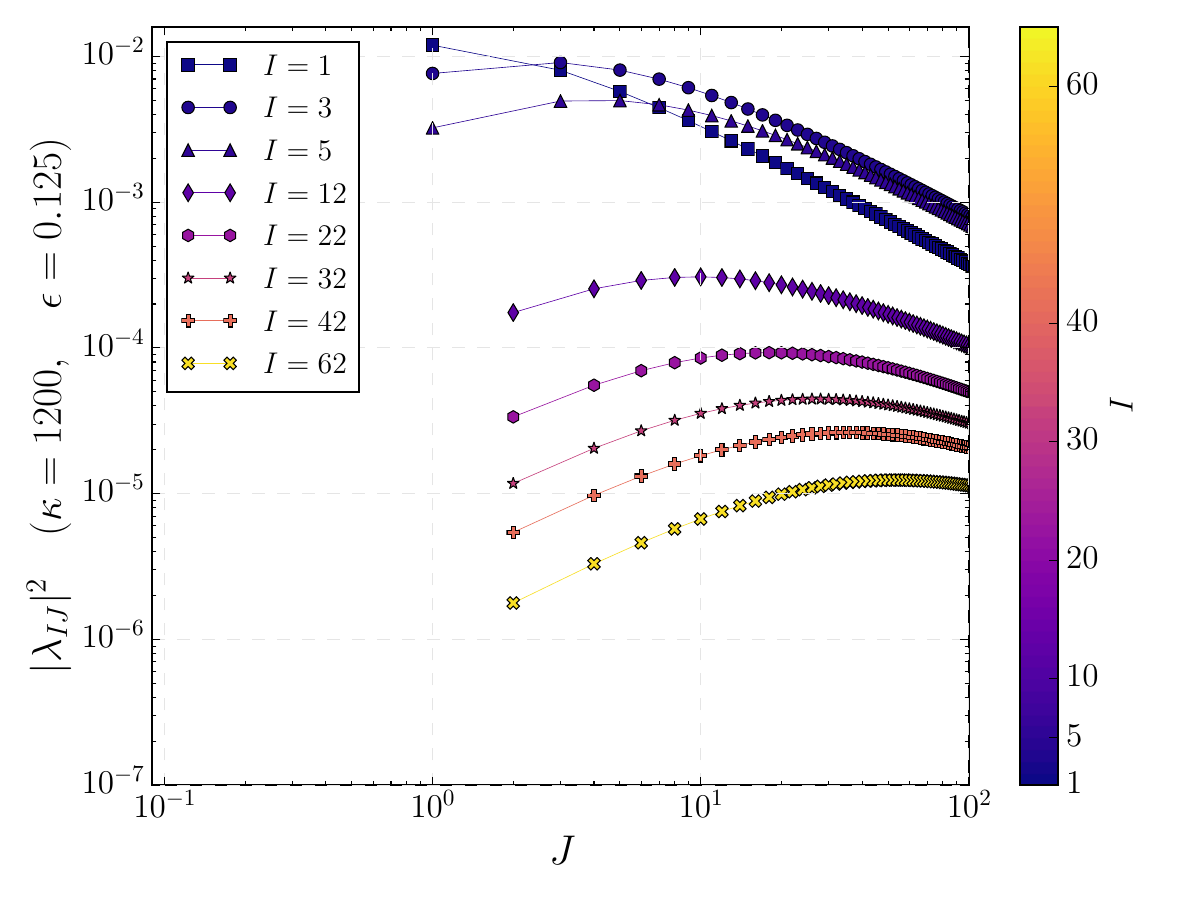}
}
\caption{\justifying Bogoliubov coefficients: These plots correspond to the trajectories given by a time reversed Eq. \eqref{eq:Traj-expand}, but now with $\epsilon=0.125$ and $\kappa=1200$. The lines interpolating points do not correspond to any fitting expression. The plots also show different markers for each value of $I$. }
\label{fig:expand-collapse-sharp}
\end{figure}

We consider several values of {\it in} modes $I$ for $|\sigma_{IJ}|^2$ and $|\lambda_{IJ}|^2$. As we see in Fig. \ref{fig:expand-collapse-sharp}, there is no mixing between even and odd modes. In addition, $|\sigma_{IJ}|^2$ behaves qualitatively similarly to the case of small accelerations, namely, it grows as a function of $J$ until it reaches a peak at $J\simeq I$, and then decreases. We do not see oscillations in the extrapolated data. However, we do see oscillations in the UV sector (large values of $J$) for each simulation (fixed number of modes $N$). The behavior of $|\sigma_{IJ}|^2$ indicates a strong mode mixing. On the other hand, $|\lambda_{IJ}|^2$ has a nontrivial behavior as functions of $J$ ({\it out} modes) for given values of $I$. We also see that increasing the value of $I$ implies smaller values of $|\lambda_{IJ}|^2$. Hence, there is less particle creation in the UV sector, as expected. We also see that they reach a maximum value around $J\simeq I$. Finally, for this particular trajectory we have observed that \emph{in} modes $J=10,20,30,\ldots$ behave differently from the rest. Concretely, these modes suffer very little mode mixing and particle production in the infrared sector, only at UV frequencies. This is why we show other closer modes in Fig. \ref{fig:expand-collapse-sharp} which turn out to be more representative.

Regarding the Richardson extrapolation for $|\sigma_{IJ}|^2$ and $|\lambda_{IJ}|^2$, we do see quite good convergence even for high values of both $I$ and $J$.

\section{Numerical analysis for expanding-collapsing cavities}
\label{Sec:expand-collap-cav}
%------------------------------------------------
In this section, we explore the last family of configurations we will consider in this paper. Concretely, we can consider an expanding cavity followed by a subsequent collapsing cavity in such a way that the cavity reaches its original configuration in the end. These configurations have been recently discussed in Ref. \cite{Mujtaba:2024vmf} for Minkowski spacetimes with one moving boundary. 

In order to proceed, we can either carry out the whole simulation numerically, or take advantage of the simulations we have performed for the expanding cavity, as we did for the case of collapsing cavities, along with a composition of two Bogoliubov transformations. We will proceed this way here. Specifically, we need to compose two Bogoliubov transformations, one consisting of the \emph{in} vacuum evolved to a time where the cavity has expanded (or collapsed), and the one of the \emph{out} vacuum state evolved backwards in time from the final collapsed (expanded) cavity to the time where we match with the evolved \emph{in} state. We discuss this in more detail in the following. 

Let us express the \emph{in} basis in some auxiliary basis at some intermediate instant where the cavity is neither expanding nor collapsing:
\begin{equation}\label{eq:bogou2}
   {}^{(in)}{\bf u}^{(I)}(t) = \sum_{J=1}^{\infty} \alpha_{IJ}\;{}^{(aux)}{\bf u}^{(J)}(t)+\beta_{IJ}\;{}^{(aux)}\bar {\bf u}^{(J)}(t),
\end{equation}
Now, in order to deal with the collapsing evolution, the auxiliary basis is evolved to times where the \emph{out} basis is the standard one and has the form
\begin{equation}\label{eq:bogow2}
   {}^{(aux)}{\bf u}^{(I)}(t) = \sum_{J=1}^{\infty} \gamma_{IJ}\;{}^{(out)}{\bf u}^{(J)}(t)+\rho_{IJ}\;{}^{(out)}\bar {\bf u}^{(J)}(t).
\end{equation}
We can compose both transformations so that we get
\begin{eqnarray}\label{eq:bogotot}\nonumber
   && {}^{(tot)} \alpha_{IJ}= \sum_{K=1}^{\infty} \alpha_{IK}\gamma_{KJ}+\beta_{IK}\bar\rho_{KJ},\\
   && {}^{(tot)} \beta_{IJ}= \sum_{K=1}^{\infty} \alpha_{IK}\rho_{KJ}+\beta_{IK}\bar\gamma_{KJ}.
\end{eqnarray}
However, we must note that $\gamma_{IJ}$ and $\rho_{IJ}$ are not independent of $\alpha_{IJ}$ and $\beta_{IJ}$. Actually, we can consider the {\it out} basis in the asymptotic future evolved backwards to the intermediate time where the cavity is neither expanding nor collapsing, and express it in the auxiliary basis. The corresponding transformation will be given by 
%------------------------------------------------
\begin{equation}\label{eq:bogow3}
    {}^{(out)}\bar{\bf u}^{(I)}(t) = \sum_{J=1}^{\infty} \alpha_{IJ}\;{}^{(aux)}\bar{\bf u}^{(J)}(t)+\beta_{IJ}\;{}^{(aux)} {\bf u}^{(J)}(t).
\end{equation}
Note that the elements of the \emph{out} basis, ${}^{(out)}\bar{\bf u}^{(I)}$, (and also those of the \emph{aux} basis) have positive norm when evolved backward in time, while ${}^{(out)}{\bf u}^{(I)}$ have negative norm in this backwards evolution, since the time direction is reversed. 

Now, we can invert Eq. \eqref{eq:bogow3} getting
\begin{equation}\label{eq:bogow4}
    {}^{(aux)}{\bf u}^{(I)}(t) = \sum_{J=1}^{\infty} \alpha_{JI}\;{}^{(out)}\bar{\bf u}^{(J)}(t)-\bar \beta_{JI}\;{}^{(out)} {\bf u}^{(J)}(t).
\end{equation}
Hence, identifying Eqs. \eqref{eq:bogow2} and \eqref{eq:bogow4} we conclude that
\begin{equation}\label{eq:gamma-rho-vs-a-b}
   \gamma_{IJ}=\alpha_{JI},\quad \rho_{IJ}=-\bar\beta_{JI}.
\end{equation}
In total, after one cycle, the \emph{in} basis can be written in terms of the \emph{out} basis as 
\begin{equation}\label{eq:bogou2}
   {}^{(in)}{\bf u}^{(I)}(t) = \sum_{J=1}^{\infty} {}^{(tot)} \alpha_{IJ}\;{}^{(out)}{\bf u}^{(J)}(t)+{}^{(tot)} \beta_{IJ}\;{}^{(out)}\bar {\bf u}^{(J)}(t),
\end{equation}
with the Bogoliubov coefficients given in Eq. \eqref{eq:bogotot} with $\gamma$ and $\rho$ expressed as in Eq.\eqref{eq:gamma-rho-vs-a-b}.

%------------------------------------------------
\subsection{One (symmetrically) expanding and collapsing mirror}
\label{Sec:1plt}
%------------------------------------------------

The trajectories of the boundaries of the cavity are given by Eq. \eqref{eq:Traj-1plt} from some time $t \simeq t_0$ up to some time $t \simeq t_f$, where the mirror is (nearly) static, and then from $t \simeq t_f$ up to some time $t \simeq 2t_f$ following the time-reversed trajectory \eqref{eq:Traj-1plt}. Hence, the mirror reaches its original position following a time-symmetric trajectory. Besides, the left boundary remains at $x^f=0$ at all times. In summary, the right mirror expands from $x^g_{in}=1$ to $x^g_{int} =(1+\epsilon)$ and then collapses to $x^g_{out}=1$. Here, we fix Eqs. \eqref{eq:betas-fit} and \eqref{eq:alphas-fit} with normalization $N_\beta=2=N_\alpha$ and we have the same \emph{in} and \emph{out} frequencies given by $\Delta\omega_I=\pi/L_0$. 

\begin{figure}[ht]
{\centering     
  \includegraphics[width = 0.48\textwidth]{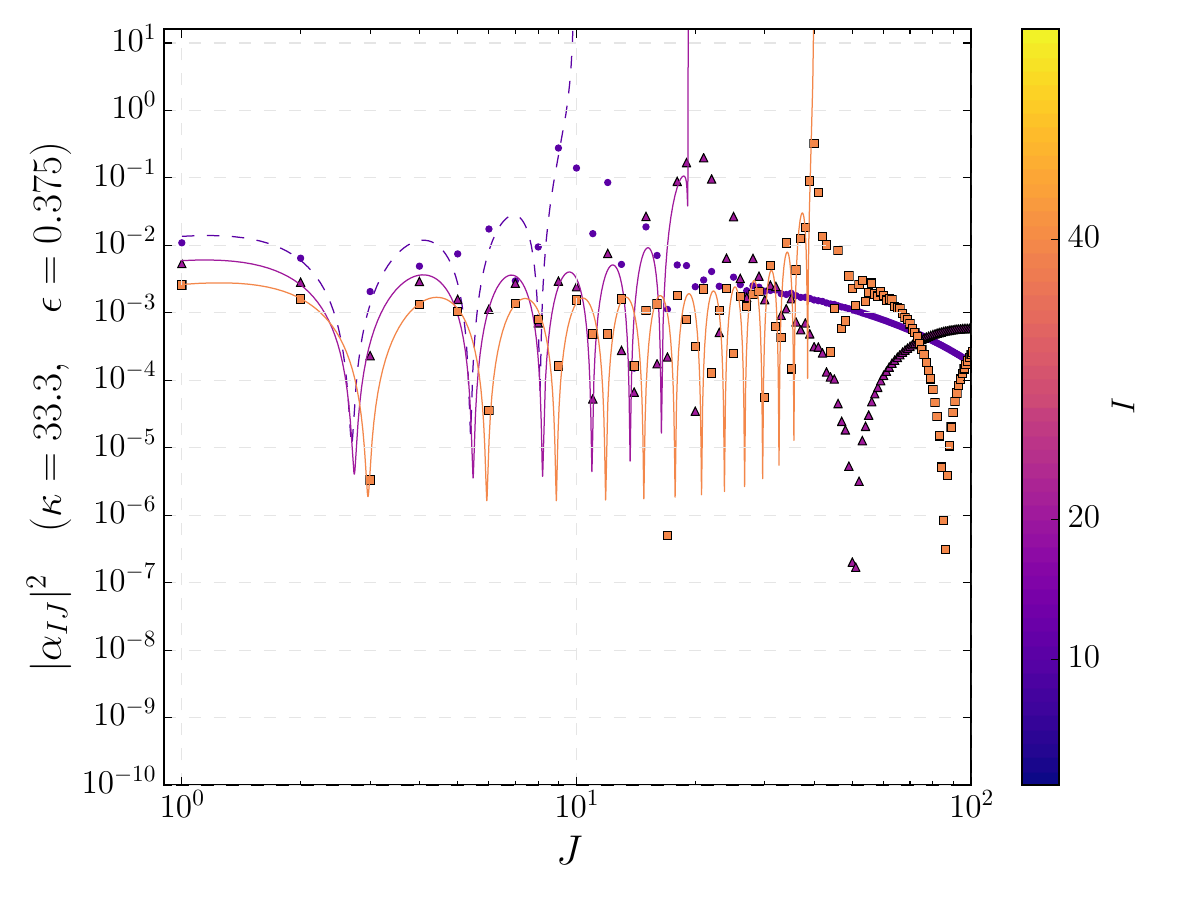}\includegraphics[width = 0.48\textwidth]{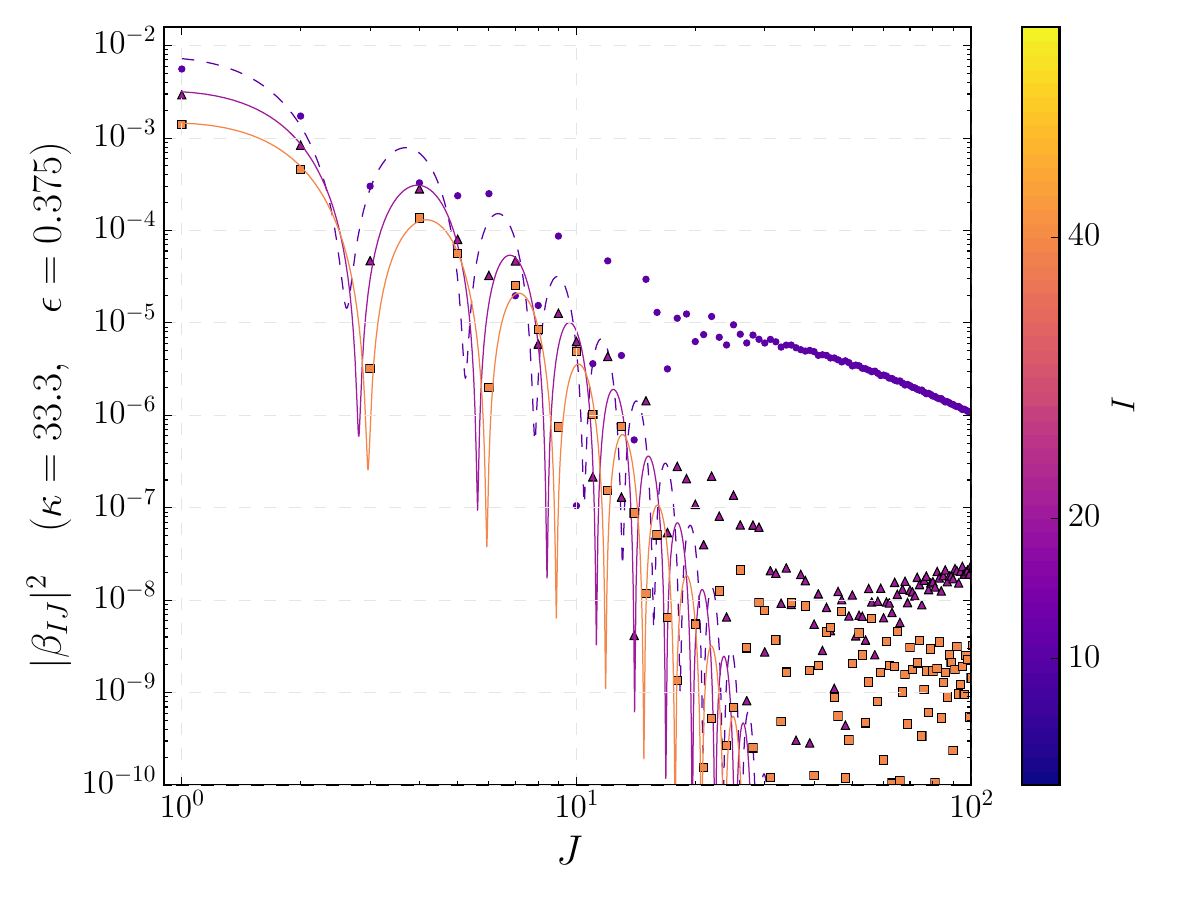}
  %\\\includegraphics[width = 0.48\textwidth]{plots/alphas2-1plt-k33-dL0375-40-115.pdf}
  %\includegraphics[width = 0.48\textwidth]{plots/betas2-1plt-k33-dL0375-40-115.pdf}
}
\caption{\justifying Bogoliubov coefficients: These plots correspond to a boundary undergoing a trajectory given by Eq. \eqref{eq:Traj-1plt} and then its time reversed trajectory returning to its initial position, with $\epsilon=0.375$ and $\kappa=33.3$. The lines interpolating points correspond to the fitting expressions in Eqs. \eqref{eq:betas-fit} and \eqref{eq:alphas-fit} for the right and left panels, respectively.}
\label{fig:1plt-expcoll}
 \end{figure}

For the sake of brevity, we will show here a concrete realization given by $\epsilon=0.375$ and $\kappa=33.3$. In Fig. \ref{fig:1plt-expcoll} we show the behavior of the Bogoliubov coefficients for the same frequency bands as in Fig. \ref{fig:1plt}. We also compare with the fitting expressions of Eqs. \eqref{eq:betas-fit} and \eqref{eq:alphas-fit}, only for infrared modes (UV modes show a different behavior, and we did not find a simple fitting expression). \footnote{We fix $A_\alpha=1=A_\beta$, and $B_\alpha=10^{-3}=B_\beta$. The other parameters $D_1$, $C_\alpha=C_\beta$ and $F$ are suitable functions of $\omega_J$ similar to the ones shown in Fig. \ref{fig:1plt-CDF-plot}. Hence, we do not show them here.} We surprisingly see that there is qualitative agreement between the two cases. This demonstrates that the thermal spectrum shows robustness against particle production because of the return of the right mirror to its original position following the time-reversed trajectory. On the other hand, the UV sector is sensitive to this new configuration of the cavity, with a stronger particle production in the UV modes.

%------------------------------------------------
\subsection{One (asymmetrically) expanding and collapsing mirror}
\label{Sec:1plt}
%------------------------------------------------

\begin{figure}[ht]
{\centering     
  \includegraphics[width = 0.48\textwidth]{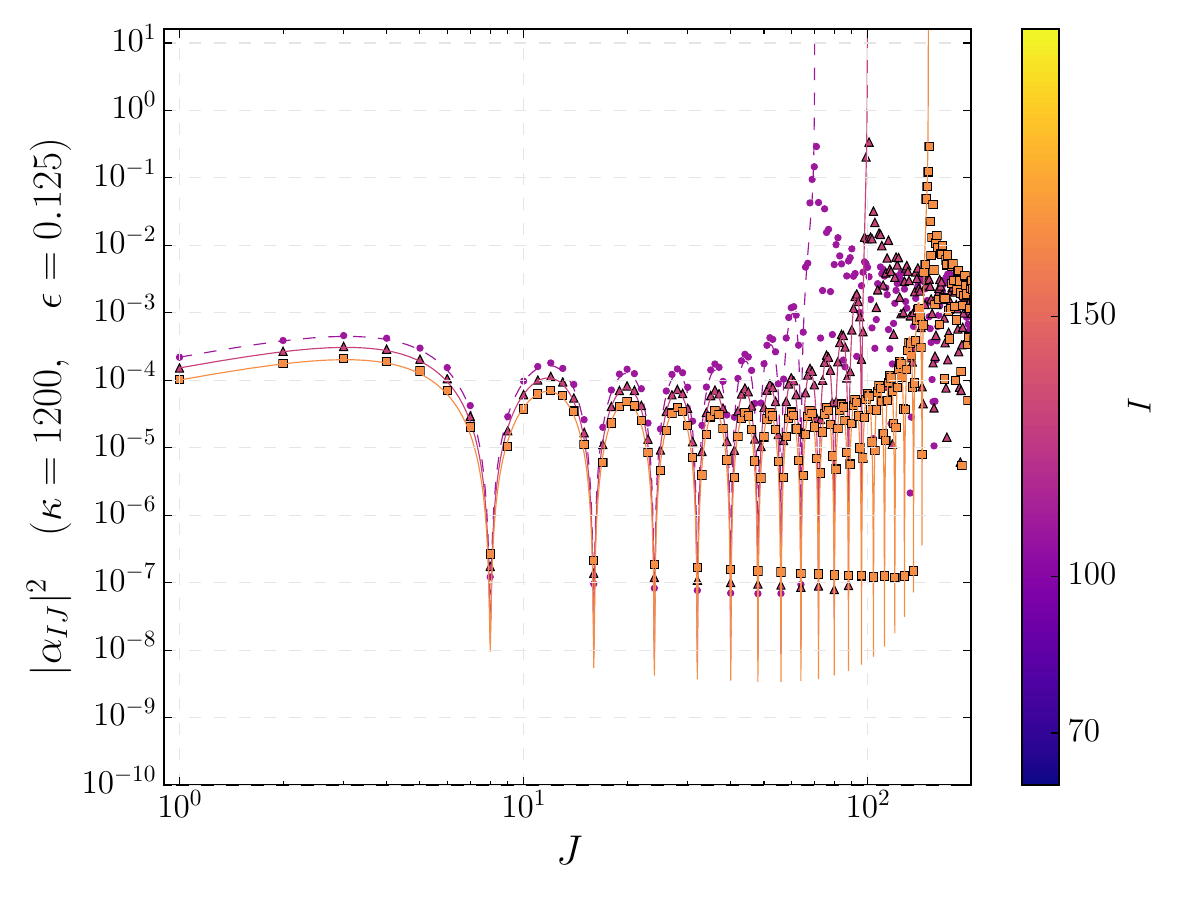}\includegraphics[width = 0.48\textwidth]{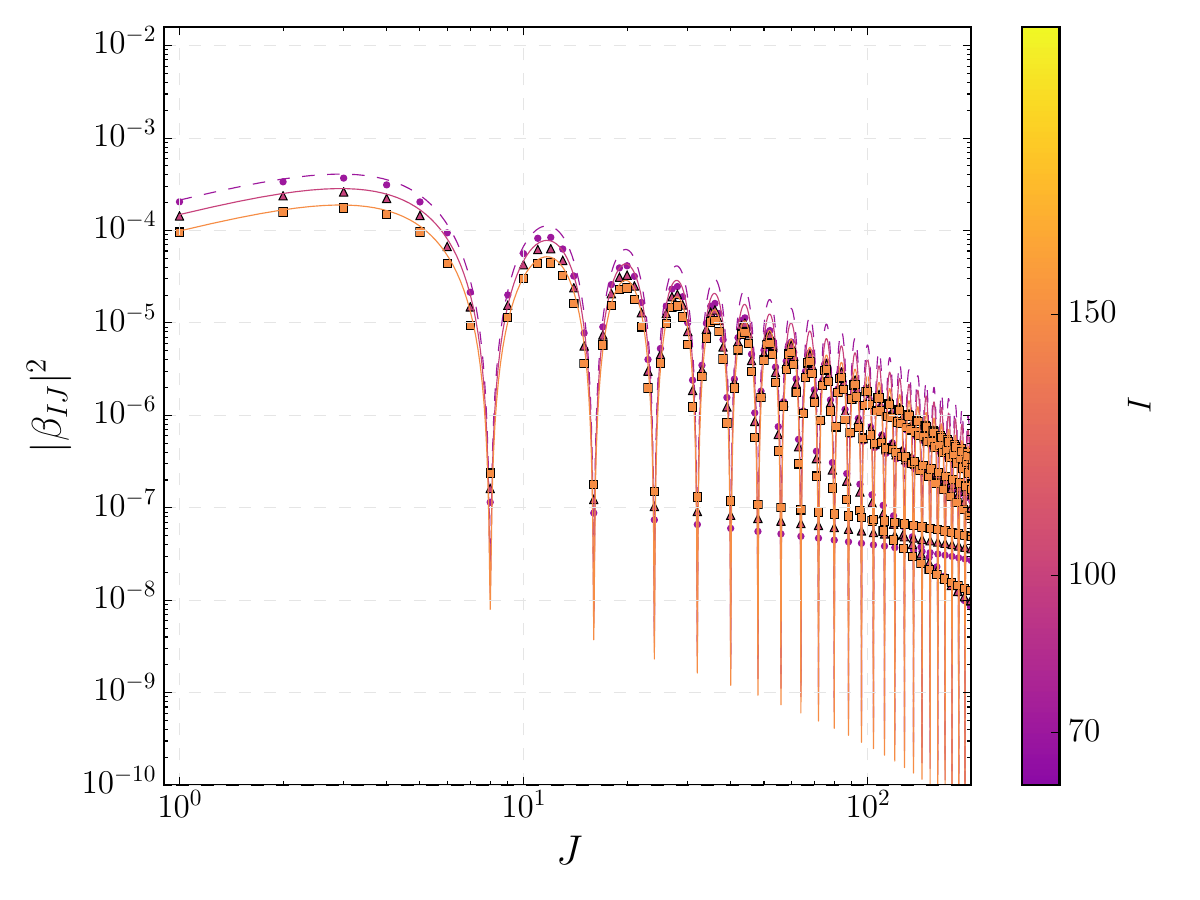}
  \\\includegraphics[width = 0.48\textwidth]{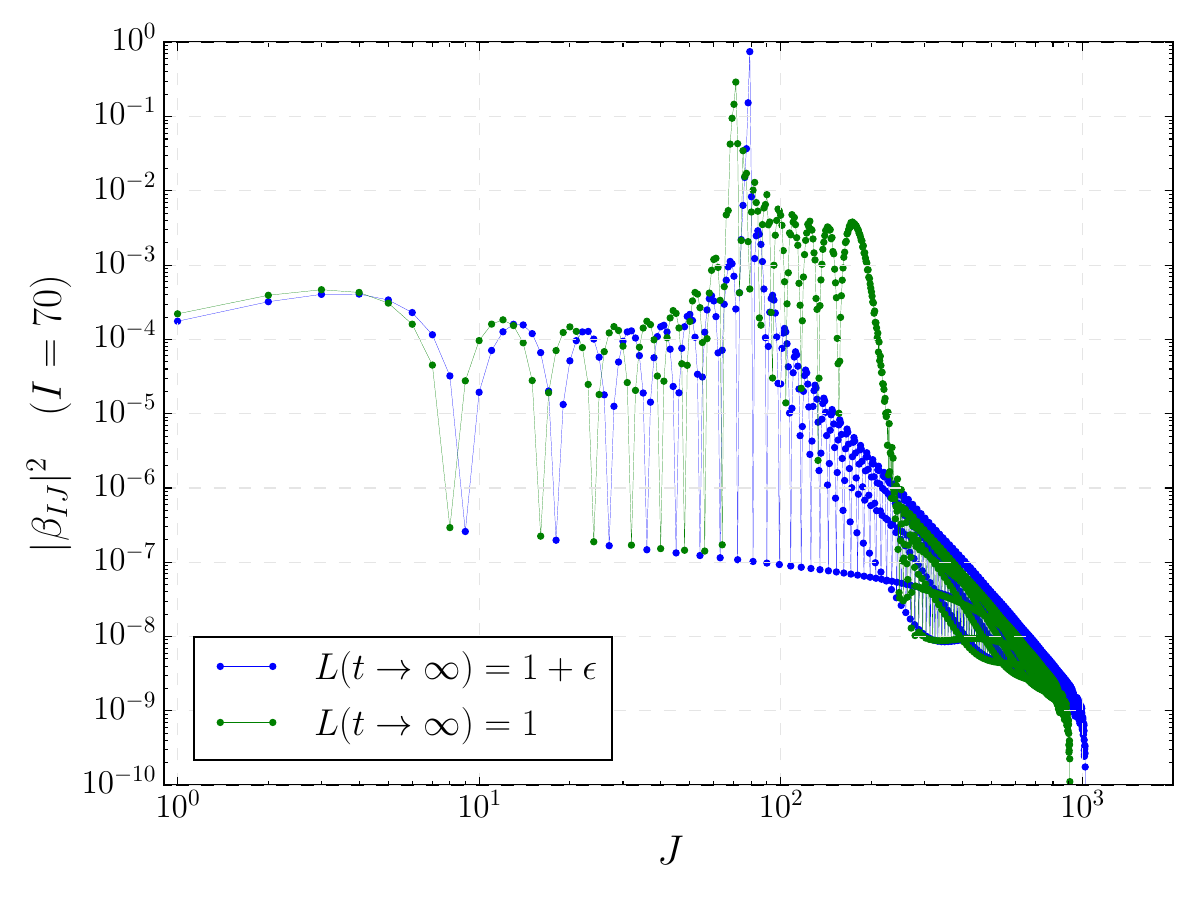}
  \includegraphics[width = 0.48\textwidth]{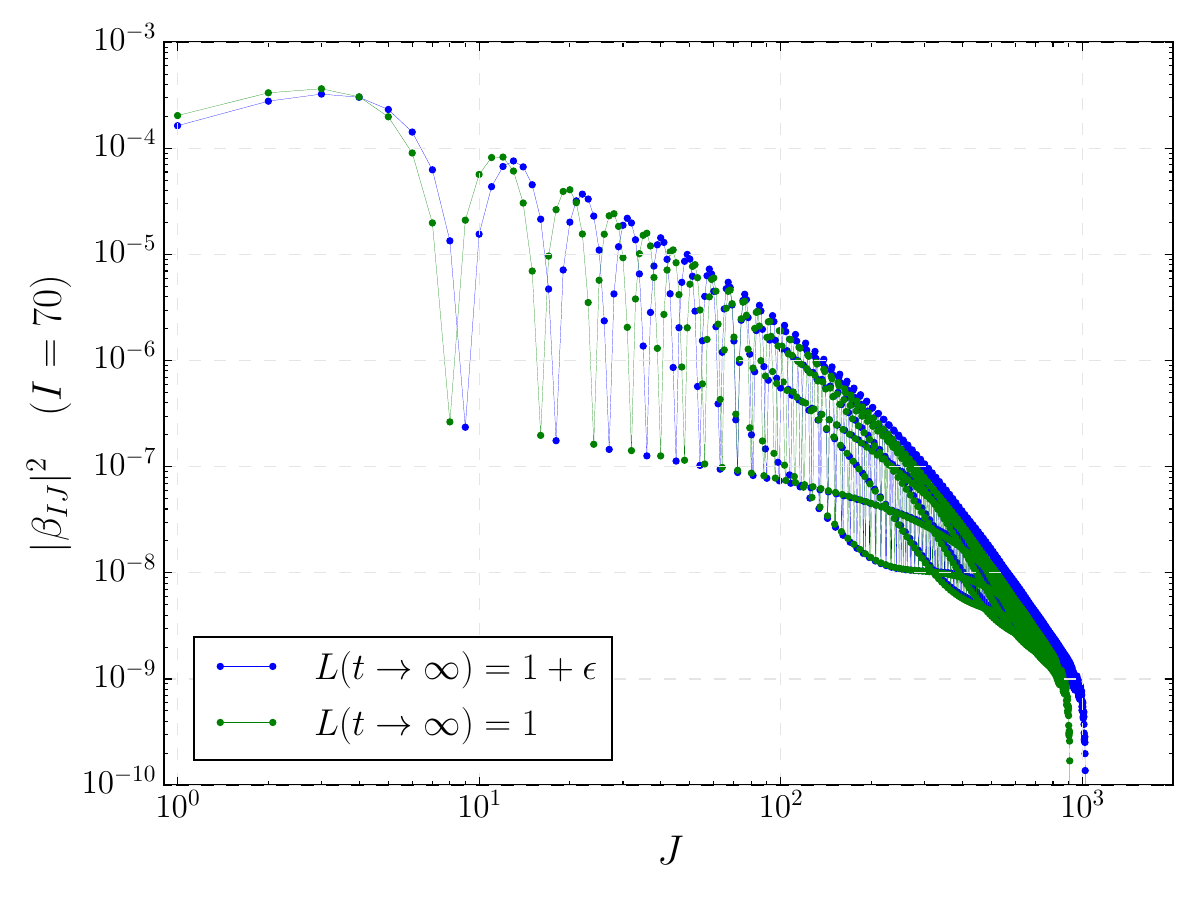}
}
\caption{\justifying Bogoliubov coefficients: These plots correspond to a boundary undergoing a trajectory given by Eq. \eqref{eq:Traj-1plt} with with $\epsilon=0.125$ and $\kappa=150$ and then it returns to its initial position following the same trajectory but with with $\epsilon=0.125$ and $\kappa=8$. Upper panel: We show the Bogoliubov coefficients in the limit $N\to\infty$. The lines interpolating points correspond to the fitting expressions in Eqs. \eqref{eq:betas-fit} and \eqref{eq:alphas-fit} for the right and left panels, respectively. Lower panel: We show the Bogoliubov coefficients for $N\to\infty$ and compare the cases where the cavity undergoes an expansion and stops at $L_f=1+\epsilon$ and when the cavity eventually returns to its original size with $L_f=1$}. 
\label{fig:1plt-expcoll2}
 \end{figure}

Now that we have seen that the infrared sector is robust against a time-symmetric collapse of the cavity returning to its original position, we will analyze what happens if the cavity returns to its position but following a different collapsing trajectory. Concretely, we are interested in a cavity that undergoes a sharp expansion and then an adiabatic collapse such that it returns to its original position, namely, the returning trajectory follows a small acceleration trajectory. As we see in the upper panels of Fig. \ref{fig:1plt-expcoll2}, the Bogoliubov coefficients in the limit $N\to\infty$ present a few (expected) qualitative differences with respect to the case of a cavity that expands and stops, without returning to its original position ---see Fig. \ref{fig:1plt-sharp}. Concretely, the parameters of the fitting expression remain the same. We must just take into account that the \emph{out} frequencies are different. This is translated into a small change in the amplitude of the coefficients and the corresponding modification in the oscillations of the gray-body factor. In addition, the strongest modifications appear in the ultraviolet part of the spectrum. Unfortunately, we cannot take the limit $N\to\infty$ via Richardson extrapolation, due to the appearance of spurious effects in the UV regime.
Therefore, in the lower panels of Fig. \ref{fig:1plt-expcoll2}, we show the Bogoliubov coefficients for a simulation with $N=1024$, for the two cases: a cavity that expands and does not return to its original position and a cavity that expands and returns adiabatically to its initial position. Here, we clearly see how the $\alpha$-coefficients get strong modifications in an intermediate band frequency, whereas we do not see considerable particle production since the $\beta$-coefficients remain nearly the same. Hence, the following question arises naturally: \emph{How robust is the thermal spectrum against several cycles of expansion and collapse of the cavity?} We will address this question in the next subsection.

%------------------------------------------------
\subsection{One (asymmetrically) expanding and collapsing mirror after several cycles}
\label{Sec:1plt}
%------------------------------------------------

For the sake of simplicity, we compute the evolution by composing the Bogoliubov transformation in Eq.~\eqref{eq:bogou2} a number of times. Concretely, two cycles will give the Bogoliubov transformation
\begin{equation}\label{eq:bogou3}
   {}^{(in)}{\bf u}^{(I)}(t) = \sum_{J=1}^{\infty} {}^{(tot,2)} \alpha_{IJ}\;{}^{(out)}{\bf u}^{(J)}(t)+{}^{(tot,2)} \beta_{IJ}\;{}^{(out)}\bar {\bf u}^{(J)}(t),
\end{equation}
with
\begin{eqnarray}\label{eq:bogotot3}\nonumber
   && {}^{(tot,2)} \alpha_{IJ}= \sum_{K=1}^{\infty} {}^{(tot)}\alpha_{IK}\;{}^{(tot)}\alpha_{KJ}+{}^{(tot)}\beta_{IK}\;{}^{(tot)}\bar\beta_{KJ},\\
   && {}^{(tot,2)} \beta_{IJ}= \sum_{K=1}^{\infty} {}^{(tot)}\alpha_{IK}\;{}^{(tot)}\beta_{KJ}+{}^{(tot)}\beta_{IK}\;{}^{(tot)}\bar\alpha_{KJ}.
\end{eqnarray}
Any desired number of cycles can be computed recursively from this expression. In general, the Bogoliubov transformation of a cycle $n+1$ on the left-hand side can be calculated from that of a cycle $n$ with $n\geq 2$ by plugging it on the right-hand side of Eq. \eqref{eq:bogotot3}. 

\begin{figure}[ht]
{\centering     
  \includegraphics[width = 0.48\textwidth]{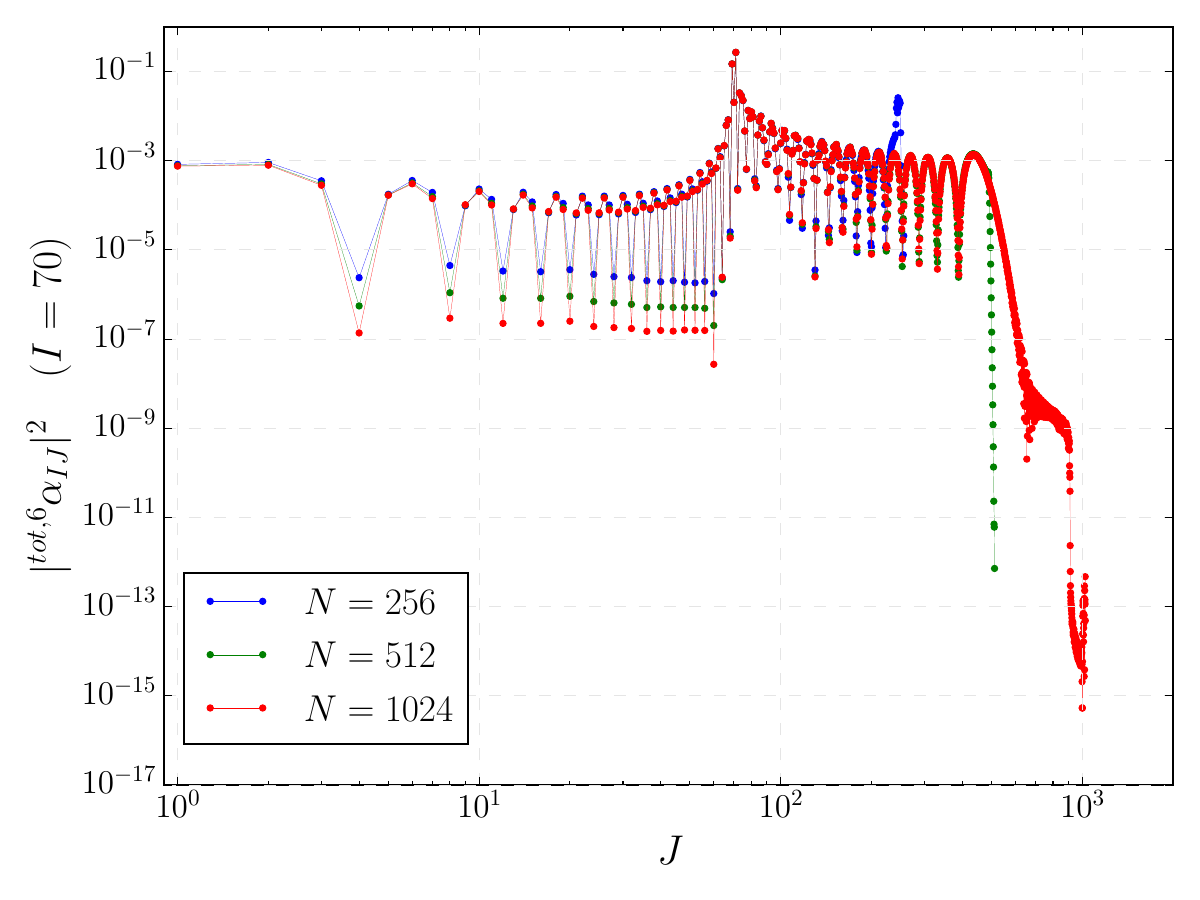}\includegraphics[width = 0.48\textwidth]{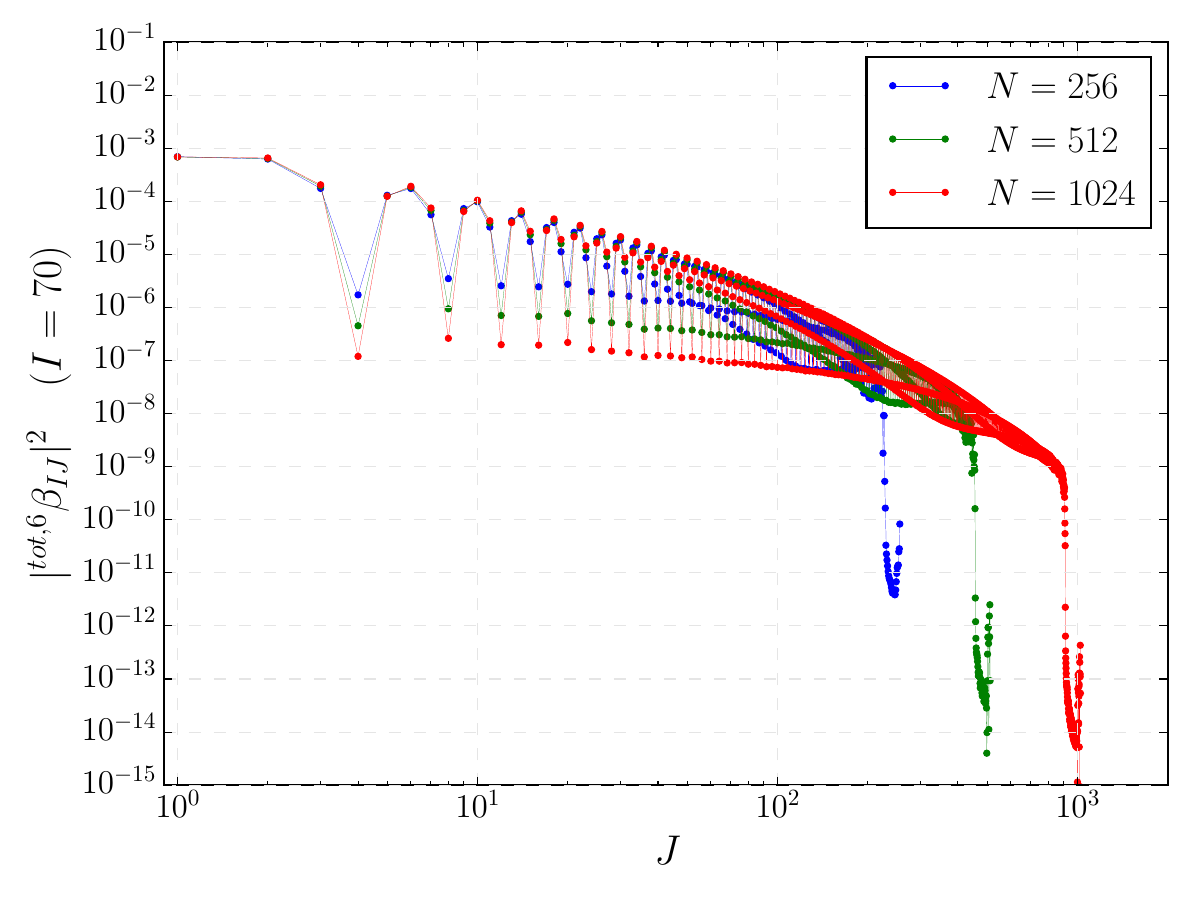}
  \\\includegraphics[width = 0.48\textwidth]{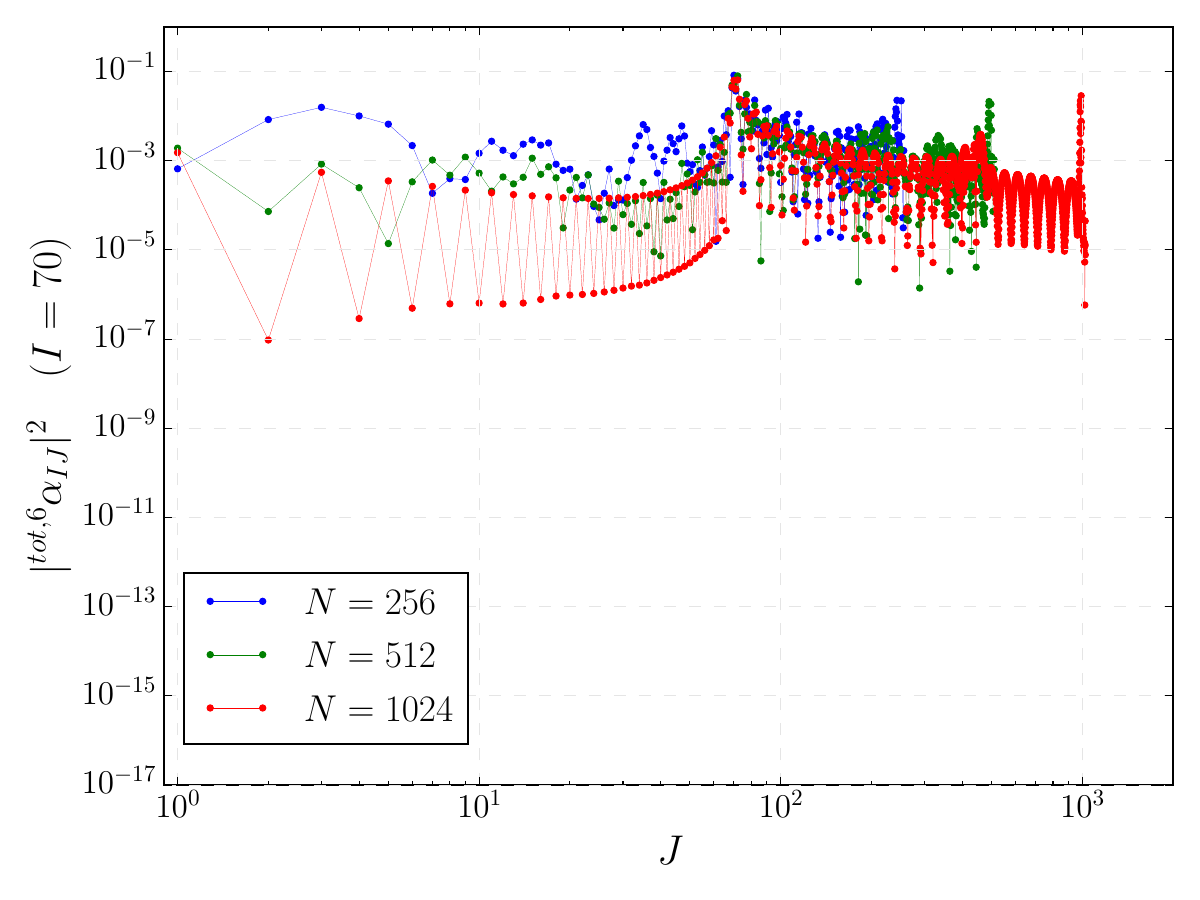}
  \includegraphics[width = 0.48\textwidth]{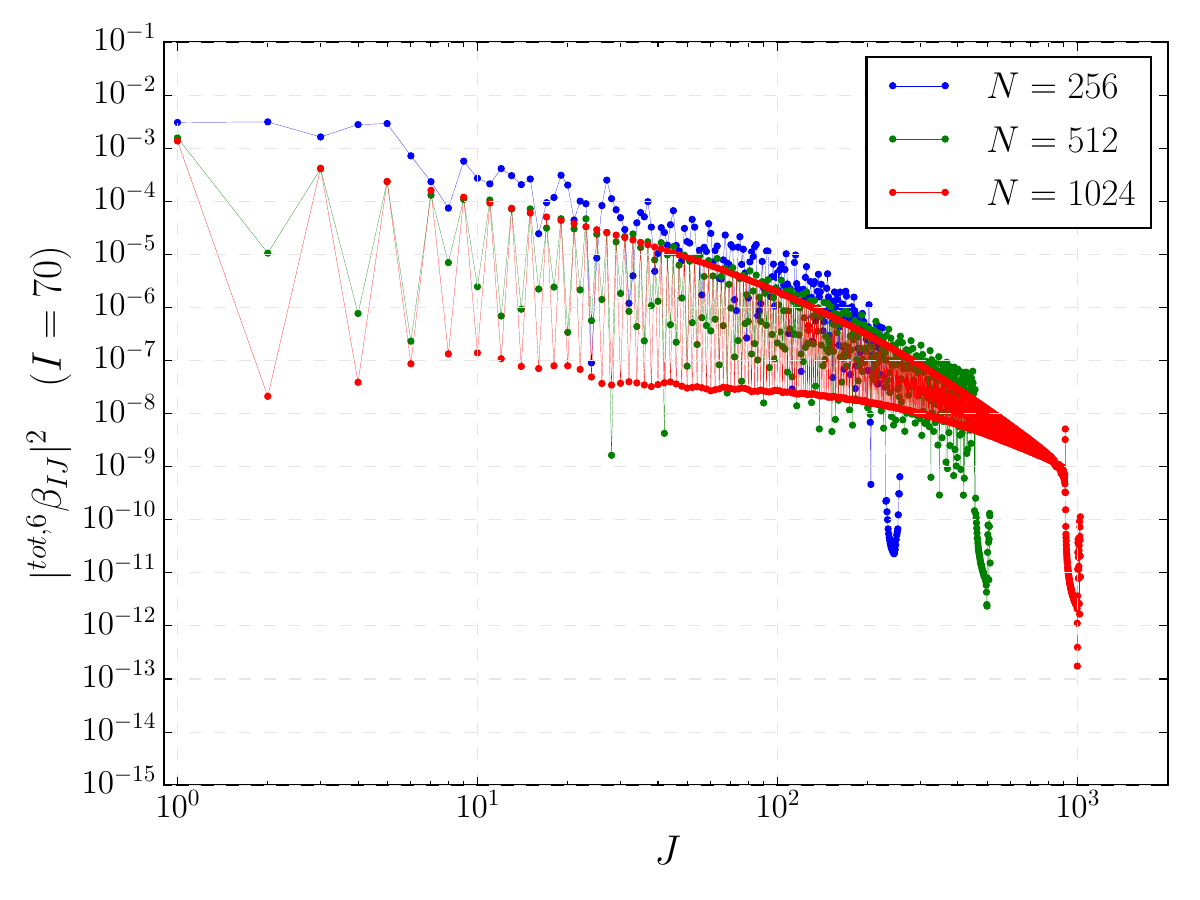}
}
\caption{\justifying Bogoliubov coefficients: These plots correspond to a boundary undergoing a trajectory given by Eq. \eqref{eq:Traj-1plt} with with $\epsilon=0.125$ and $\kappa=150$ and then it returns to its initial position following the same trajectory but with with $\epsilon=0.125$ and $\kappa=8$, repeating this cycle two times (upper panels) and three times (lower panels). On the left panels we can find the $\alpha$-coefficients and in the right panels the $\beta$-coefficients.}  
\label{fig:1plt-expcolln}
 \end{figure}

In Fig. \ref{fig:1plt-expcolln} we show the Bogoliubov coefficients for the mirror undergoing two cycles of expansion and contraction in the upper panels and four cycles in the lower panels. As we can see, the thermal structure is qualitatively present if the number of cycles is less than three. However, we can see that we lost convergence in the infrared sector for different values of $N$ already at the third cycle. Actually, although we do not show the results here, we have seen that as one increases the number of cycles, the amplitudes of both types of Bogoliubov coefficients grow very rapidly to the point that we lose all numerical precision, indicating a numerical (perhaps physical) instability that requires a more detailed analysis. 

\section{Conclusions}
\label{Sec:conclusions}

In this work, we have presented a rigorous and exhaustive numerical analysis on the particle production in cavities with moving reflective boundaries, using the dynamical Casimir effect as an analog to Hawking radiation. Our analysis encompasses multiple physically relevant configurations, including expanding, collapsing, concatenations of expanding/collapsing configurations, and rigidly accelerating cavities. 

We have demonstrated that even though we can describe the finite size and transient effects of the resulting production spectrum, there are situations more suitable for thermal production than others. In particular, scenarios with one mirror and two mirrors that symmetrically accelerate are more prone to describe a final thermal state of the field. Besides, we have seen that returning the cavity to its original size after expansion does not considerably modify the thermal aspects of the spectrum in these situations. Moreover, we can repeat the process of expansion and collapse a few times without qualitatively spoiling it. Other configurations do not exhibit robust thermal character.

In all the configurations studied, the emergence of a thermal spectrum is sensitive to the frequency range, as the infrared region generically exhibits a more suitable thermal distribution, and specifically to the parameters that govern the acceleration of the boundaries --namely, the acceleration strength $\kappa$ and the duration through the parameter $\varepsilon$. For all of the situations under consideration, we can see that the most energetic modes present substantial deviations with respect to the ideal thermal distribution, even for the most favorable ones. This has an impact on the potential experimental implementations of this idea, as only a certain range of frequencies will exhibit this thermal signature.

To quantitatively characterize the most important deviations from exact thermality, we introduced a set of fitting expressions for the Bogoliubov coefficients, incorporating gray-body factors that account for finite-size and transient effects. These factors exhibit oscillatory behavior as functions of frequency, arising from the finite duration of acceleration, and enable us to distinguish genuine thermal contributions from spurious particle creation due to the finite trajectories of the boundaries. We have also checked the robustness of thermal spectra under modifications such that a cavity that undergoes a collapse to its original size and have seen that the spectrum is robust provided the accelerations involved in the collapse are sufficiently small. In this case, the fitting expressions are valid for the infrared modes, provided that the $out$ modes correspond to those of the final size of the cavity. Actually, if the cavity undergoes a few cycles of expansion and collapse, most of the qualitative properties of the spectrum are preserved, but the fitting expressions are not valid anymore. If the number of cycles increases beyond three, the thermal properties of the spectrum are lost.   

This study contributes towards the possible validation of experimental implementations of Hawking-like radiation in analog models by presenting a robust and quantitative methodology to evaluate whether a given spectrum is indeed thermal, distinguishing between the genuine physical signal and the experimental spurious artifacts. This kind of robust methodology is crucial in the context of analog gravity, as it provides an agnostic method to validate the experimental results. By identifying the roles played by acceleration parameters, mode selection, and boundary configurations, we contribute to the design of more precise analog gravity experiments and deepen the understanding of how quantum field theoretic effects manifest in non-stationary backgrounds.
%------------------------------------------------

\acknowledgments

The authors would like to thank Gerardo Garc\'ia-Moreno for helpful discussions. Financial support is provided by the Spanish Government through the projects PID2020-119632GB-I00, and PID2019-105943GB-I00 (with FEDER contribution). 

\appendix

%------------------------------------------------
\section{Fitting coefficients}
\label{Appendix:fit}
%------------------------------------------------
In this section, we show the behavior of several coefficients in the fitting expressions of $|\alpha^{\text{(f)}}_{IJ}|^2$ and $|\beta^{\text{(f)}}_{IJ}|^2$. We start with the trajectory in Eq. \eqref{eq:Traj-1plt} with $\kappa=33.3$ and $\epsilon = 0.375$. We show the parameters that change as functions of $\omega_I$, that is, $C_\alpha = C_\beta$ and $F$. The other parameters are nearly constant, and their value is given in the main text of the corresponding section. 
\begin{figure}[ht]
{\centering     
  \includegraphics[width = 0.48\textwidth]{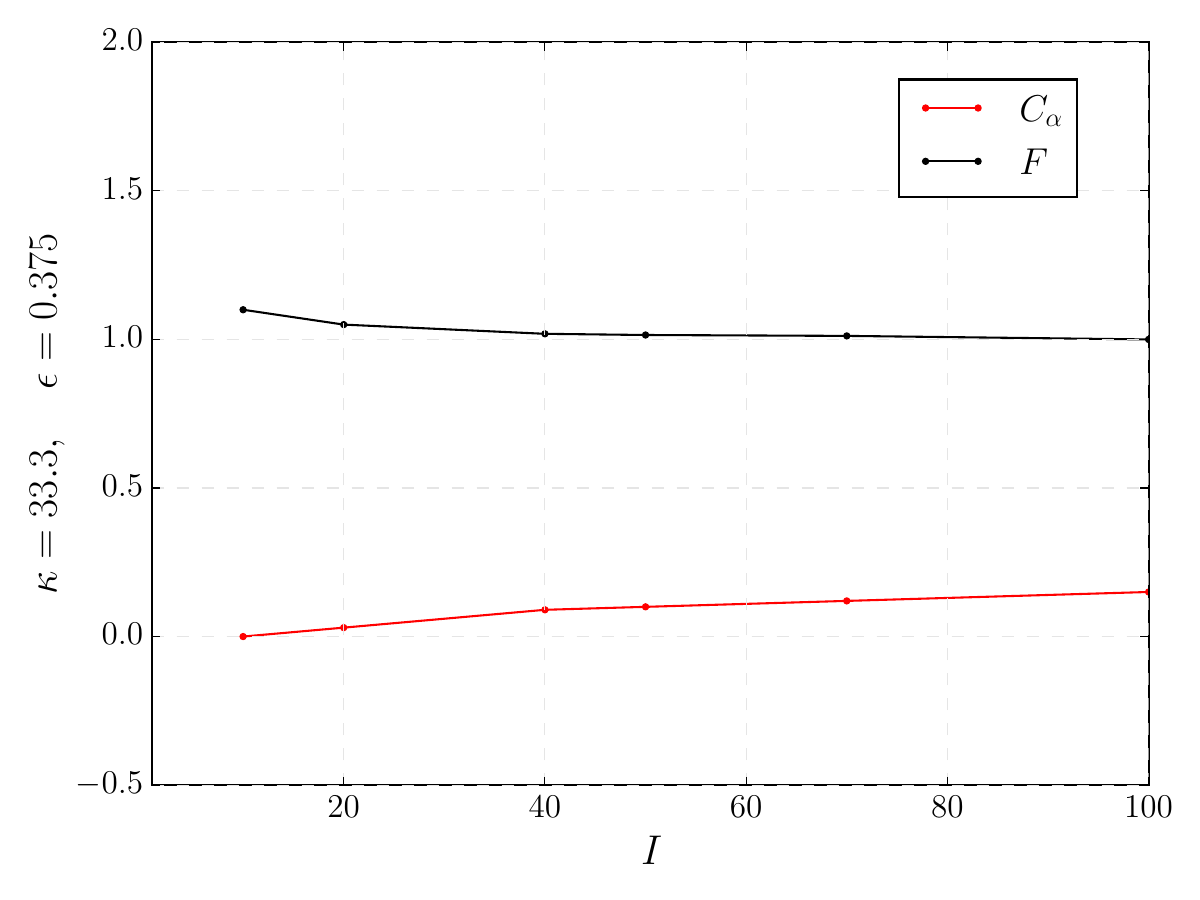}
}
\caption{\justifying Fitting parameters $C_\alpha$  and $F$ as functions of $\omega_I$. }
\label{fig:1plt-CDF-plot}
\end{figure}

For the trajectory in Eq. \eqref{eq:Traj-1plt} with $\kappa=1200$ and $\epsilon = 0.125$, again we show the parameters that change as functions of $\omega_I$. In this case, they correspond to $D_1$, $D_2$, $F$ and $\tilde \kappa$. The other parameters are nearly constant, and their value is given in the main text of the corresponding section. 
\begin{figure}[ht]
{\centering     
  \includegraphics[width = 0.48\textwidth]{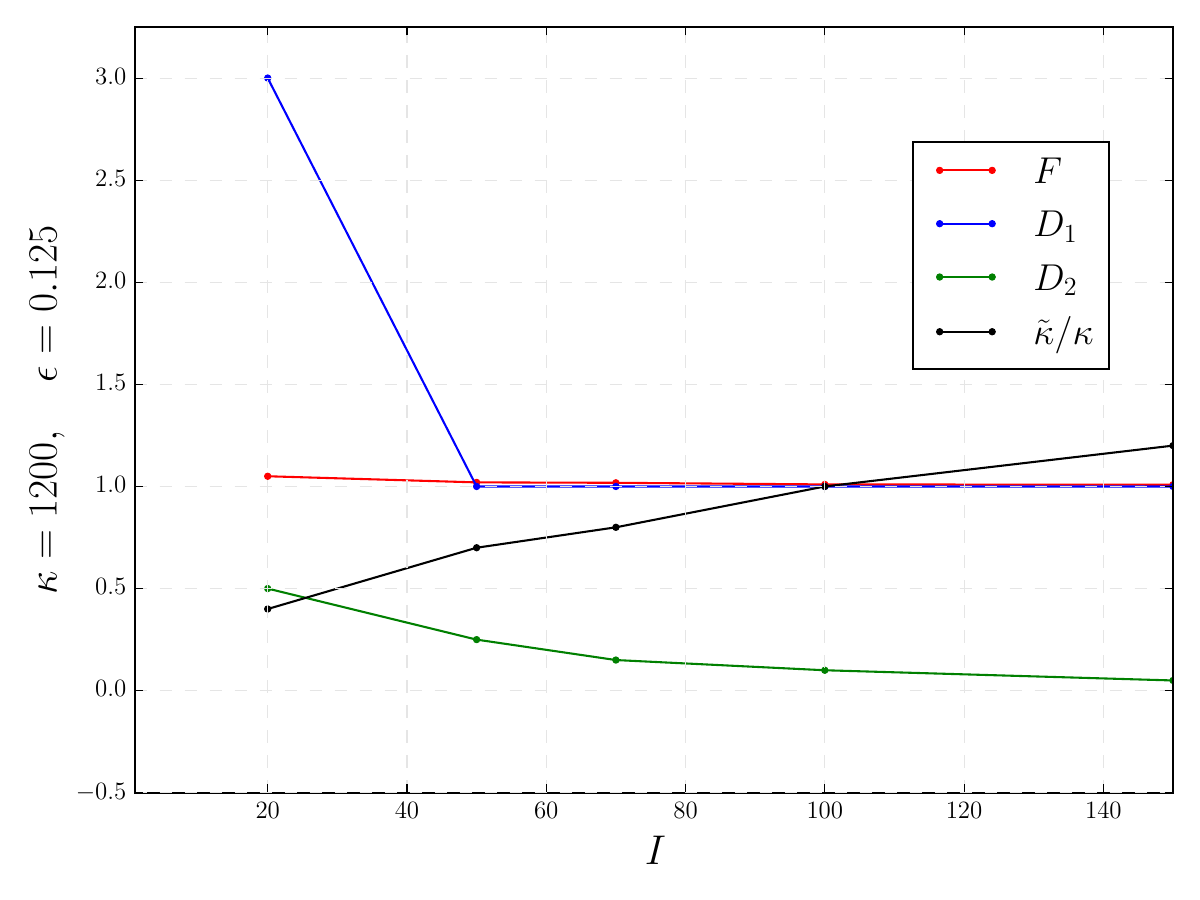}
}
\caption{\justifying Fitting parameters $D_!$, $D_2$, $F$ and $\tilde \kappa$ as functions of the $in$ modes $\omega_J$. }
\label{fig:1plt-CDF-plot-b}
\end{figure}

We now consider the trajectory in Eq. \eqref{eq:Traj-expand} with $\kappa=33.3$ and $\epsilon = 0.375$. We show the parameters that change as functions of $\omega_I$, namely, $C_\alpha = C_\beta$, $\tilde \kappa$, $D_1$, $D_2$, and $F$. The other parameters are nearly constant, and their value is given in the main text of the corresponding section. 
\begin{figure}[ht]
{\centering     
  \includegraphics[width = 0.48\textwidth]{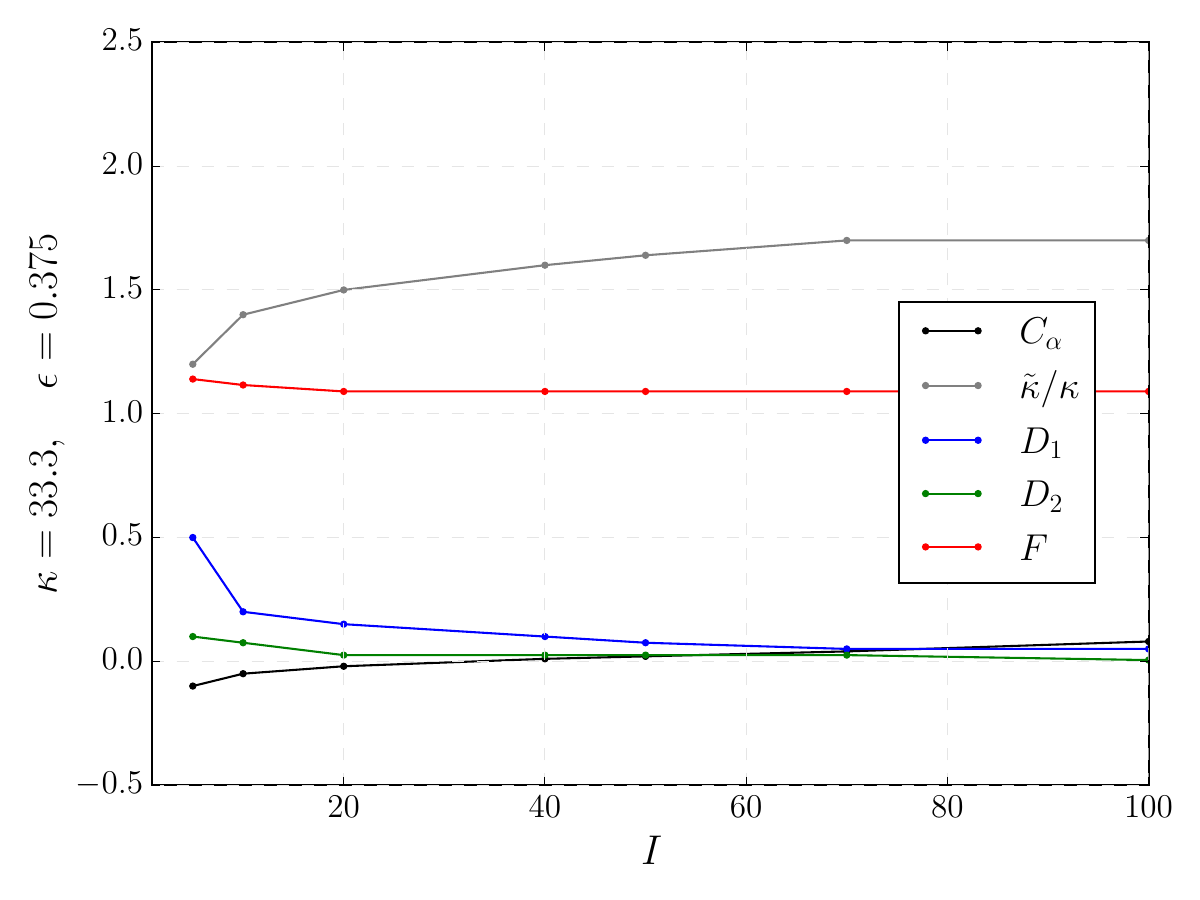}
}
\caption{\justifying Fitting parameters $C_\alpha = C_\beta$, $\tilde \kappa$, $D_1$, $D_2$, and $F$ as functions of $\omega_I$. }
\label{fig:expand-CDF-plot}
\end{figure}

For the same trajectory in Eq. \eqref{eq:Traj-expand} but with $\kappa=1200$ and $\epsilon = 0.125$ we show the parameters that change as functions of $\omega_I$, in this case $\tilde \kappa$, $D_1$, $D_2$, and $F$. The other parameters are nearly constant, and their value is given in the main text of the corresponding section. 
\begin{figure}[ht]
{\centering     
  \includegraphics[width = 0.48\textwidth]{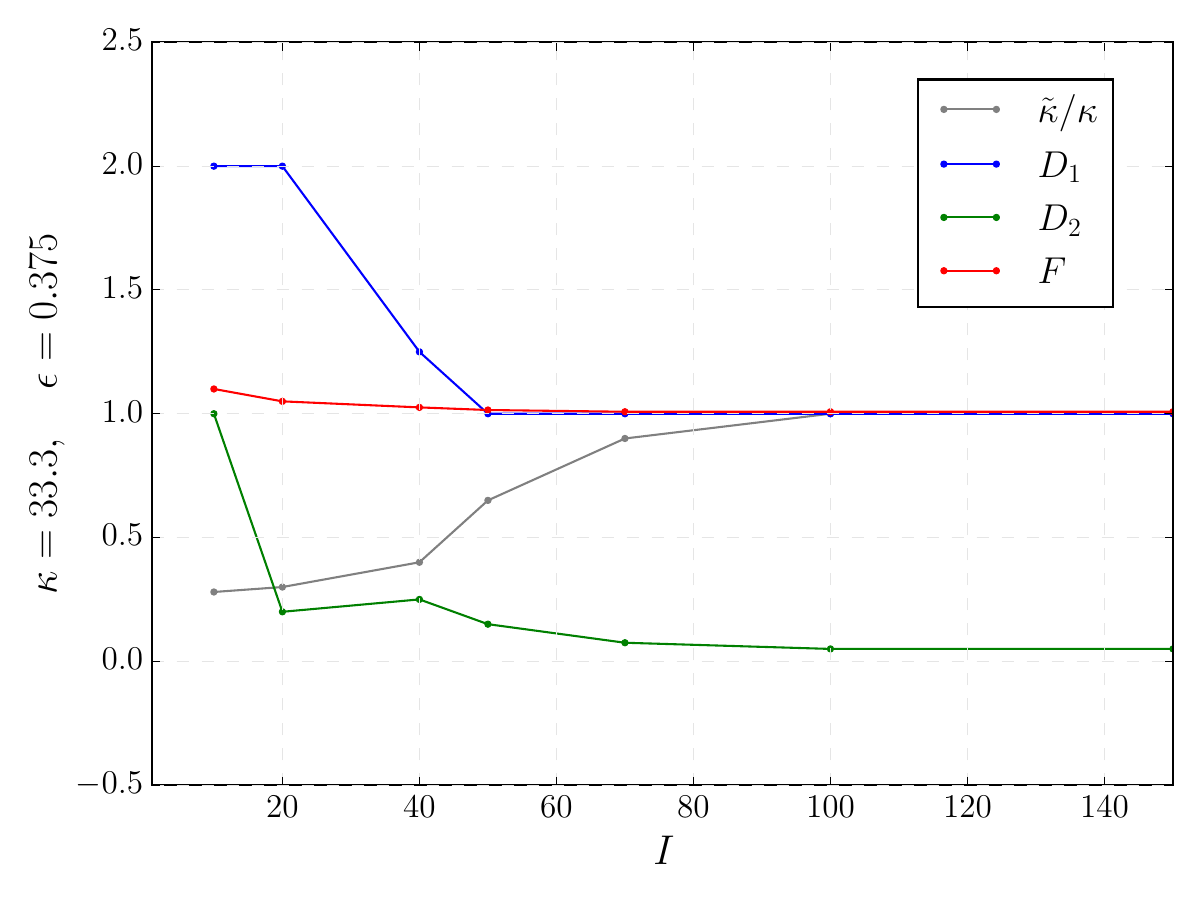}
}
\caption{\justifying Fitting parameters $\tilde \kappa$, $D_1$, $D_2$, and $F$ as functions of $\omega_I$. }
\label{fig:expand-CDF-plot-b}
\end{figure}

%------------------------------------------------
\newpage
%------------------------------------------------

%------------------------------------------------
\bibliography{draft_biblio}
%------------------------------------------------

%------------------------------------------------

%________________________________________________
\end{document}